# Settling the complexity of computing approximate two-player Nash equilibria

Aviad Rubinstein[*]

August 26, 2016

### Abstract

We prove that there exists a constant $\epsilon > 0$ such that, assuming the Exponential Time Hypothesis for PPAD, computing an $\epsilon$-approximate Nash equilibrium in a two-player $n \times n$ game requires time $n^{\log^{1-o(1)} n}$. This matches (up to the $o(1)$ term) the algorithm of Lipton, Markakis, and Mehta [54].

Our proof relies on a variety of techniques from the study of probabilistically checkable proofs (PCP); this is the first time that such ideas are used for a reduction between problems inside PPAD.

En route, we also prove new hardness results for computing Nash equilibria in games with many players. In particular, we show that computing an $\epsilon$-approximate Nash equilibrium in a game with $n$ players requires $2^{\Omega(n)}$ *oracle queries* to the payoff tensors. This resolves an open problem posed by Hart and Nisan [43], Babichenko [13], and Chen et al. [28]. In fact, our results for $n$-player games are stronger: they hold with respect to the $(\epsilon, \delta)$-WeakNash relaxation recently introduced by Babichenko et al. [15].

---

[*]UC Berkeley. I am grateful to Yakov Babichenko, Jonah Brown-Cohen, Karthik C.S., Alessandro Chiesa, Elad Haramaty, Christos Papadimitriou, Muli Safra, Luca Trevisan, Michael Viderman, and anonymous reviewers for inspiring discussions and suggestions. This research was supported by Microsoft Research PhD Fellowship. It was also supported in part by NSF grant CCF1408635 and by Templeton Foundation grant 3966. This work was done in part at the Simons Institute for the Theory of Computing.

# 1 Introduction

For the past decade, the central open problem in equilibrium computation has been whether two-player Nash equilibrium admits a PTAS. We had good reasons to be hopeful: there was a series of improved approximation ratios [51, 34, 33, 23, 67] and several approximation schemes for special cases [48, 35, 3, 17]. Yet most interesting are two inefficient algorithms for two-player Nash:

- the classic Lemke-Howson algorithm [53] finds an exact Nash equilibrium in exponential time; and

- a simple algorithm by Lipton, Markakis, and Mehta [54] finds an $\epsilon$-Approximate Nash Equilibrium in time $n^{O(\log n)}$.

Although the Lemke-Howson algorithm takes exponential time, it has a special structure which places the problem inside the complexity class PPAD [57]; i.e. it has a polynomial time reduction to the canonical problem ENDOFALINE[1]:

**Definition 1.1** (ENDOFALINE [32]). *Given two circuits $S$ and $P$, with $m$ input bits and $m$ output bits each, such that $P(0^m) = 0^m \neq S(0^m)$, find an input $x \in \{0,1\}^m$ such that $P(S(x)) \neq x$ or $S(P(x)) \neq x \neq 0^m$.*

Proving hardness for problems in PPAD is notoriously challenging because they are *total*, i.e. they always have a solution, so the standard techniques from NP-hardness do not apply. By now, however, we know that exponential and polynomial approximations for two-player Nash are PPAD-complete '[32, 29], and so is $\epsilon$-approximation for games with $n$ players [62].

However, $\epsilon$-approximation for two-player Nash is unlikely to have the same fate: otherwise, the quasi-polynomial algorithm of [54] would refute the Exponential Time Hypothesis for PPAD:

**Hypothesis 1** (ETH for PPAD [15]). *Solving ENDOFALINE requires time $2^{\tilde{\Omega}(n)}$.*[2]

Thus the strongest hardness result we can hope to prove (given our current understanding of complexity[3]) is a quasi-polynomial hardness that sits inside PPAD:

**Theorem 1.2** (Main Theorem). *There exists a constant $\epsilon > 0$ such that, assuming ETH for PPAD, finding an $\epsilon$-Approximate Nash Equilibrium in a two-player $n \times n$ game requires time $T(n) = n^{\log^{1-o(1)} n}$.*

## 1.1 Techniques

Given an ENDOFALINE instance of size $n$, we construct a two-player $N \times N$ game for $N = 2^{n^{1/2+o(1)}}$ whose approximate equilibria correspond to solutions to the ENDOFALINE instance. Thus, assuming the "ETH for PPAD", finding an approximate equilibrium requires time $2^n = N^{\log^{1-o(1)} N}$.

---

[1] In the literature the problem has been called ENDOFTHELINE; we believe that the name ENDOFALINE is a more accurate description.

[2] As usual, $n$ is the size of the description of the instance, i.e. the size of the circuits $S$ and $P$.

[3] Given our current understanding of complexity, refuting ETH for PPAD seems unlikely: there are matching black-box lower bounds [45, 19]. Recall that the NP-analogue ETH [47] is widely used (e.g. [49, 55, 1, 25, 30]), often in stronger variants such as SETH [46, 26] and NSETH [27].

The main steps of the final construction are: (i) reducing ENDOFALINE to a new discrete problem which we call LOCALENDOFALINE; (ii) reducing LOCALENDOFALINE to a problem of finding an approximate Brouwer fixed point; (iii) reducing from Brouwer fixed point to finding an approximate Nash equilibrium in a multiplayer game over $n^{1/2+o(1)}$ players with $2^{n^{1/2+o(1)}}$ actions each; and (iv) reducing to the two-player game.

The main novelty in the reduction is the use of techniques such as error correcting codes and probabilistically checkable proofs (PCPs) inside PPAD. In particular, the way we use PCPs in our proof is very unusual.

**Constructing the first gap: showing hardness of $\epsilon$-SUCCINCTBROUWER$_2$**

The first step in all known PPAD-hardness results for (approximate) Nash equilibrium is reducing ENDOFALINE to the problem of finding an (approximate) Brouwer fixed point of a continuous, Lipschitz function $f\colon [0,1]^n \to [0,1]^n$. Let $\epsilon > 0$ be an arbitrarily small constant. Previously, the state of the art for computational hardness of approximation of Brouwer fixed points was:

**Theorem 1.3** ([62], informal). *It is PPAD-hard to find an* $\mathbf{x} \in [0,1]^n$ *such that* $\|f(\mathbf{x}) - \mathbf{x}\|_\infty \leq \epsilon$.

Here and for the rest of the paper, all distances are relative; in particular, for $\mathbf{x} \in [0,1]^n$ and $p < q$, we have $\|\mathbf{x}\|_p \leq \|\mathbf{x}\|_q$.

Theorem 1.3 implied that it is hard to find an $\mathbf{x}$ such that $f(\mathbf{x})$ is approximately equal to $\mathbf{x}$ *on every coordinate*. The first step in our proof is to strengthen this result to obtain hardness of approximation with respect to 2-norm:

**Theorem 1.4** (Informal). *It is PPAD-hard to find an* $\mathbf{x} \in [0,1]^n$ *such that* $\|f(\mathbf{x}) - \mathbf{x}\|_2 \leq \epsilon$.

Now, even finding an $\mathbf{x}$ such that $f(\mathbf{x})$ is approximately equal to $\mathbf{x}$ on *most of the coordinates* is already PPAD-hard.

Theorem 1.3 was obtained by adapting a construction due to Hirsch, Papadimitriou, and Vavasis [45]. The main idea is to partition the $[0,1]^n$ into $2^n$ subcube, and consider the grid formed by the subcube-centers; then embed a path (in fact, many paths and cycles when reducing from ENDOFALINE) along an arbitrary sequence of neighboring grid-points/subcube-centers. The function is carefully defined along the embedded path, guaranteeing both Lipschitz continuity and that approximate endpoints occur only near subcube-centers corresponding to ends of paths.

Here we observe that if we want a larger displacement (in particular, constant relative 2-norm) we actually want the consecutive vertices on the path to be as far as possible from each other. We thus replace the neighboring grid-points with their encoding by an *error correcting code*.

The first obstacle to using PCP-like techniques for problems in PPAD is their totality (i.e. a solution always exists). For NP-hard problems, the PCP verifier expects the proof to be encoded in some error correcting code. If the proof is far from any codeword, the verifier detects that (with high probability), and immediately rejects. For problems in PPAD (more generally, in TFNP) this is always tricky because it is not clear what does it mean "to reject". Hirsch et al.'s construction has the following useful property: for the vast majority of $\mathbf{x}$'s (in particular, all $\mathbf{x}$'s far from the embedding of the paths) the displacement $f(\mathbf{x}) - \mathbf{x}$ is the same default displacement. Thus, when an $\mathbf{x}$ is too far from any codeword to faithfully decode it, we can simply apply the default displacement.

We note that Theorem 1.4 is already significant enough to obtain new results for many-player games (see discussion in Subsection 1.2). Furthermore, its proof is relatively simple (in particular, PCP-free) and is self-contained in Section 3.

**The main challenge: locality.**

Our ultimate goal is to construct a two-player game that simulates the Brouwer function from Theorem 1.4. This is done via an *imitation gadget*: Alice's mixed strategy induces a point $\mathbf{x}^{(\mathcal{A})} \in [0,1]^n$; Bob's strategy induces $\mathbf{x}^{(\mathcal{B})} \in [0,1]^n$; Alice wants to minimize $\left\|\mathbf{x}^{(\mathcal{A})} - \mathbf{x}^{(\mathcal{B})}\right\|_2$, whereas Bob wants to minimize $\left\|f\left(\mathbf{x}^{(\mathcal{A})}\right) - \mathbf{x}^{(\mathcal{B})}\right\|_2$. Alice and Bob are both satisfied at a fixed point, where $\mathbf{x}^{(\mathcal{A})} = \mathbf{x}^{(\mathcal{B})} = f\left(\mathbf{x}^{(\mathcal{A})}\right)$.

The main obstacle is that we want to incentivize Bob to minimize $\left\|f\left(\mathbf{x}^{(\mathcal{A})}\right) - \mathbf{x}^{(\mathcal{B})}\right\|_2$ via *local constraints* (payoffs - each depends on one pure strategy), while $f\left(\mathbf{x}^{(\mathcal{A})}\right)$ has a *global dependency* on Alice's entire mixed strategy.

Our goal is thus to construct a hard Brouwer function that can be *locally computed*. How local does the computation need to be? In a game of size $2^{\sqrt{n}} \times 2^{\sqrt{n}}$, each strategy can faithfully store information about $\sqrt{n}$ bits. Specifically, our construction will be $n^{1/2+o(1)}$-local.

We haven't yet defined exactly what it means for our construction to be "$n^{1/2+o(1)}$-local"; the exact formulation is quite cumbersome as the query access needs to be partly adaptive, robust to noise, etc. Eventually (Section 7), we formalize the "locality" of our Brouwer function via a statement about multiplayer games. On a high level, however, our goal is to show that for any $j \in \{1, \ldots, n\}$, the $j$-th output $f_j(\mathbf{x})$ can be approximately computed, with high probability, by accessing $\mathbf{x}$ at only $n^{1/2+o(1)}$ coordinates.

This is a good place to note that achieving any sense of "local computation" in our setting is surprising, even if we consider just the error correcting encoding for our Brouwer function: in order to maintain constant relative distance, an average bit of the output must depend on a constant fraction of the input bits!

LOCALENDOFALINE

In order to introduce locality, we go back to the ENDOFALINE problem. "Wishful thinking": imagine that we could replace the arbitrary predecessor and successor circuits in ENDOFALINE with $\mathsf{NC}^0$ (constant depth and constant fan-in) circuits $S^{\text{LOCAL}}, P^{\text{LOCAL}} : \{0,1\}^n \to \{0,1\}^n$, so that each output bit only depends on a constant number of input bits. Imagine further that we had the guarantee that for each input, the outputs of $S^{\text{LOCAL}}, P^{\text{LOCAL}}$ differ from the input on just a constant number of bits. Additionally, it would be really nice if we had a succinct pointer that immediately told us which bits are about to be replaced. (We later call this succinct pointer the *counter*, because it also cycles through its possible values in a fixed order.)

Suppose all our wishes came true, and furthermore the hard Brouwer function from Theorem 1.4 used a *linear* error correcting code. Then, we could use the encoding of the counter, henceforth $C(u)$, to read only the bits that are about to be replaced, and the inputs that determine the new values of those bits. Thus, using only local access to a tiny fraction of the bits ($|C(u)| + O(1)$), we can construct a difference vector $u - S^{\text{LOCAL}}(u)$ (which is 0 almost everywhere). As we discussed above, the encodings $E(u), E(S^{\text{LOCAL}}(u))$ must differ on a constant fraction of the bits - but because the code is linear, we can also locally construct the difference vector $E(u) - E(S^{\text{LOCAL}}(u)) = E(u - (S^{\text{LOCAL}}(u)))$. Given $E(u) - E(S^{\text{LOCAL}}(u))$, we can locally compute any bit of $E(S^{\text{LOCAL}}(u))$ by accessing only the corresponding bit of $E(u)$.

Back to reality: unfortunately we do not know of a reduction to such a restricted variant of EndOfALine. Surprisingly, we can almost do that. The problem LocalEndOfALine (formally defined in Section 5) satisfies all the guarantees defined above, is linear-time reducible from EndOfALine, but has one caveat: it is only defined on a strict subset $V^{\text{LOCAL}}$ of the discrete hypercube ($V^{\text{LOCAL}} \subsetneq \{0,1\}^n$). Verifying that a vertex belongs to $V$ is quite easy - it can be done in $\mathsf{AC}^0$. Let us take a brief break to acknowledge this new insight about the canonical problem of PPAD:

**Theorem 1.5.** *The predecessor and successor circuits of* EndOfALine *are, wlog,* $\mathsf{AC}^0$ *circuits.*

The class $\mathsf{AC}^0$ is quite restricted, but the outputs of its circuits are not local functions of the inputs. Now, we want to represent $u$ in a way that will make it possible to locally determine whether $u \in V^{\text{LOCAL}}$ or not. To this end we augment the linear error correcting encoding $E(u)$ with a *probabilistically checkable proof* (PCP) $\pi(u)$ of the statement $(u \in V^{\text{LOCAL}})$.

**Our holographic proof system**

Some authors distinguish between PCPs and holographic proofs[4]: a PCP verifier has unrestricted access to the instance, and queries the proof locally; whereas the holographic proof verifier has restricted, local access to both the proof and (an error correcting encoding of) the instance. In this sense, what we actually want is a holographic proof.

We construct a holographic proof system with some very unusual properties. We are able to achieve them thanks to our modest locality desideratum: $n^{1/2+o(1)}$, as opposed to the typical $\mathsf{polylog}(n)$ or $O(1)$. We highlight here a few of those properties; see Section 6 for details.

- **(Local proof construction)** The most surprising property of our holographic proof system is that the proof $\pi(u)$ can be constructed from local access to the encoding $E(u)$. In particular, note that we can locally compute $E(S^{\text{LOCAL}}(u))$ because $E(\cdot)$ is linear - but $\pi(\cdot)$ is not. Once we obtain $E(S^{\text{LOCAL}}(u))$, we can use local proof construction to compute $\pi(S^{\text{LOCAL}}(u))$ locally.

- **(Very low random-bit complexity)** Our verifier is only allowed to use $(1/2 + o(1))\log_2 n$ random bits - this is much lower even than the $\log_2 n$ bits necessary to choose one entry at random. In related works, similar random-bit complexity was achieved by bundling the entries together via "birthday repetition". To some extent, something similar happens here, but our locality is already $n^{1/2+o(1)}$ so no bundling (or repetition) is necessary. To achieve nearly optimal random-bit complexity, we use $\lambda$-biased sets over large finite fields together with the Sampling Lemma of Ben-Sasson et al. [21].

- **(Tolerant verifier)** Typically, a verifier must reject (with high probability) whenever the input is far from valid, but it is allowed to reject even if the input is off by only one bit. Our verifier, however, is *required* to accept (with high probability) inputs that are close to valid proofs. (This is related to the notion of "tolerant testing", which was defined in [58] and discussed in [42] for locally testable codes.)

---

[4]In a nutshell, PCPs or holographic proofs are proofs that can be verified "locally" (with high probability) by reading only a small (random) portion of the proof; see e.g. [7, Chapter 18] for many more details.

- **(Local decoding)** We make explicit use of the property that our holographic proof system is also a locally decodable code. While the relations between PCPs and locally testable codes have been heavily explored (see e.g. Goldreich's survey [41]), the connection to locally decodable codes is not as immediate. Nevertheless, related ideas of Locally Decode/Reject Codes [56] and decodable PCP [38] have been used before in order to facilitate composition of tests (our holographic proof system, in contrast, is essentially composition-free). Fortunately, as noted by [38] many constructions of PCPs are already implicitly locally decodable.

- **(Robust everything)** Ben-Sasson et al. [20] introduce a notion of *robust soundness*, where on an invalid proof, the string read by the verifier must be far from any acceptable string. (Originally the requirement is far in expectation, but we want far with high probability.) Another way of looking at the same requirement, is that even if a malicious prover *adaptively* changes a small fraction of the bits queried by the verifier, the test is still sound. In this sense, we require that all our guarantees, not just soundness, continue to hold (with high probability) even if a malicious entity adaptively changes a small fraction of the bits queried by the verifier.

How is local proof construction possible? At a high level, our holographic proof system expects an encoding of $u$ as a low-degree $t$-variate polynomial, and a few more low-degree $t$-variate polynomials, that encode the proof of $u \in V^{\text{LOCAL}}$. (This is essentially the standard "arithmetization", dating back at least to [11, 63], although our construction is most directly inspired by [59, 65].) In our actual proof, $t$ is a small super-constant, e.g. $t \triangleq \sqrt{\log n}$; but for our exposition here, let us consider $t = 2$, i.e. we have bivariate polynomials.

The most interesting part of the proof verification is testing that a certain low-degree polynomial $\Psi \colon \mathcal{G}^2 \to \mathcal{G}$, for some finite field $\mathcal{G}$ of size $|\mathcal{G}| = \Theta\left(n^{1/2+o(1)}\right)$, is identically zero over all of $\mathcal{F}^2$, for some subset $\mathcal{F} \subsetneq \mathcal{G}$ of cardinality $|\mathcal{F}| = |\mathcal{G}|/\text{polylog}(n)$. This can be done by expecting the prover to provide the following low-degree polynomials:

$$\Psi'(x,y) \triangleq \sum_{f_i \in \mathcal{F}} \Psi(x, f_i) y^i$$

$$\Psi''(x,y) \triangleq \sum_{f_j \in \mathcal{F}} \Psi'(f_j, y) x^j.$$

Then, $\Psi''(x,y) = \sum_{f_i, f_j \in \mathcal{F}} \Psi(f_j, f_i) x^j y^i$ is the zero polynomial if and only if $\Psi$ is indeed identically zero over all of $\mathcal{F}^2$. $\Psi(x,y)$ can be computed by accessing $E(u)$ on just a constant number of entries. Thus, computing $\Psi'(x,y)$ requires $\Psi(x, f_i)$ for all $f_i \in \mathcal{F}$, so a total of $\Theta\left(n^{1/2+o(1)}\right)$ queries to $E(u)$. However, computing even one entry of $\Psi''(\cdot)$ requires $\Omega(n)$ queries to $E(u)$. The crucial observation is that we don't actually need the prover to provide $\Psi''$. Instead, it suffices that the prover provide $\Psi'$, and the verifier checks that $\sum_{f_j \in \mathcal{F}} \Psi'(f_j, y) x^j = 0$ for sufficiently many $(x, y)$.

#### Putting it all together via polymatrix games

The above arguments suffice to construct a hard Brouwer function (in the sense of Theorem 1.4) that can be computed "$n^{1/2+o(1)}$-locally". We formalize this statement in terms of approximate Nash equilibria in a polymatrix game.

**Definition 1.6** (Polymatrix games)**.** In a polymatrix game, each pair of players simultaneously plays a separate two-player subgame. Every player has to play the same strategy in every two-player subgame, and her utility is the sum of her subgame utilities. The game is given in the form of the payoff matrix for each two-player subgame.

We construct a bipartite polymatrix game between $n^{1/2+o(1)}$ players with $2^{n^{1/2+o(1)}}$ actions each. By "bipartite", we mean that each player on Alice's side only interacts with players on Bob's side and vice versa. The important term here is "polymatrix": it means that when we compute the payoffs in each subgame, they can only depend on the $n^{1/2+o(1)}$ coordinates described by the two players' strategies. It is in this sense that we guarantee "local computation".

The mixed strategy profile $\mathcal{A}$ of all the players on Alice's side of the bipartite game induces a vector $\mathbf{x}^{(\mathcal{A})} \in [0,1]^m$, for some $m = n^{1+o(1)}$. The mixed strategy profile $\mathcal{B}$ of all the players on Bob's side induces a vector $\mathbf{x}^{(\mathcal{B})} \in [0,1]^m$. Our main technical result is:

**Proposition 1.7** (Informal). *If all but an $\epsilon$-fraction of the players play $\epsilon$-optimally, then* $\left\| \mathbf{x}^{(\mathcal{A})} - \mathbf{x}^{(\mathcal{B})} \right\|_2^2 = O(\epsilon)$ *and* $\left\| f\left( \mathbf{x}^{(\mathcal{A})} \right) - \mathbf{x}^{(\mathcal{B})} \right\|_2^2 = O(\epsilon)$.

Each player on Alice's side corresponds to one of the PCP verifier's random string. Her strategy corresponds to an assignment to the bits queried by the verifier given this random string. On Bob's side, we consider a partition of $\{1, \ldots, m\}$ into $n^{1/2+o(1)}$ tuples of $n^{1/2+o(1)}$ indices each. Each player on Bob's side assigns values to one such tuple.

On each two-player subgame, the player on Alice's side is incentivized to imitate the assignment of the player on Bob's side on the few coordinates where they intersect. The player on Bob's side, uses Alice's strategy to locally compute $f_j\left(\mathbf{x}^{(\mathcal{A})}\right)$ on a few $j$'s in his $\left(n^{1/2+o(1)}\right)$-tuple of coordinates. This computation may be inaccurate, but we can guarantee that for most coordinates it is approximately correct most of the time.

**From polymatrix to bimatrix**

The final reduction from the polymatrix game to two-player game follows more or less from known techniques for hardness of Nash equilibria [4, 32, 15]. We let each of Alice and Bob control one side of the bipartite polymatrix game. In particular, each strategy in the two-player game corresponds to picking a player of the polymatrix game, and a strategy for that player. We add a gadget due to Althofer [4] to guarantee that Alice and Bob mix approximately uniformly across all their players. See Section 8 for details.

## 1.2 Results for multiplayer relaxations of Nash equilibrium

Our hardness for norm-2 approximate Brouwer fixed point (Theorem 1.4) has some important consequences for multiplayer games. All our results in this regime (as well as much of the existing literature) are inspired by a paper of Babichenko [13] and a blog post of Shmaya [64].

For multiplayer games, there are several interesting questions one can ask. First, note that the normal form representation of the game is exponential in the number of players, so it is difficult to talk about computational complexity. To alleviate this, different restricted classes of multiplayer games have been studied. We have already seen *polymatrix games* (Definition 1.6), which have a succinct description in terms of the normal forms of the $\binom{n}{2}$ bimatrix subgames. Another interesting class is *graphical games* where we are given a (low-degree) graph over the players, and each player's utility is only affected by the actions of its neighbors. The graph of the game constructed in Proposition 1.7 has a high degree, but it is important that it is bipartite, as are all the games described below. Most generally, we can talk about the class of *succinct games*, which are described via a circuit that computes any entry of the payoff tensors (this includes polymatrix and graphical games). Finally, there has been recent significant progress on the *query complexity* of finding approximate Nash equilibria in arbitrary $n$-player games where the payoff tensors are given via a black-box oracle [14, 43, 13, 28].

There are also a few different notions of approximation of Nash equilibrium (all are defined formally in Subsection 2.2). The strictest notion, $\epsilon$-*Well-Supported Nash Equilibrium*, requires that for *every action* in the support of *every player*, the expected utility, given other players' mixed strategies, is within (additive) $\epsilon$ of the optimal strategy for that player. For this notion, Babichenko [13] showed a $2^{\Omega(n)}$ lower bound on query complexity for any (possibly randomized) algorithm. In followup work, [60] showed PPAD-completeness for succinct games, and soon after [62] extended this PPAD-completeness to games that are both polymatrix degree-3-graphical.

The most central model in the literature, $\epsilon$-*Approximate Nash Equilibrium*, requires that *every player's* expected utility from her mixed strategy is within $\epsilon$ of the optimum she can achieve (given other players' strategies). I.e., any player is allowed to assign a small probability to poor strategies, as long as in expectation she does well. The last PPAD-completeness result extends immediately to this model (the two notions of approximation are equivalent, up to constant factors, for constant degree graphical games). For query complexity, Hart and Nisan [43] and Babichenko [13] asked whether the latter's exponential lower bound can be extended to $\epsilon$-Approximate Nash Equilibrium. Very recently Chen et al. [28] solved it almost entirely, showing a lower bound of $2^{\Omega(n/\log n)}$; and they asked whether the $\Theta(\log n)$ gap in the exponent can be resolved. Here, we obtain a tight $2^{\Omega(n)}$ lower bound on the query complexity, as well as stronger inapproximability guarantees.

Finally, the most lenient notion is that of $(\epsilon, \delta)$-*WeakNash*: it only requires that a $(1 - \delta)$-fraction of the players play $\epsilon$-optimally ("Can almost everybody be almost happy?"). This notion was recently defined in [15] who conjectured that it is also PPAD-complete for polymatrix, graphical games. Their conjecture remains an interesting open problem (see Subsection 1.3). Here, as a consequence of Theorem 1.4, we prove that $(\epsilon, \delta)$-WeakNash is PPAD-complete for the more general class of succinct games.

**Corollary 1.8.** *There exist constants $\epsilon, \delta > 0$, such that finding an $(\epsilon, \delta)$-WeakNash is* PPAD-*hard for succinct multiplayer games where each player has two actions.*

Furthermore, as we hinted earlier, our proof also extends to giving truly exponential lower bounds on the query complexity:

**Corollary 1.9.** *There exist constants $\epsilon, \delta > 0$, such that any (potentially randomized) algorithm for finding an $(\epsilon, \delta)$-WeakNash for multiplayer games where each player has two actions requires $2^{\Omega(n)}$ queries to the players' payoffs tensors.*

### 1.3 The PCP Conjecture for PPAD

Rather than posing new open problems, let us restate the following conjecture due to [15]:

**Conjecture 1.10** (PCP for PPAD; [15])**.** *There exist constants $\epsilon, \delta > 0$ such that finding an $(\epsilon, \delta)$-WeakNash in a bipartite, degree three polymatrix game with two actions per player is* PPAD-*complete.*

The main original motivation was an approach to prove our main theorem given this conjecture. As pointed out by [15], it turns out that resolving this conjecture would also have interesting consequences for relative approximations of two-player Nash equilibrium, as well as applications to inapproximabability of market equilibrium.

More importantly, this question is interesting in its own right: how far can we extend the ideas from the PCP Theorem (for NP) to the wold of PPAD? The PCP $[r(n), q(n)]$ characterization [8] is mainly concerned with two parameters: $r(n)$, the number of random bits, and $q(n)$, the number of bits read from the proof. A major tool in all proofs of the PCP Theorem is *verifier composition*: in the work of Polishchuk and Spielman [59, 65], for example, it is first shown that NP $\subseteq$ PCP $\left[O(\log n), n^{1/2+o(1)}\right]$, and then via composition it is eventually obtained that NP = PCP $[O(\log n), O(1)]$. In some **informal sense**, one may think of our main technical result as something analogous[5] to PPAD $\subseteq$ PCP $\left[(1/2 + o(1))\log_2 n, n^{1/2+o(1)}\right]$. Furthermore, our techniques in Section 6 build on many existing ideas from the PCP literature [12, 59, 65, 21, 20] that have been used to show similar statements for NP. It is thus natural to ask: is there a sense in which our "verifier" can be composed? can such composition eventually resolve the PCP Conjecture for PPAD?

More generally, some of the tools we use here, even as simple as error correcting codes, have been the basic building blocks in hardness of approximation for decades, yet to the best of our knowledge have not been used before for any problem in PPAD. We hope to see other applications of similar ideas in this regime[6].

## 1.4 Additional related work

Aside from [15], previous attempts to show lower bounds for approximate Nash in two player games have mostly focused on limited models of computation [35] and lower bounding the support required for obtaining approximate equilibria [4, 39, 6, 5] (in contrast, [54]'s algorithm runs in quasi-polynomial time because there exist approximate equilibria with support size at most $O\left(\frac{\log n}{\epsilon^2}\right)$).

**Birthday repetition and related quasi-polynomial lower bounds** Hazan and Krauthgamer [44] showed that finding an $\epsilon$-Approximate Nash Equilibrium with $\epsilon$-optimal welfare is as hard as the PLANTED-CLIQUE problem; Austrin et al. [9] later showed that the optimal-welfare constraint can be replaced by other decision problems. Braverman et al. [25] recently showed that the hardness PLANTED-CLIQUE can be replaced by the Exponential Time Hypothesis, the NP-analog of the ETH for PPAD we use here. The work of Braverman et al, together with an earlier paper by Aaronson et al. [1] inspired a line of works on quasi-polynomial hardness results via the technique of "birthday repetition" [15, 24, 61, 22]. In particular [15] investigated whether birthday repetition can give quasi-polynomial hardness for finding any $\epsilon$-Approximate Nash Equilibrium (our main theorem). As we discussed in Subsection 1.3 the main obstacle is that we don't have a PPAD-analogue for the PCP Theorem.

**Multiplicative hardness of approximation** Daskalakis [31] and our recent work [60] show that finding an $\epsilon$-relative Well-Supported Nash Equilibrium in two-player games is PPAD-hard. The case of $\epsilon$-relative Approximate Nash Equilibrium is still open: our main theorem implies that it requires at least quasi-polynomial time, but it is not known whether it is PPAD-hard, or even if it requires a large support (see also discussion in [15]).

---

[5]We stress that our analogy is very loose. For example, we are not aware of any formal extension of PCP to function problems, and it is well known that NP $\subseteq$ PCP $\left[(1/2 + o(1))\log_2 n, n^{1/2+o(1)}\right]$.

[6]In fact, in the few months since our paper first appeared, our techniques already found applications for lower bounds on the *communication complexity* of Nash equilibrium [16].

**Approximation algorithms** The state of the art for games with arbitrary payoffs is $\approx$ 0.339 for two-player games due to Tsaknakis and Spirakis [67] and $0.5+\epsilon$ for polymatrix games due to Deligkas et al. [37]. For two-player games, PTAS have been given for the special cases of constant rank games by Kannan and Theobald [48], small-probability games by Daskalakis and Papadimitriou [35], positive semi-definite games by Alon et al. [3], and sparse games by Barman [17]. For games with many players and a constant number of strategies, PTAS were given for the special cases of anonymous games by Daskalakis and Papadimitriou [36] and polymatrix games on a tree by Barman et al. [18]. Finally, let us return to the more general class of succinct $n$-player games, and mention an approximation algorithm due to Goldberg and Roth [40]; their algorithm runs in exponential time, but uses only a polynomial number of oracle queries.

**Communication complexity** The hard instance of Brouwer we construct here (Theorem 1.4) has already been useful in followup work [16], for proving lower bounds on the communication complexity of approximate Nash equilibrium in $N \times N$ two-player games, as well as binary-action $n$-player games.

## 1.5 Organization

In Section 3, we prove the hardness of finding a 2-norm-approximate Brouwer fixed point (Theorem 1.4). In Section 4, we use this result to obtain our resutls for many-player games (Corrolaries 1.8 and 1.9). In Section 5 we introduce the restricted variant LOCALENDOFA-LINE and prove that it is also PPAD-complete. In Section 6 we construct our holographic proof system. In Section 7 we bring together ideas from Sections 3, 5, and 6 to prove our hardness for polymatrix games of subexponential size. Finally, in Section 8 we reduce from polymatrix to two-player games.

## 2 Preliminaries

We use $\mathbf{0}_n$ (respectively $\mathbf{1}_n$) to denote the length-$n$ vectors whose value is 0 (1) in every coordinate.

**Small constants**

Our proof uses several arbitrary small constants that satisfy:

$$0 < \epsilon_{\text{Precision}} \ll \epsilon_{\text{Nash}} \ll \delta \ll h \ll \epsilon_{\text{Complete}} \ll \epsilon_{\text{Sound}} \ll \epsilon_{\text{Decode}} \ll 1.$$

By $\ll$ we mean that we pick them such that $\epsilon_{\text{Precision}}$ is arbitrarily smaller than any polynomial in $\epsilon_{\text{Nash}}$, and $\epsilon_{\text{Nash}}$ is arbitrarily smaller than any polynomial in $\delta$, etc. Although their significance will be fully understood later in the paper, we briefly mention that $\epsilon_{\text{Precision}}$ is the precision with which the players can specify real values; $\epsilon_{\text{Nash}}$ is the approximation factor of Nash equilibrium in Proposition 6.1; $\delta$ and $h$ are parameters of the Brouwer function construction; finally, $\epsilon_{\text{Complete}}, \epsilon_{\text{Sound}}, \epsilon_{\text{Decode}}$, are parameters of our holographic proof system.

**Relative norms**

Throughout the paper, we use normalized $p$-norms: for a vector $\mathbf{x} = (x_1, \ldots, x_n) \in \mathbb{R}^n$ we define
$$\|\mathbf{x}\|_p^p \triangleq \mathsf{E}_{i \in [n]} \left[ (x_i)^p \right],$$
where the expectation over coordinates is taken from uniform distribution, until otherwise specified.

Similarly, for a binary string $\pi \in \{0,1\}^n$, we denote
$$|\pi| \triangleq \mathsf{E}_{i \in [n]} [\pi_i].$$

**How to catch a far-from-uniform distribution**

**Lemma 2.1** (Lemma 3 in the full version of [35]). *Let $\{a_i\}_{i=1}^n$ be real numbers satisfying the following properties for some $\theta > 0$: (1) $a_1 \geq a_2 \geq \cdots \geq a_n$; (2) $\sum a_i = 0$; (3) $\sum_{i=1}^{n/2} a_i \leq \theta$. Then $\sum_{i=1}^n |a_i| \leq 4\theta$.*

### 2.1 $\lambda$-biased sets

**Definition 2.2** ($\lambda$-biased sets). Let $\mathcal{G}$ be a finite field, and $t > 0$ an integer. A multiset $S \subseteq \mathcal{G}^t$ is $\lambda$-biased if for every nontrivial character $\chi$ of $\mathcal{G}^t$,
$$|\mathsf{E}_{\mathbf{y} \sim S} [\chi(\mathbf{y})]| \leq \lambda.$$

**Lemma 2.3** ([10, Theorem 3.2]). *A randomly chosen multiset $S \subseteq \mathcal{G}^t$ of cardinality $\Theta\left(t \log |\mathcal{G}| / \lambda^2\right)$ is $\lambda$-biased with high probability.*

For many applications, an explicit construction is necessary. In our case, however, we can enumerate over all sets $S$ of sufficient cardinality in quasi-polynomial time[7]. The following Sampling Lemma due to Ben-Sasson et al. [21] allows us to estimate the average of any function over $\mathcal{G}^t$ using only one line and $(1 + o(1)) \log_2 |\mathcal{G}^t|$ randomness:

**Lemma 2.4** (Sampling Lemma: [21, Lemma 4.3]). *Let $B : \mathcal{G}^t \to [0,1]$. Then, for any $\epsilon > 0$,*
$$\Pr_{\substack{\mathbf{x} \in \mathcal{G}^t, \\ \mathbf{y} \in S}} [|\mathsf{E}_{\beta \in \mathcal{G}} [B(\mathbf{x} + \beta \mathbf{y})] - \mathsf{E}_{\mathbf{z} \in \mathcal{G}^t} [B(\mathbf{z})]| > \epsilon] \leq \left(\frac{1}{|\mathcal{G}|} + \lambda\right) \frac{\mathsf{E}_{\mathbf{z} \in \mathcal{G}^t} [B(\mathbf{z})]}{\epsilon^2}.$$

### 2.2 Different notions of approximate Nash equilibrium

A mixed strategy of player $i$ is a distribution $x_i$ over $i$'s set of actions, $A_i$. We say that a vector of mixed strategies $\mathbf{x} \in \times_j \Delta A_j$ is a *Nash equilibrium* if every strategy $a_i$ in the support of every $x_i$ is a best response to the actions of the mixed strategies of the rest of the players, $x_{-i}$. Formally,
$$\forall a_i \in \mathrm{supp}(x_i) \quad \mathsf{E}_{a_{-i} \sim x_{-i}} [u_i(a_i, a_{-i})] = \max_{a' \in A_i} \mathsf{E}_{a_{-i} \sim x_{-i}} [u_i(a', a_{-i})].$$

---
[7] Note that we need an $\epsilon$-biased set for a large field $\mathcal{G} = \mathbb{F}_{2^\ell}$. Such constructions are not as common in the literature which mostly focuses on the field $\mathbb{F}_2$. To the best of our knowledge, existing explicit constructions for larger fields require much larger cardinality. Nevertheless, for our modest pseudorandomness desiderata, we could actually use the explicit construction from [2]. For ease of presentation, we prefer to brute-force derandomize the construction from [10].

Equivalently, **x** is a Nash equilibrium if each mixed strategy $x_i$ is a best response to $x_{-i}$:

$$\mathsf{E}_{\mathbf{a}\sim\mathbf{x}}[u_i(\mathbf{a})] = \max_{x_i' \in \Delta A_i} \mathsf{E}_{\mathbf{a}\sim(x_i'; x_{-i})}[u_i(\mathbf{a})].$$

Each of those equivalent definitions can be generalized to include approximation in a different way.

**Definition 2.5** ($\epsilon$-Approximate Nash Equilibrium). We say that **x** is an $\epsilon$-*Approximate Nash Equilibrium* ($\epsilon$-*ANE*) if each $x_i$ is an $\epsilon$-best response to $x_{-i}$:

$$\mathsf{E}_{\mathbf{a}\sim\mathbf{x}}[u_i(\mathbf{a})] \geq \max_{x_i' \in \Delta A_i} \mathsf{E}_{\mathbf{a}\sim(x_i'; x_{-i})}[u_i(\mathbf{a})] - \epsilon.$$

On the other hand, we generalize the first definition of Nash equilibrium in the following stricter definition:

**Definition 2.6** ($\epsilon$-Well-Supported Nash Equilibrium). **x** is a $\epsilon$-*Well-Supported Nash Equilibrium* ($\epsilon$-*WSNE*) if every $a_i$ in the support of $x_i$ is an $\epsilon$-best response to $x_{-i}$:

$$\forall a_i \in \mathrm{supp}(x_i) \quad \mathsf{E}_{a_{-i}\sim x_{-i}}[u_i(a_i, a_{-i})] \geq \max_{a' \in A_i} \mathsf{E}_{a_{-i}\sim x_{-i}}\left[u_i\left(a', a_{-i}\right)\right] - \epsilon.$$

**WeakNash**

We can further relax both of the above definitions by requiring that the corresponding condition only hold for most of the players (rather than all of them).

**Definition 2.7** (($\epsilon, \delta$)-WeakNash [15]). We say that **x** is an ($\epsilon, \delta$)-*WeakNash* if for a $(1 - \delta)$-fraction of $i$'s, $x_i$ is an $\epsilon$-best mixed response to $x_{-i}$:

$$\Pr_i\left[\mathsf{E}_{\mathbf{a}\sim\mathbf{x}}[u_i(\mathbf{a})] \geq \max_{x_i' \in \Delta A_i} \mathsf{E}_{\mathbf{a}\sim(x_i'; x_{-i})}[u_i(\mathbf{a})] - \epsilon\right] \geq 1 - \delta.$$

**Definition 2.8** (($\epsilon, \delta$)-Well-Supported WeakNash). **x** is a ($\epsilon, \delta$)-*Well-Supported WeakNash* if for a $(1 - \delta)$-fraction of $i$'s, every $a_i$ in the support of $x_i$ is an $\epsilon$-best response to $x_{-i}$:

$$\Pr_i\left[\forall a_i \in \mathrm{supp}(x_i) \; \mathsf{E}_{a_{-i}\sim x_{-i}}[u_i(a_i, a_{-i})] \geq \max_{a' \in A_i} \mathsf{E}_{a_{-i}\sim x_{-i}}\left[u_i\left(a', a_{-i}\right)\right] - \epsilon\right] \geq 1 - \delta.$$

### 2.3 Classes of multiplayer games

**Definition 2.9** (Polymatrix games). In a *polymatrix game*, each pair of players simultaneously plays a separate two-player game, which we call the *subgame*. Every player has to play the same strategy in every two-player subgame, and her utility is the sum of her subgame utilities. In other words, the $i$-th player utility $u_i$ decomposes as

$$u_i(a_i, a_{-i}) = \sum_j v_i^j(a_i, a_j).$$

The game is given in the form of the payoff matrix for each two-player subgame.

**Definition 2.10** (Graphical games). In a graphical game [50], the utility of each player depends only on the action chosen by a few other players. This game now naturally induces a directed graph: we say that $(i, j) \in E$ if the utility of player $j$ depends on the strategy chosen by player $i$. When the maximal incoming degree is bounded, the game has a representation polynomial in the number of players and strategies.

**Definition 2.11** (Succinct games). }

A *succinct game* has a polynomial (in the number of players and strategies) size circuit that computes the entries of the payoff tensors. The input to the computational problem of finding (approximate) Nash equilibria to succinct games is the circuit.

## 3 2-norm approximate Brouwer fixed point

In this section we obtain a qualitative strengthening of previous hardness of approximation for Brouwer (Theorem 1.3) where we replace the $\infty$-norm of the inapproximability with $p$-norm for a constant $p$ (in particular, $p = 2$).

**Definition 3.1** ($\epsilon$-SUCCINCTBROUWER$_p$). Given an arithmetic circuit that computes a function $f \colon [0, 1]^n \to [0, 1]^n$ that is $1/\epsilon$-Lipschitz with respect to $p$-norm, $\epsilon$-SUCCINCTBROUWER$_p$ is the problem of computing an $\mathbf{x}$ such that $\|f(\mathbf{x}) - \mathbf{x}\|_p \leq \epsilon$.

**Theorem 3.2.** *There exist a constant $\epsilon > 0$, such that $\epsilon$-SUCCINCTBROUWER$_2$ is PPAD-hard. Furthermore, there is a linear time reduction from ENDOFALINE to $\epsilon$-SUCCINCTBROUWER$_2$.*

The proof has two main arguments: in Lemma 3.3 we show how to embed the ENDOFALINE graph as a collection of continuous paths in $[-1, 2]^{O(n)}$; Lemma 3.4 describes how to embed a continuous Brouwer function whose fixed points correspond to endpoints of the paths constructed in Lemma 3.3.

### 3.1 Embedding with an error correcting code

Let $m = \Theta(n)$, and let $\alpha, \eta > 0$ be some sufficiently small constants. For convenience of notation we will construct a function $f \colon [-1, 2]^{3m} \to [-1, 2]^{3m}$ (instead of $[0, 1]$); in particular, now the vertices of the discrete hypercube $\{0, 1\}^{3m}$ are interior points of our domain.

**Lemma 3.3.** *We can efficiently embed an ENDOFALINE graph $G$ over $\{0, 1\}^n$ as a collection of continuous paths and cycles in $[-1, 2]^{3m}$, such that the following hold:*

- *Each edge in $G$ corresponds to a concatenation of a few line segments between vertices of $\{0, 1\}^{3m}$; we henceforth call them* Brouwer line segments *and* Brouwer vertices.

- *The points on any two non-consecutive Brouwer line segments are $\eta$-far.*

- *The points on any two consecutive Brouwer line segments are also $\eta$-far, except near the point $\mathbf{y} \in \{0, 1\}^{3m}$ where the two Brouwer line segments connect;*

- *The (unsigned) angle between every two consecutive Brouwer line segments is at least $\alpha$ (and at most $\pi/4$).*

- *Given any point $\mathbf{x} \in [-1, 2]^{3m}$, we can use the ENDOFALINE predecessor and successor circuits to determine in polynomial time whether $\mathbf{x}$ is $\eta$-close to any Brouwer line segment, and if so what is the distance to this Brouwer line segment, and what are its endpoints.*

- *There is a one-to-one correspondence between endpoints of the embedded paths and solutions of the ENDOFALINE instance.*

*Proof.* We think of the $3m$ coordinates as partitioned into three parts, representing the current vertex in $G$, the next vertex in $G$, and an auxiliary compute-next vs copy bit. Let $E_\mathcal{C}(\cdot)$ denote the encoding in a binary, linear error correcting code $\mathcal{C}$ with message length $n$, block length $m$, constant relative distance, and an efficient decoding scheme (e.g. [66]). The current and next vertex are encoded with $\mathcal{C}$, whereas the Compute-vs-Copy bit is encoded with a repetition code.

For each edge $(u \to v)$ in $G$, we have Brouwer line segments connecting following points (in this order): $(E_\mathcal{C}(u), E_\mathcal{C}(u), \mathbf{0}_m)$, $(E_\mathcal{C}(u), E_\mathcal{C}(v), \mathbf{0}_m)$, $(E_\mathcal{C}(u), E_\mathcal{C}(v), \mathbf{1}_m)$, $(E_\mathcal{C}(v), E_\mathcal{C}(v), \mathbf{1}_m)$, $(E_\mathcal{C}(v), E_\mathcal{C}(v), \mathbf{0}_m)$. Notice that in each Brouwer line segment, only one subset of $m$ coordinates change. Thus whenever we are close to a line, we can successfully decode the $2m$ fixed coordinates. Once we decode the $2m$ fixed coordinates, we can compute what should be the values on the other $m$ coordinates using the ENDOFALINE predecessor and successor circuits, and determine the endpoints of the Brouwer line segment, and then also the distance to it.

Because the code has a constant relative distance, the lower bounds on distances and angles between Brouwer line segments follow. Finally the angles are at most $\pi/4$ because vertices of a hypercube cannot form obtuse angles (the dot product of any two vertices of $\{0,1\}^{3m}$ is non-negative). □

### 3.2 Constructing a continuous function

Let $m$ be as before, and let $0 < \delta \ll h < 1$ be sufficiently small constants. Theorem 3.2 follows from the next lemma by setting $f(\mathbf{x}) \triangleq \mathbf{x} + g(\mathbf{x})$.

**Lemma 3.4.** *We can efficiently embed an ENDOFALINE graph $G$ over $\{0,1\}^n$ as a displacement function $g: [-1,2]^{4m} \to [-\delta, \delta]^{4m}$ such that:*

1. *$g(\cdot)$ does not send any point outside the hypercube, i.e. $\mathbf{x} + g(\mathbf{x}) \in [-1, 2]^{4m}$.*

2. *$g(\cdot)$ is $O(1)$-Lipschitz (thus, $f(\cdot)$ is also $O(1)$-Lipschitz).*

3. *$\|g(\mathbf{x})\|_2 = \Omega(\delta)$ for every $\mathbf{x}$ that does not correspond to an endpoint of a path.*

4. *The value of $g$ at any point $\mathbf{x}$ whose first $3m$ coordinates are $2\sqrt{h}$-close to a Brouwer line segment (resp. two consecutive Brouwer line segments) from the embedding in Lemma 3.3, depends only on its location relative to the endpoints of the Brouwer line segment(s).*

5. *The value of $g$ at any point $\mathbf{x}$ whose first $3m$ coordinates are $2\sqrt{h}$-far from all Brouwer line segments from Lemma 3.3, does not depend on the graph $G$.*

*Proof.* We think of the $4m$ coordinates as partitioned into four parts: the first three are used as in Lemma 3.3 (current, next, and compute-vs-copy), and the last $m$ coordinates represent a special default direction in which the displacement points when far from all Brouwer line segments (similarly to the single special coordinate in [45]).

We consider a path starting at $(\mathbf{0}_{3m}, 2 \cdot \mathbf{1}_m)$, i.e. the concatenation of 0 on the first $3m$ coordinates, and 2 on the last $m$ coordinates. The path first goes to $\mathbf{0}_{4m}$ (in a straight line), and thereafter the last $m$ coordinates remain constant (0), while the first $3m$ coordinates follow the path from Lemma 3.3. We say that a point $\mathbf{x}$ is in the *picture* if $\mathsf{E}_{i \in \{3m+1,\ldots,4m\}} x_i < 1/2$. We construct $g$ separately inside and outside the picture (and make sure that the construction agrees on the hyperplane $\mathsf{E}_{i \in \{3m+1,\ldots,4m\}} x_i = 1/2$).

**Truncation** In order for $g(\cdot)$ to be a displacement function, we must ensure that it never sends any points outside the hypercube, i.e. $\forall \mathbf{x} \in [-1,2]^{4m}$, we require that also $\mathbf{x} + g(\mathbf{x}) \in [-1,2]^{4m}$. Below, it is convenient to first define an *untruncated* displacement function $\hat{g} : [-1,2]^{4m} \to [-\delta, \delta]^{4m}$ which is not restricted by the above condition. We then truncate each coordinate to fit in $[-1,2]$: $[g(\mathbf{x})]_i = \max\{-1, \min\{2, x_i + [\hat{g}(\mathbf{x})]_i\}\} - x_i$. It is clear that the truncation is easy to compute and if $\hat{g}(\cdot)$ is $(M-1)$-Lipschitz, then $g(\cdot)$ is $M$-Lipschitz. It is, however, important to make sure that the the magnitude of the displacement is not compromised. Typically, some of the coordinates may need to be truncated, but we design the displacement so that most coordinates, say 99%, are not truncated. If $\hat{g}(\mathbf{x})$ has a non-negligible component in at least 5% of the coordinates, then in total $g(\mathbf{x})$ maintains a non-negligible magnitude.

**Inside the picture**

We have to define the displacement for $\mathbf{x}$ far from every line, for $\mathbf{x}$ near one line, and for $\mathbf{x}$ near two consecutive lines (notice that by Lemma 3.3 no point is close to two non-consecutive lines). For $\mathbf{x}$ far from every line, we use the default displacement, which points in the positive special direction: $\hat{g}(\mathbf{x}) = (\mathbf{0}_{3m}, \delta \cdot \mathbf{1}_m)$. Because $\mathbf{x}$ is inside the picture, the truncated displacement $g(\mathbf{x})$ is close to $\hat{g}(\mathbf{x})$, and therefore satisfies $\|g(\mathbf{x})\|_2 = \Omega(\delta)$.

For $\mathbf{x}$ which is close to one line, we construct the displacement as follows: on the line, the displacement points in the direction of the path; at distance $h$ from the line, the displacement points in towards the line; at distance $2h$ from the line, the displacement points against the direction of the path; at distance $3h$, the displacement points in the default direction.

Formally, let $\beta_{(\mathbf{s} \to \mathbf{t})}(\mathbf{x})$ denote the magnitude of the component of $\mathbf{x} - \mathbf{s}$ in the direction of line $(\mathbf{s} \to \mathbf{t})$,

$$\beta_{(\mathbf{s} \to \mathbf{t})}(\mathbf{x}) \triangleq \frac{(\mathbf{t} - \mathbf{s})}{\|\mathbf{s} - \mathbf{t}\|_2} \cdot (\mathbf{x} - \mathbf{s}),$$

where $\cdot$ denotes the (in-expectation) dot product. Let $\mathbf{z} = \mathbf{z}(\mathbf{x})$ be the point nearest to $\mathbf{x}$ on the line; notice that $\mathbf{z}$ satisfies

$$\mathbf{z} = \beta_{(\mathbf{s} \to \mathbf{t})}(\mathbf{x}) \mathbf{t} + (1 - \beta_{(\mathbf{s} \to \mathbf{t})}(\mathbf{x})) \mathbf{s}.$$

For points near the line ($\|\mathbf{x} - \mathbf{z}\|_2 \leq 3h$), but far from its endpoints ($\beta_{(\mathbf{s} \to \mathbf{t})}(\mathbf{x}) \in \left[\sqrt{h}, 1 - \sqrt{h}\right]$), we define the displacement:

$$\hat{g}(\mathbf{x}) \triangleq \begin{cases} \delta \frac{(\mathbf{t} - \mathbf{s})}{\|\mathbf{t} - \mathbf{s}\|_2} & \|\mathbf{x} - \mathbf{z}\|_2 = 0 \\ \delta \frac{(\mathbf{z} - \mathbf{x})}{h} & \|\mathbf{x} - \mathbf{z}\|_2 = h \\ \delta \frac{(\mathbf{s} - \mathbf{t})}{\|\mathbf{t} - \mathbf{s}\|_2} & \|\mathbf{x} - \mathbf{z}\|_2 = 2h \\ \delta (\mathbf{0}_{3m}, \mathbf{1}_m) & \|\mathbf{x} - \mathbf{z}\|_2 = 3h \end{cases} \quad (1)$$

At intermediate distances from the line, we interpolate: at distance $\|\mathbf{x} - \mathbf{z}\|_2 = \frac{1}{3}h$, for example, we have $\hat{g}(\mathbf{x}) = \frac{2}{3}\delta \frac{(\mathbf{t} - \mathbf{s})}{\|\mathbf{t} - \mathbf{s}\|_2} + \frac{1}{3}\delta \frac{(\mathbf{z} - \mathbf{x})}{h}$. Notice that every two of $(\mathbf{t} - \mathbf{s})$, $(\mathbf{z} - \mathbf{x})$, and $(\mathbf{0}_{3m}, \mathbf{1}_m)$ are orthogonal, so the interpolation does not lead to cancellation. Also, every point $\mathbf{z}$ on the line is $\Omega(1)$-far in every coordinate from $\{-1, 2\}$, so the truncated displacement $g(\mathbf{x})$ still satisfies $\|g(\mathbf{x})\|_2 = \Omega(\delta)$. For each case in (1), $\hat{g}(\cdot)$ is either constant, or (in the case of $\|\mathbf{x} - \mathbf{z}\|_2 = h$) $O(\delta/h)$-Lipschitz; by choice of $\delta \ll h$, it follows that $\hat{g}(\cdot)$ is in particular $O(1)$-Lipschitz. Furthermore, notice that $\|\mathbf{x} - \mathbf{z}\|_2$ is 1-Lipschitz, so after interpolating for intermediate distances, $\hat{g}(\cdot)$ continues to be $O(1)$-Lipschitz. Notice also that at distance $3h$ the displacement defined in (1) agrees with the displacements for points far from every line, so Lipschitz continuity is preserved.

**Close to a vertex**

At distance $O\left(\sqrt{h}\right)$ from a Brouwer vertex (recall $\sqrt{h} \gg h$), we use a different displacement that interpolates between the incoming and outgoing Brouwer line segments. Consider $\mathbf{x}$ which is close to the line from $\mathbf{s}$ to $\mathbf{y}$, and also to the line from $\mathbf{y}$ to $\mathbf{t}$. Notice that every two consecutive Brouwer line segments change disjoint subsets of the coordinates, so $(\mathbf{s} \to \mathbf{y})$ and $(\mathbf{y} \to \mathbf{t})$ are orthogonal. Let $\mathbf{z}_{(\mathbf{s} \to \mathbf{y})}$ be the point on line $(\mathbf{s} \to \mathbf{y})$ that is at distance $\sqrt{h}$ from $\mathbf{y}$; similarly, let $\mathbf{z}_{(\mathbf{y} \to \mathbf{t})}$ be the point on line $(\mathbf{y} \to \mathbf{t})$ that is at distance $\sqrt{h}$ from $\mathbf{y}$.

The high level idea is to "cut the corner" and drive the flow along the line segment $L_{\mathbf{y}}$ that connects $\mathbf{z}_{(\mathbf{s} \to \mathbf{y})}$ and $\mathbf{z}_{(\mathbf{y} \to \mathbf{t})}$. In particular, we consider points $\mathbf{x}$ that are within distance $3h$ of $L_{\mathbf{y}}$. For all points further away (including $\mathbf{y}$ itself), we use the default displacement.

Our goal is to interpolate between the line displacement for $(\mathbf{s} \to \mathbf{y})$ (which is defined up to $\beta_{(\mathbf{s} \to \mathbf{y})}(\mathbf{x}) = 1 - \sqrt{h}$), and the line displacement for $(\mathbf{y} \to \mathbf{t})$ (which begins at $\beta_{(\mathbf{y} \to \mathbf{t})}(\mathbf{x}) = \sqrt{h}$). Let $\Delta_{(\mathbf{s} \to \mathbf{y})}(\mathbf{x}) \triangleq \beta_{(\mathbf{s} \to \mathbf{y})}(\mathbf{x}) - \left(1 - \sqrt{h}\right)$, and $\Delta_{(\mathbf{y} \to \mathbf{t})}(\mathbf{x}) \triangleq \sqrt{h} - \beta_{(\mathbf{y} \to \mathbf{t})}(\mathbf{x})$. We set our interpolation parameter $\alpha = \alpha(\mathbf{x}) \triangleq \frac{\Delta_{(\mathbf{y} \to \mathbf{t})}(\mathbf{x})}{\Delta_{(\mathbf{y} \to \mathbf{t})}(\mathbf{x}) + \Delta_{(\mathbf{s} \to \mathbf{y})}(\mathbf{x})}$, and set

$$\mathbf{z} \triangleq \alpha \mathbf{z}_{(\mathbf{s} \to \mathbf{y})} + (1 - \alpha) \mathbf{z}_{(\mathbf{y} \to \mathbf{t})}. \tag{2}$$

For points $\mathbf{x}$ near $\mathbf{y}$ such that $\Delta_{(\mathbf{s} \to \mathbf{y})}(\mathbf{x}), \Delta_{(\mathbf{y} \to \mathbf{t})}(\mathbf{x}) \geq 0$, we can now define the displacement analogously to (1):

$$\hat{g}(\mathbf{x}) \triangleq \begin{cases} \delta \cdot \left[\alpha \frac{(\mathbf{y}-\mathbf{s})}{\|\mathbf{y}-\mathbf{s}\|_2} + (1-\alpha) \frac{(\mathbf{t}-\mathbf{y})}{\|\mathbf{t}-\mathbf{y}\|_2}\right] & \|\mathbf{x}-\mathbf{z}\|_2 = 0 \\ \delta \frac{(\mathbf{z}-\mathbf{x})}{h} & \|\mathbf{x}-\mathbf{z}\|_2 = h \\ \delta \cdot \left[\alpha \frac{(\mathbf{s}-\mathbf{y})}{\|\mathbf{y}-\mathbf{s}\|_2} + (1-\alpha) \frac{(\mathbf{y}-\mathbf{t})}{\|\mathbf{t}-\mathbf{y}\|_2}\right] & \|\mathbf{x}-\mathbf{z}\|_2 = 2h \\ \delta \left(\mathbf{0}_{3m}, \mathbf{1}_m\right) & \|\mathbf{x}-\mathbf{z}\|_2 \geq 3h \end{cases}. \tag{3}$$

At intermediate distances, interpolate according to $\|\mathbf{x} - \mathbf{z}\|_2$. Notice that for each fixed choice of $\alpha \in [0, 1]$ (and $\mathbf{z}$), $\hat{g}$ is $O(\delta/h) = O(1)$-Lipschitz. Furthermore, $\Delta_{(\mathbf{s} \to \mathbf{y})}$ and $\Delta_{(\mathbf{y} \to \mathbf{t})}$ are 1-Lipschitz in $\mathbf{x}$. For any $\mathbf{z} \in L_{\mathbf{y}}$, $\Delta_{(\mathbf{y} \to \mathbf{t})}(\mathbf{z}) + \Delta_{(\mathbf{s} \to \mathbf{y})}(\mathbf{z}) = \sqrt{h}$. For general $\mathbf{x}$, we have

$$\Delta_{(\mathbf{y} \to \mathbf{t})}(\mathbf{x}) + \Delta_{(\mathbf{s} \to \mathbf{y})}(\mathbf{x}) \geq \Delta_{(\mathbf{y} \to \mathbf{t})}(\mathbf{z}) + \Delta_{(\mathbf{s} \to \mathbf{y})}(\mathbf{z}) - 2\|\mathbf{x} - \mathbf{z}\|_2 = \sqrt{h} - 2\|\mathbf{x} - \mathbf{z}\|_2; \tag{4}$$

so $\alpha$ is $O\left(1/\sqrt{h}\right)$-Lipschitz whenever $\|\mathbf{x} - \mathbf{z}\|_2 < 3h$, and otherwise has no effect on $\hat{g}(\mathbf{x})$. We conclude that $\hat{g}$ is still $O(1)$-Lipschitz when interpolating across different values of $\alpha$. At the interface with (1), $\alpha$ is 1 (0 near $\mathbf{z}_{(\mathbf{y} \to \mathbf{t})}$), so (1) and (3) are equal. Therefore $\hat{g}$ is $O(1)$-Lipschitz on all of $[-1, 2]^{4m}$.

To lower bound the magnitude of the displacement, we argue that $(\mathbf{z} - \mathbf{x})$ is orthogonal to $\left[\alpha \frac{(\mathbf{y}-\mathbf{s})}{\|\mathbf{y}-\mathbf{s}\|_2} + (1-\alpha) \frac{(\mathbf{t}-\mathbf{y})}{\|\mathbf{t}-\mathbf{y}\|_2}\right]$. First, observe that we can restrict our attention to the component of $(\mathbf{z} - \mathbf{x})$ that belongs to the plane defined by $\mathbf{s}, \mathbf{y}, \mathbf{t}$ (in which $\mathbf{z}$ also lies). Let $P_{\mathbf{s},\mathbf{y},\mathbf{t}}(\mathbf{x})$ denote the projection of $\mathbf{x}$ to this plain. We can write points in this plane in terms of their $\Delta(\cdot) \triangleq \left(\Delta_{(\mathbf{s} \to \mathbf{y})}(\cdot), \Delta_{(\mathbf{y} \to \mathbf{t})}(\cdot)\right)$ values. (Recall that $(\mathbf{s} \to \mathbf{y})$ and $(\mathbf{y} \to \mathbf{t})$ are orthogonal.)

First, observe that $\Delta\left(\mathbf{z}_{(\mathbf{s}\to\mathbf{y})}\right) = \left(0, \sqrt{h}\right)$, $\Delta\left(\mathbf{z}_{(\mathbf{y}\to\mathbf{t})}\right) = \left(\sqrt{h}, 0\right)$ and $\Delta(\mathbf{y}) = \left(\sqrt{h}, \sqrt{h}\right)$. Notice also that

$$\left[\alpha \frac{(\mathbf{y}-\mathbf{s})}{\|\mathbf{y}-\mathbf{s}\|_2} + (1-\alpha) \frac{(\mathbf{t}-\mathbf{y})}{\|\mathbf{t}-\mathbf{y}\|_2}\right] = \left[\alpha \frac{(\mathbf{y}-\mathbf{z}_{(\mathbf{s}\to\mathbf{y})})}{\sqrt{h}} + (1-\alpha) \frac{(\mathbf{z}_{(\mathbf{y}\to\mathbf{t})}-\mathbf{y})}{\sqrt{h}}\right].$$

Putting those together, we have that

$$\Delta\left(\left[\alpha \frac{\mathbf{y}}{\|\mathbf{y}-\mathbf{s}\|_2} + (1-\alpha) \frac{\mathbf{t}}{\|\mathbf{t}-\mathbf{y}\|_2}\right]\right) - \Delta\left(\left[\alpha \frac{\mathbf{s}}{\|\mathbf{y}-\mathbf{s}\|_2} + (1-\alpha) \frac{\mathbf{y}}{\|\mathbf{t}-\mathbf{y}\|_2}\right]\right) = (\alpha, 1-\alpha). \tag{5}$$

For $\mathbf{z}$, we have

$$\Delta(\mathbf{z}) = \alpha\Delta\left(\mathbf{z}_{(\mathbf{s}\to\mathbf{y})}\right) + (1-\alpha)\Delta\left(\mathbf{z}_{(\mathbf{y}\to\mathbf{t})}\right) = \sqrt{h}(1-\alpha, \alpha).$$

Finally, for $P_{\mathbf{s},\mathbf{y},\mathbf{t}}(\mathbf{x})$, we can write

$$\Delta(P_{\mathbf{s},\mathbf{y},\mathbf{t}}(\mathbf{x})) = \left(\Delta_{(\mathbf{y}\to\mathbf{t})}(\mathbf{x}), \Delta_{(\mathbf{s}\to\mathbf{y})}(\mathbf{x})\right)$$
$$= \frac{1}{\Delta_{(\mathbf{y}\to\mathbf{t})}(\mathbf{x}) + \Delta_{(\mathbf{s}\to\mathbf{y})}(\mathbf{x})}(1-\alpha, \alpha).$$

Therefore $\Delta(\mathbf{z}) - \Delta(P_{\mathbf{s},\mathbf{y},\mathbf{t}}(\mathbf{x}))$ is orthogonal to (5).

**Outside the picture**

The displacement outside the picture is constructed by interpolating the displacement at $\mathsf{E}_{i\in\{3m+1,\ldots,4m\}}x_i = 1/2$, and the displacement at points in the "top" of the hypercube, where $x_i = 2$ for every $i$ in the last $m$ coordinates. The former displacement, where $\mathsf{E}_{i\in\{3m+1,\ldots,4m\}}x_i = 1/2$ is defined to match the displacement inside the picture. Namely, it is the default displacement everywhere except near the first Brouwer line segment which goes "down" from $\mathbf{s} = (\mathbf{0}_{3m}, 2\cdot\mathbf{1}_m)$ to $\mathbf{t} = (\mathbf{0}_{4m})$. Near this line, it is defined according to (1). (Notice that $\|\mathbf{t}-\mathbf{s}\|_2 = 1$.)

Formally, let $\mathbf{z}_{1/2} = \left(\mathbf{0}_{3m}, \frac{1}{2}\cdot\mathbf{1}_m\right)$; for $\mathbf{x}$ on the boundary of the picture, we have:

$$\hat{g}(\mathbf{x}) \triangleq \begin{cases} \delta(\mathbf{0}_{3m}, -\mathbf{1}_m) & \|\mathbf{x}-\mathbf{z}_{1/2}\|_2 = 0 \\ \delta\frac{(\mathbf{z}_{1/2}-\mathbf{x})}{h} & \|\mathbf{x}-\mathbf{z}_{1/2}\|_2 = h \\ \delta(\mathbf{0}_{3m}, \mathbf{1}_m) & \|\mathbf{x}-\mathbf{z}_{1/2}\|_2 \geq 2h \end{cases} \tag{6}$$

For points $\mathbf{x}$ such that $\mathsf{E}_{i\in\{3m+1,\ldots,4m\}}x_i$ is very close to 2, the displacement $\delta(\mathbf{0}_{3m}, \mathbf{1}_m)$ is not helpful because it points outside the hypercube, i.e. it would get completely erased by the truncation. Instead, we define the displacement as follows:

$$\hat{g}(\mathbf{x}) \triangleq \begin{cases} \delta(\mathbf{0}_{3m}, -\mathbf{1}_m) & \|\mathbf{x}-\mathbf{z}_2\|_2 = 0 \\ \delta\frac{(\mathbf{z}_1-\mathbf{x})}{h} & \|\mathbf{x}-\mathbf{z}_2\|_2 \geq h, \end{cases} \tag{7}$$

where $\mathbf{z}_2 = (\mathbf{0}_{3m}, 2\cdot\mathbf{1}_m)$. When $\theta \triangleq \mathsf{E}_{i\in\{3m+1,\ldots,4m\}}x_i \in (1/2, 2)$, we interpolate between (6) and (7) according to $\frac{\theta-1/2}{3/2}$. □

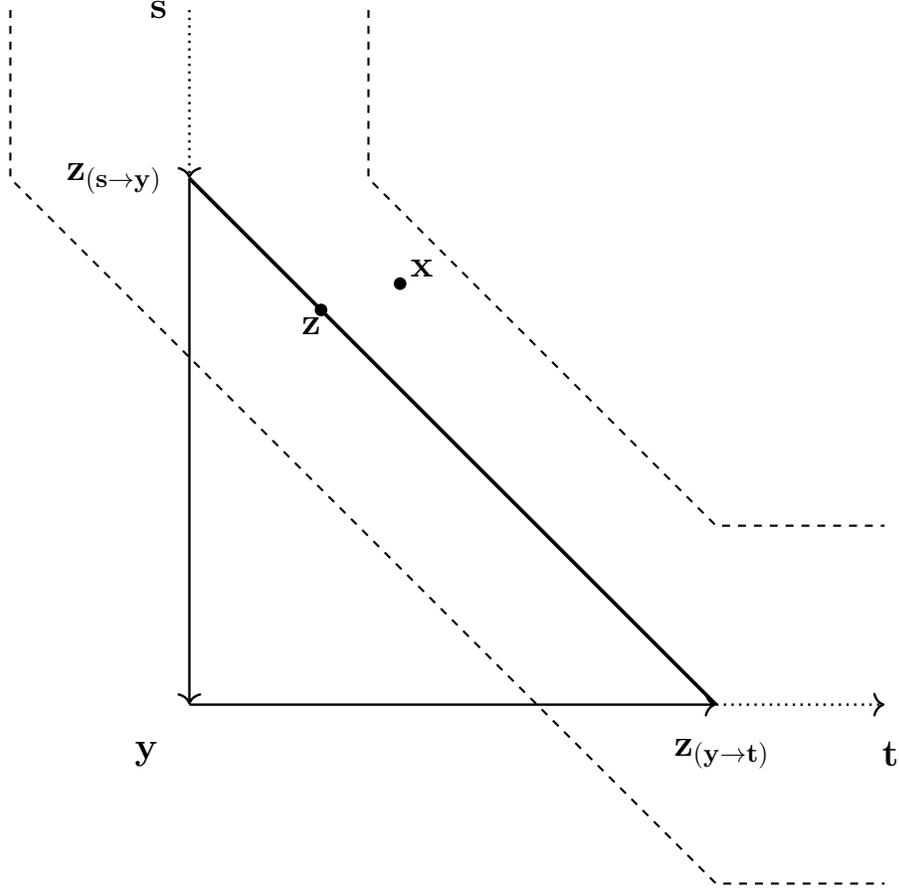

Figure 1: Geometry near a Brouwer vertex

The figure (not drawn to scale) shows some of the important points near a Brouwer vertex $\mathbf{y}$: There is an incoming Brouwer line segment from $\mathbf{s}$ through $\mathbf{z}_{(\mathbf{s}\to\mathbf{y})}$, and an outgoing Brouwer line segment to $\mathbf{t}$ through $\mathbf{z}_{(\mathbf{y}\to\mathbf{t})}$. For each point $\mathbf{x}$ between the dashed lines, we assign a point $\mathbf{z}$ on the line $L_{\mathbf{y}}$ as in (2), and define the displacement according to (3). Outside the dashed lines (including at $\mathbf{y}$ itself), we use the default displacement $\delta\left(\mathbf{0}_{3m}, \mathbf{1}_m\right)$.

## 3.3 Locally computing the Brouwer function

In the Brouwer function above, to compute all of $f(\mathbf{x})$ exactly, we essentially need to know $\mathbf{x}$ in every coordinate. However, in order to compute $f_i(\mathbf{x})$ (the $i$-th coordinate of $f(\mathbf{x})$), it suffices to know $x_i$ and that $\mathbf{x}$ is:

- inside the picture, but far from every line;

- close to some point $\mathbf{z}$ on line $(\mathbf{s} \to \mathbf{t})$ (but far from $\mathbf{s}$ and $\mathbf{t}$),
    - also need to know $s_i, t_i, z_i, \|\mathbf{x} - \mathbf{z}\|_2$ and $\|\mathbf{t} - \mathbf{s}\|_2$;

- close to some point $\mathbf{z}$ on line $L_\mathbf{y}$ for vertex $\mathbf{y}$ on the intersection of lines $(\mathbf{s} \to \mathbf{y})$ and $(\mathbf{y} \to \mathbf{t})$,
    - also need to know $s_i, y_i, t_i, z_i, \|\mathbf{x} - \mathbf{z}\|_2, \|\mathbf{y} - \mathbf{s}\|_2, \|\mathbf{t} - \mathbf{y}\|_2$, and $\alpha$; or

- outside the picture,
    - also need to know $\mathsf{E}_{i \in \{3m+1,\ldots,4m\}} x_i$ and $\|\mathbf{x} - \mathbf{z}\|_2$, where $\mathbf{z}$ is the $\left(\mathsf{E}_{i \in \{3m+1,\ldots,4m\}} x_i\right)$-weighted average of $\mathbf{z}_{1/2}$ and $\mathbf{z}_2$.

By Lipschitz continuity, if we only want to compute $f_i(\mathbf{x})$ to within $\pm \epsilon$, it suffices to know all the quantities above to within $\pm \epsilon$. Furthermore, at distance $\pm \epsilon$ near interfaces between the different cases (inside/outside the picture, close to 0/1/2 lines), we can use the wrong displacement, and still be within $\pm \epsilon$ of $f_i(\mathbf{x})$.

## 4 Multiplayer games

In this section we establish hardness results for multiplayer games. We first show PPAD-hardness of $(\epsilon, \delta)$-WeakNash in multiplayer games where each player has a large constant number of actions. This is superseded by Corollary 4.2, which proves the same for games where each player has only two actions. Nevertheless, we begin with the proof of the weaker Corollary 4.1, as it is the simplest working example of the imitation gadget argument used both in Corollary 4.2, and in the proof our main theorem (in particular, the construction in Section 7). All the PPAD-hardness results in this section have query complexity hardness analogues, as described in Subsection 4.2.

**Corollary 4.1.** *There exist constants $\epsilon, \delta > 0$, such that finding an $(\epsilon, \delta)$-WeakNash is PPAD-hard for succinct multiplayer games where each player has a constant number of actions.*

*Proof.* Let $f$ be the hard function guaranteed by Theorem 3.2. We construct a game with two groups of $n$ players each. The action set of each player corresponds to $\{0, 1/k, 2/k \ldots, 1\}$ for a sufficiently large constant $k > 0$. We denote the choice of strategies for the first group $\mathbf{a} \triangleq (a_1 \ldots a_n)$, and $\mathbf{b} \triangleq (b_1, \ldots b_n)$ for the second group.

Each player $(A, i)$ in the first group attempts to imitate the behavior of the corresponding player in the second group. Her utility is given by

$$u_i(a_i, b_i) \triangleq -|a_i - b_i|^2.$$

The second group players attempt to imitate the value of $f$, when applied to the vector of actions taken by all the players in the first group. The utility of the $j$-th player, $(B, j)$, is

$$v_j(b_j, \mathbf{a}) \triangleq -|f_j(\mathbf{a}) - b_j|^2,$$

where $f_j$ denotes the $j$-th output of $f$.

Observe that the expected utility for $(A, i)$ is given by:

$$\mathsf{E}[u_i(a_i, b_i)] = -|a_i - \mathsf{E}(b_i)|^2 - \mathrm{Var}(b_i).$$

For any value of $\mathsf{E}(b_i)$, player $(A, i)$ has an action $\alpha_i \in [\mathsf{E}(b_i) - 1/2k, \mathsf{E}(b_i) + 1/2k]$. Her utility when playing $\alpha_i$ is lower bounded by:

$$\mathsf{E}[u_i(\alpha_i, b_i)] \geq -\frac{1}{4k^2} - \mathrm{Var}(b_i).$$

On the other hand, for any $\widehat{a_i} \notin \{\alpha_i, \alpha_i + 1/k\}$,

$$\mathsf{E}[u_i(\widehat{a_i}, b_i)] \leq -\frac{1}{k^2} - \mathrm{Var}(b_i).$$

Therefore in every $(\delta, \delta/k^2)$-WeakNash, it holds for all but a $2\delta$-fraction of $i$'s, that player $(A, i)$ assigns probability at least $(1 - O(\delta))$ to strategies $\{\alpha_i, \alpha_i + 1/k\}$. In particular, with high probability over the randomness of the players, it holds that all but an $O(\delta)$-fraction of the players play one of those strategies. Therefore with high probability

$$\|\mathbf{a} - \mathsf{E}(\mathbf{b})\|_2 = O\left(\sqrt{\delta} + 1/k\right). \tag{8}$$

Similarly, player $(B, j)$ (the $j$-th player in the second group) has an action $\beta_j \in [\mathsf{E}(f_j(\mathbf{a})) - 1/2k, \mathsf{E}(f_j(\mathbf{a})) + 1/2k]$. Therefore in every $(\delta, \delta/k^2)$-WeakNash, it holds for all but a $2\delta$-fraction of $j$'s, that player $(B, j)$ assigns probability at least $(1 - O(\delta))$ to strategies $\{\beta_j, \beta_j + 1/k\}$. In particular, with high probability over the randomness of the players, it holds that all but an $O(\delta)$-fraction of the players play one of those strategies. Therefore with high probability

$$\|\mathbf{b} - \mathsf{E}(f(\mathbf{a}))\|_2 = O\left(\sqrt{\delta} + 1/k\right).$$

Now, by the Lipschitz condition on $f$, whenever (8) holds,

$$\|f(\mathbf{a}) - f(\mathsf{E}(\mathbf{b}))\|_2 = O\left(\sqrt{\delta} + 1/k\right).$$

Therefore $\mathsf{E}(\mathbf{a})$ and $\mathsf{E}(\mathbf{b})$ are both solutions to the $\Theta\left(\sqrt{\delta} + 1/k\right)$-SUCCINCTBROUWER$_2$ instance. □

### 4.1 Binary actions

**Corollary 4.2.** *There exist constants $\epsilon, \delta > 0$, such that finding an $(\epsilon, \delta)$-WeakNash is PPAD-hard for succinct multiplayer games where each player has two actions.*

*Proof.* We replace each player $(A, i)$ (respectively, $(B, j)$) in the proof of Corollary 4.1 with $k + 1$ players, denoted: $(A, i, 0), (A, i, 1/k), \ldots, (A, i, 1)$. Each new player has two actions:

$\{+, -\}$. Given a $(k+1)$-tuple of actions for players $\{(A, i, \cdot)\}$, we define the *realized value* $r_i \in [0, 1]$ to be
$$r_i \triangleq \max \{x \text{ s.t. } (A, i, x) \text{ plays action } +\}.$$
Let the realized value $q_j$ for players $(B, j, 0), \ldots, (B, j, 1)$ be defined analogously.

We let player $(A, i, x)$'s utility be:
$$u_{i,x}(+, \mathbf{b}) \triangleq -\left|\left(x + \frac{1}{k}\right) - q_i\right|^2$$
$$u_{i,x}(-, \mathbf{b}) \triangleq -\left|\left(x - \frac{1}{k}\right) - q_i\right|^2.$$

Similarly, for player $(B, j, y)$, we have
$$u_{j,y}(\pm, \mathbf{a}) \triangleq -\left|\left(y \pm \frac{1}{k}\right) - f_i(\mathbf{r})\right|^2,$$
where $\mathbf{r} \triangleq (r_1, \ldots, r_n)$ is the vector of realized values on the first group.

For any $(A, i, x)$, we have
$$\mathsf{E}[(u_{i,x}(\pm, \mathbf{b}))] = -\left|\left(x \pm \frac{1}{k}\right) - \mathsf{E}[q_i]\right|^2 - \mathrm{Var}[q_i].$$

Subtracting her two possible payoffs, we have
$$\mathsf{E}[(u_{i,x}(+, \mathbf{b})) - (u_{i,x}(-, \mathbf{b}))] = \left|(x - \mathsf{E}[q_i]) - \frac{1}{k}\right|^2 - \left|(x - \mathsf{E}[q_i]) + \frac{1}{k}\right|^2$$
$$= -\frac{4}{k}(x - \mathsf{E}[q_i]).$$

In particular, if $\mathsf{E}[q_i] > x + k\sqrt{\epsilon}$, player $(A, i, x)$ must assign probability at most $\sqrt{\epsilon}$ to action $\{-\}$ in order to play $\epsilon$-optimally. For any $i$ such that all players $(A, i, \cdot)$ use $\epsilon$-optimal mixed strategies, we have that
$$|\mathsf{E}[q_i] - \mathsf{E}[r_i]| = O\left(k\sqrt{\epsilon} + 1/k\right). \tag{9}$$

In any $(\epsilon, \delta)$-WeakNash, for all but a $2\delta k$-fraction of $i$'s it holds that all players $(A, i, \cdot)$ use $\epsilon$-optimal mixed strategies; thus (9) holds for all but a $2\delta k$-fraction of $i$'s. Therefore,
$$\|\mathsf{E}[\mathbf{q}] - \mathsf{E}[\mathbf{r}]\|_2 = O\left(\sqrt{\delta k} + k\sqrt{\epsilon} + 1/k\right).$$

By the Lipschitz condition on $f$, the latter also implies,
$$\|f(\mathsf{E}[\mathbf{q}]) - f(\mathsf{E}[\mathbf{r}])\|_2 = O\left(\sqrt{\delta k} + k\sqrt{\epsilon} + 1/k\right).$$

Similarly, for every $j$ such that all players $(B, j, \cdot)$ use $\epsilon$-optimal mixed strategies, we have that
$$|\mathsf{E}[q_j] - f_j(\mathsf{E}[\mathbf{r}])| = O\left(k\sqrt{\epsilon} + 1/k\right).$$

Thus in any $(\epsilon, \delta)$-WeakNash,
$$\|\mathsf{E}[\mathbf{q}] - f(\mathsf{E}[\mathbf{r}])\|_2 = O\left(\sqrt{\delta k} + k\sqrt{\epsilon} + 1/k\right).$$

Therefore $\mathsf{E}(\mathbf{q})$ and $\mathsf{E}(\mathbf{r})$ are both solutions to the $\Theta\left(\sqrt{\delta k} + k\sqrt{\epsilon} + 1/k\right)$-SUCCINCTBROUWER$_2$ instance. □

## 4.2 Query complexity

Notice that the proofs of Theorem 3.2 and Corollary 4.2 access the successor and predecessor circuits as black box. In particular, we can replace the circuits with access to a black box oracle, and ask how many oracle queries are necessary to compute an approximate Brouwer fixed point or an $(\epsilon, \delta)$-WeakNash. The reductions in Theorem 3.2 and Corollary 4.2 imply that both require at least as many queries as solving the ENDOFALINE instance with oracle access to the successor and predecessor functions: $2^{\Omega(n)}$ queries (**e.g. [13, Proposition 4]**). We therefore have the following corollaries.

**Definition 4.3** ($\epsilon$-ORACLEBROUWER$_p$). Given black-box oracle access to a function $f\colon [0,1]^n \to [0,1]^n$ that is $1/\epsilon$-Lipschitz with respect to $p$-norm, $\epsilon$-ORACLEBROUWER$_p$ is the problem of computing an $\mathbf{x}$ such that $\|f(\mathbf{x}) - \mathbf{x}\|_p \leq \epsilon$.

**Corollary 4.4.** *There exist a constant $\epsilon > 0$, such that any (potentially randomized) algorithm for $\epsilon$-ORACLEBROUWER$_2$ requires $2^{\Omega(n)}$ queries.*

**Corollary 4.5.** *There exist constants $\epsilon, \delta > 0$, such that any (potentially randomized) algorithm for finding an $(\epsilon, \delta)$-WeakNash for multiplayer games where each player has two actions requires $2^{\Omega(n)}$ queries to the players' payoffs tensors.*

## 5 ENDOFALINE with local computation

In this section we introduce a local variant of ENDOFALINE with very simple successor and predecessor circuits.

**Definition 5.1** (LOCALENDOFALINE). The problem LOCALENDOFALINE is similar to ENDOFALINE, but the graph is defined on a subset $V^{\text{LOCAL}} \subseteq \{0,1\}^n$. The input consists of a membership circuit $M_{V^{\text{LOCAL}}}\colon \{0,1\}^n \to \{0,1\}$ that determines whether a string $u \in \{0,1\}^n$ corresponds to a vertex of the graph (i.e. $M_{V^{\text{LOCAL}}}(u) = \begin{cases} 1 & u \in V^{\text{LOCAL}} \\ 0 & u \notin V^{\text{LOCAL}} \end{cases}$), a special vertex $u_\mathbf{0} \in V^{\text{LOCAL}}$, and successor and predecessor circuits $S^{\text{LOCAL}}, P^{\text{LOCAL}}\colon \{0,1\}^n \to \{0,1\}^n$ with the promise that every output bit depends only on a constant number of input bits. (I.e. $S^{\text{LOCAL}}$ and $P^{\text{LOCAL}}$ are in $\mathsf{NC}^0$.) Furthermore, we require that the outputs of $S^{\text{LOCAL}}$ and $P^{\text{LOCAL}}$ are identical to the respective inputs, except at a constant number of coordinates. Similarly to ENDOFALINE, we are guaranteed that $P^{\text{LOCAL}}(u_\mathbf{0}) = u_\mathbf{0} \neq S^{\text{LOCAL}}(u_\mathbf{0})$.

The goal is to find an input $u \in V^{\text{LOCAL}}$ that satisfies any of the following:

- End-of-a-line: $P^{\text{LOCAL}}(S^{\text{LOCAL}}(u)) \neq u$ or $S^{\text{LOCAL}}(P^{\text{LOCAL}}(u)) \neq u \neq u_\mathbf{0}$; or
- Boundary conditions: $S^{\text{LOCAL}}(u) \notin V^{\text{LOCAL}}$ or $P^{\text{LOCAL}}(u) \notin V^{\text{LOCAL}}$.

Notice that each vertex $u \in V^{\text{LOCAL}}$, only defers from $S^{\text{LOCAL}}(u)$ and $P^{\text{LOCAL}}(u)$ at a constant number of bits, and the value of each of those only depends on a constant number of bits of $u$. We will refer to all those bits as the *critical bits* of $u$, and denote their cardinality $q_{\text{CRITICAL}}$.

We now reduce ENDOFALINE to LOCALENDOFALINE, proving that the latter is PPAD-complete. In the following sections we use this reduction "white-box", and revisit some of its specific properties.

**Theorem 5.2.** *There is a linear-time reduction from* ENDOFALINE *to* LOCALENDOFALINE.

Notice that there is a trivial reduction in the other direction: add self-loops to all strings not in $V^{\text{LOCAL}}$.

*Proof of Theorem 5.2.* Given circuits $S, P\colon \{0,1\}^n \to \{0,1\}^n$, we construct new circuits $S^{\text{LOCAL}}, P^{\text{LOCAL}} \colon \{0,1\}^m \to \{0,1\}^m$ that satisfy Definition 5.1. We assume wlog that the circuits have fan-out 2 (otherwise, replace gates with larger fan-outs with a binary tree of equality gates; this only blows up the size of the circuit by a constant factor). We set $m = 4(|S| + |P|)$, where by $|S|$ and $|P|$ we mean the number of lines (i.e. number of inputs + number of logic gates) in each circuit, respectively. We think of the $m$ bits as representing two copies of each circuit ($S_1, P_1$ and $S_2, P_2$), with each line represented by two bits, representing three possible states: value 0, value 1, and inactive.

The path begins with circuit $S_1$ containing the computation from $\mathbf{0}_n$ to $S(\mathbf{0}_n)$, and circuit $P_1$ containing the reverse computation back to $\mathbf{0}_n$; the lines of circuits $S_2, P_2$ are inactive. This is the special vertex, $u_{\mathbf{0}}$.

Over the next $n$ steps, the output bits of $S_1$ are copied (one-by-one) to the input bits of $S_2$ (and the corresponding lines are activated). Over the next $|S| - n$ steps, the values on the lines of $S_2$ are updated -one-by-one, starting from the inputs and propagating to the outputs in a fixed order- until all the output bits are activated and set to $S(S(\mathbf{0}_n))$. Then, over the next $|P|$ steps, the input of $P_2$ is set to $S(S(\mathbf{0}_n))$, the values on the gates are updated (propagating in a fixed order from output to input), and finally setting the output of $P_2$ to $S(\mathbf{0}_n)$.

Notice that so far, if we want to trace back a step (a.k.a. implement $P^{\text{LOCAL}}$), we simply need to deactivate the last activated line. We now start erasing values when going forward, but we must do so carefully so that we can reconstruct them when going backward. We proceed by deactivating the lines in $S_1$: line by line, in reverse order from outputs to inputs. Indeed, in order to go back we can reconstruct the output of any gate from its inputs. If we want to reconstruct an input to $S_1$ - this is no problem as it is saved as the output of $P_1$. Once we finish deactivating $S_1$, we deactivate $P_1$, also in reverse order. Notice again that the input to $P_1$ is saved as the input to $S_2$ and output of $P_2$. Now, we copy the values from $S_2$ and $P_2$ to $S_1$ and $P_1$ in the same order that we used to deactivate the lines, starting from the outputs of $S_1$, to inputs, to outputs of $P_1$, to inputs. Then we deactivate the lines in $S_2$ and $P_2$; again, we use the same order in which they were activated, starting from inputs of $S_2$, to outputs, to inputs of $P_2$, to outputs. So now we have $S_1$ as a witness to the computation of $S(S(\mathbf{0}_n))$ from $S(\mathbf{0}_n)$ and $P_1$ goes in the reverse order, while all the lines of $S_2$ and $P_2$ are inactive.

We repeat the process from the last two paragraphs to find $S(S(S(\mathbf{0}_n)))$, etc.

We let $V^{\text{LOCAL}}$ to be the set of strings that correspond to a legal partial computation as described above. $M_{V^{\text{LOCAL}}}$ verifies that a string corresponds to one of the following scenarios:

- $S_1$ holding the computation from some $x \in \{0,1\}^n$ to $S(x)$, $P_1$ holding the computation from $S(x)$ back to $x$, and $S_2$ and $P_2$ holding part of the computation from $S(x)$ to $S(S(x))$ and back to $S(x)$;

- $S_2$ and $P_2$ holding the computation from some $x$ to $S(x)$ and back to $x$, and $P_1$ and $S_1$ are in the process of erasing the computation from $x$ to $P(x)$ and back to $x$;

- $S_2$ and $P_2$ holding the computation from some $x$ to $S(x)$ and back to $x$, and $S_1$ and $P_1$ are in the process of copying it; or

- $S_1$ and $P_1$ holding the computation from some $x$ to $S(x)$ and back to $x$, are $S_2$ and $P_2$ in the process of erasing the same computation.

Notice that the joint active/inactive state of all the lines always belongs to one of $O(m)$ different states (that can be described by $\log_2 m + O(1)$ bits. Verifying that a string corresponds to one of the above scenarios is equivalent to checking that the joint active/inactive state is valid, and that the values on all the lines satisfy all of the following *local conditions*:

- The $i$-th input of $S_1$ is either inactive, equal to the $i$-th output of $P_1$, or, if the latter is inactive (during the copy phase), equal to the $i$-th input of $S_2$.

- The $i$-th input of $P_1$ is either inactive, equal to the $i$-th output of $S_1$, or, if the latter is inactive (during the erase phase), equal to the $i$-th input of $S_2$.

- The output of every gate in $S_1, P_1$ is either inactive, equal to the gate applied to its inputs, or, if either of the inputs is inactive (during the copy phase), equal to the respective line in $S_2, P_2$.

- The $i$-th input of $S_2$ is either inactive, equal to the $i$-th output of $P_2$, or, if the latter is inactive (during the compute phase), equal to the $i$-th input of $P_2$.

- The $i$-th input of $P_2$ is either inactive, equal to the $i$-th output of $S_2$, or, if the latter is inactive (during the erase phase), equal to the $i$-th input of $P_1$.

- The output of every gate in $S_2, P_2$ is either inactive, equal to the gate applied to its inputs, or, if either of the inputs is inactive (during the erase phase), equal to the respective line in $S_1, P_1$.

Notice that each line participates in at most three local constraints. Note also that each of those conditions only depends on a constant number of bits so $M_{V^{\text{LOCAL}}}$, which simply needs to compute their AND, is in $\mathsf{AC}^0$ (this proves Theorem 1.5).

In every step a line is either activated or deactivated, so there are no fixed points; but fixed points of the original circuit lead to violations of the form $P^{\text{LOCAL}}(S^{\text{LOCAL}}(u)) \neq u$ and $S^{\text{LOCAL}}(P^{\text{LOCAL}}(u)) \neq u \neq u_0$. In particular, every solution to the new LOCALENDOFALINE instance corresponds to a valid solution to the original ENDOFALINE instance.

Also, notice that in every step we change at most two bits (and at least one), and deciding the next value of each bit can be done by looking only at a constant number of bits (conditioned on being a legal vertex in $V^{\text{LOCAL}}$): which is the next line to be activated/deactivated? if activated, what are the inputs to the corresponding gate? □

## 5.1 The LOCALENDOFALINE counter

Consider the instance produced in the reduction above. Notice that joint active/inactive state of all the lines rotates among $O(m)$ possible vectors. I.e. there is a $(\log m + O(1))$-bit vector, henceforth called the *counter* which describes precisely which lines should be active. Furthermore, given the counter of some $u \in V^{\text{LOCAL}}$, it is easy to compute the counter of $S^{\text{LOCAL}}(u)$, i.e. we always know which line should be activated/deactivated next - regardless of the values on the lines. Similarly, for all $u \neq u_0 \in V^{\text{LOCAL}}$, it is also easy to compute the counter of $P^{\text{LOCAL}}(u)$. An additional useful property is that given the counter of $u$, we also know the coordinates of the critical bits of $u$.

# 6 Holographic Proof

In this section we construct our holographic proof system for statements of the form $u \in V^{\text{LOCAL}}$. Note that we use the term "verifier" loosely, as our verifier's responsibilities extend far beyond checking the validity of a proof - it is also expected to perform local decoding, "local proof construction", etc.

**Proposition 6.1.** *For sufficiently small constants $\epsilon_{\text{DECODE}} \gg \epsilon_{\text{SOUND}} \gg \epsilon_{\text{COMPLETE}} > 0$, the following holds. Fix any instance $(S^{\text{LOCAL}}, P^{\text{LOCAL}}, V^{\text{LOCAL}})$ of* LOCALENDOFALINE, *generated via the reduction in Section 5 from an instance $(S, P)$* ENDOFALINE. *Given a vertex $u \in V^{\text{LOCAL}}$, there is a polynomial time deterministic algorithm that takes as input $(S, P)$ and $u$, and outputs a holographic proof $\Pi(u) = (E(u), C(u), \pi(u))$ where:*

- $E(u)$ *is an **encoding** of $u$ with a **linear** error correcting code of constant relative distance.*

- $C(u)$ *is an encoding of the **counter** of $u$ (with a good error correcting code).*

- $\pi(u)$ *is a **proof** that $u \in V^{\text{LOCAL}}$.*

- *Let $n \triangleq |u|$; then the total length of $\Pi(u)$ is $n^{1+o(1)}$.*

- *Let $t \triangleq \sqrt{\log n}$ (in particular, $t = \omega(1)$, but also $t = o\left(\frac{\log n}{\log \log n}\right)$). Let $\mathcal{G}$ be a finite field of size $O\left(n^{1/t+o(1)}\right)$. $E(u)$ and $\pi(u)$ can be written as functions over domain $\mathcal{G}^t$. In particular, we can talk about accessing $E(u)$, $\pi(u)$, or $\Pi(u)$ at a point $\mathbf{g} \in \mathcal{G}^t$. ($C(u)$ is much shorter and the verifier reads it entirely.)*

- *There is a quasi-polynomial[8] time probabilistic verifier such that:*

    - *(**Local access**) The verifier's access to $\Pi(u)$ is restricted as follows:*
        * *(**Querying subspaces**) The verifier reads $n^{o(1)}$ axis-parallel $(t/2)$-dimensional subspaces of $\Pi(u)$. Each subspace is defined by a restriction of a $(t/2)$-tuple of coordinates. The tuple of coordinates which is restricted is always one of a constant number of possibilities (either $\{2, \ldots, t/2+1\}$, or the union of any two of: $\{1, \ldots, t/4\}$, $\{t/4+1, \ldots, t/2\}$, $\{t/2+1, \ldots, 3t/4\}$, and $\{3t/4+1, \ldots, t\}$). We denote the union of all the subspaces read by the verifier $G_{\text{ALL}}$. It will be useful to decompose $G_{\text{ALL}} = G_{\text{PCP}} \cup G_{\text{LTC}} \cup G_{\text{CRITICAL}}$; another subset of interest is $G_{\text{SAMPLE}} \subset G_{\text{PCP}}$.*
        * *(**Partially adaptive**) $G_{\text{SAMPLE}}, G_{\text{PCP}}, G_{\text{LTC}}$ are chosen non-adaptively, i.e. they depend only on the verifier's internal randomness. $G_{\text{CRITICAL}}$ is a union of affine translations of a subspace that is also chosen non-adaptively, but the affine translations depend on $C(u)$.*
        * *(**Randomness**) The verifier uses only $(1/2 + o(1)) \log_2 n$ bits of randomness to decide which subspaces to query.*

---

[8]The verifier of our holographic proof actually runs in polynomial time, except for the derandomization of the construction of $\lambda$-biased sets. As we discuss in Section 2.1 this could also be obtained (with slightly worse parameters) in deterministic polynomial time, but quasi-polynomial time suffices for the purpose of our main theorem.

* **(Good Sample)** For any function $B\colon \mathcal{G}^t \to [0,1]$, and any $G \in \{G_{\text{SAMPLE}}, G_{\text{PCP}}, G_{\text{LTC}}, G_{\text{CRITICAL}}\}$, with probability $1 - o(1)$ over the verifier's randomness,
$$\left|\mathsf{E}_{\mathbf{g}\in G}\left[B\left(\mathbf{g}\right)\right] - \mathsf{E}_{\mathbf{h}\in\mathcal{G}^t}\left[B\left(\mathbf{h}\right)\right]\right| = o(1).$$

- **(Soundness)** If the verifier is given a string $\Pi'$ which is $\epsilon_{\text{SOUND}}$-far from $\Pi(u)$ for every $u \in V^{\text{LOCAL}}$, the verifier rejects with probability $1 - o(1)$.

  * **(Robust soundness)** Furthermore, with probability $1 - o(1)$, the bits read by the verifier are $\epsilon_{\text{SOUND}}^2$-far from any string that would make the verifier accept. In other words, the verifier continues to reject with probability $1 - o(1)$ even in the presence of an adaptive adversary, who observes the queries made by the verifier, and then corrupts an $\epsilon_{\text{SOUND}}^2/2$-fraction of each of $G_{\text{PCP}}, G_{\text{LTC}}$.

- **(Completeness)** If the verifier is given access to a string $\widehat{\Pi}$ that is $\epsilon_{\text{COMPLETE}}$-close to a valid $\Pi(u)$ for some $u \in V^{\text{LOCAL}}$, then the verifier accepts with probability $1 - o(1)$.

  * **(Robust completeness)** Furthermore, the verifier continues to accept with probability at least $1 - o(1)$ even in the presence of an adaptive adversary, who observes the queries made by the verifier, and then corrupts a $\sqrt{\epsilon_{\text{COMPLETE}}}$-fraction of $G_{\text{PCP}}, G_{\text{LTC}}$.

- **(Decoding)** If the verifier is given access to a string $\widehat{\Pi}$ that is $\epsilon_{\text{DECODE}}$-close to a valid $\Pi(u)$ for some $u \in V^{\text{LOCAL}}$:

  * **(Error-correction on a sample)** With probability at least $1 - o(1)$, the verifier correctly decodes the entries of $E(u)$ and $\pi(u)$ on $G_{\text{SAMPLE}}$. In particular, with probability $1 - o(1)$, the verifier can estimate the distance between $\widehat{\Pi}$ and $\Pi(u)$ to within an additive error of $\pm o(1)$.

  * **(Error-correction on the critical bits)** The verifier can *adaptively* decode and correct any $q_{\text{CRITICAL}}$-tuple of symbols from the proof with success probability at least $1 - o(1)$. Let $G_{\text{CRITICAL}} \subset \mathcal{G}^t$ denote the union of all $O\left(|\mathcal{G}|^2\right)$ axis-parallel subspaces queried in this process.

  * **(Proof-construction)** Given access as above to the counter $C(u)$ and the linear code $E(u)$ (but not to the proof $\pi(u)$), the verifier can output (with probability at least $1 - o(1)$) the correct values of $\pi(u)$ on $G_{\text{SAMPLE}}$.

  * **(Robust decoding)** Both types of error-correction and the proof-construction continue to hold even in the presence of an adaptive adversary, who observes the queries made by the verifier, and then corrupts a $\sqrt{\epsilon_{\text{DECODE}}}$-fraction of each of $G_{\text{SAMPLE}}, G_{\text{PCP}}, G_{\text{LTC}}, G_{\text{CRITICAL}}$.

The rest of this section is devoted to the proof of Proposition 6.1.

## 6.1 From LOCALENDOFALINE to coloring graphs

Let $t \triangleq \sqrt{\log n}$. Set $\ell \triangleq \frac{\log_2 n}{t-1} = \Theta\left(\sqrt{\log n}\right)$, and $\ell' \triangleq \ell + c\log\log n$ for some sufficiently large constant $c$. Let $p(z)$ be an $\mathbb{F}_2$-irreducible polynomial of degree $\ell'$, and let $\mathcal{G} \triangleq \mathbb{F}_2[z]/\langle p(z)\rangle$ denote the field of size $2^{\ell'}$. We will focus our attention on a subset $\mathcal{F} \subset \mathcal{G}$ of polynomials (modulo $p(z)$) of degree $\ell$; notice that $|\mathcal{G}|/|\mathcal{F}| = \mathsf{poly}\log n$.

Let $\alpha$ be a generator of $\mathcal{G}$, and let $\mathcal{E} = \{1, \alpha, \ldots, \alpha^{5t\ell}\}$. We define the *Extended Shuffle-$\mathcal{F}$-Exchange* graph on $\mathcal{F}^{t-1} \times \mathcal{E}$ to be the directed graph such that each vertex

$(x_1, \ldots, x_{t-1}, \alpha^i) \in \mathcal{F}^{t-1} \times (\mathcal{E} \setminus \{\alpha^{5t\ell}\})$, has a *static edge* to vertex $(x_1, \ldots, x_{t-1}, \alpha^{i+1})$, a *shuffle edge* to vertex $(x_2, \ldots, x_{t-1}, x_1, \alpha^{i+1})$, and $\ell$ *exchange edges* to all vertices of the form $(x_1 + [z^j], \ldots, x_{t-1}, \alpha^{i+1})$, for every $j \in \{0, \ldots, \ell-1\}$. Here $[z^j]$ is the element of $\mathcal{G}$ representing the equivalence class $\{z^j + q(z)\, p(z) : q(z) \in \mathbb{F}_2[z]\}$. Notice that $x \in \mathcal{F}$ iff it can be written as $x = \sum_{j \in S}[z^j]$ for some subset $S \subseteq \{0, \ldots, \ell-1\}$. In particular, for any $(x_1, \ldots, x_{t-1}), (y_1, \ldots, y_{t-1}) \in \mathcal{F}^{t-1}$, the Extended Shuffle-$\mathcal{F}$-Exchange graph contains a path of length $(t-1)(\ell+1)$ from $(x_1, \ldots, x_{t-1}, 1)$ to $(y_1, \ldots, y_{t-1}, \alpha^{(t-1)(\ell+1)})$.

For each $\alpha^i \in \mathcal{E}$, we call the set of vertices $\{(x_1, \ldots, x_{t-1}, \alpha^i)\}$ a *layer*; in particular, $\{(x_1, \ldots, x_{t-1}, 1)\}$ form the *first layer*, and $\{(x_1, \ldots, x_{t-1}, \alpha^{5t\ell})\}$ form the *last layer*.

We encode each vertex $u \in V^{\text{LOCAL}}$ as a partial coloring of the Extended Shuffle-$\mathcal{F}$-Exchange graph, representing the assignments to the lines in the four circuits $S_1, P_1, S_2, P_2$. Recall that $|\mathcal{F}| = 2^\ell = n^{1/(t-1)}$. We associate each line of $S_1, P_1, S_2, P_2$ (including input, gates, and output lines) with a vector $\mathbf{x} \in \mathcal{F}^{t-1}$ arbitrarily. For the first- and last-layer vertices $(\mathbf{x}, 1)$ and $(\mathbf{x}, \alpha^{5t\ell})$, we assign colors that represent the assignments (from $\{0, 1, \bot\}$, where $\bot$ represents an inactive line) to the corresponding line. Recall from the "local conditions" in Section 5 that every line only participates in three constraints.

We would like the inputs to the line in $(x_1, \ldots, x_{t-1}, \alpha^{5t\ell})$ to be propagated by the Extended Shuffle-$\mathcal{F}$-Exchange graph from the first layer. One important feature of the Extended Shuffle-$\mathcal{F}$-Exchange graph is that we can do exactly that: By using standard packet-routing techniques, we can route the assignments from the first layer to the last layer so that every vertex on the way needs to remember at most three assignments [52, Theorem 3.16]. Fix one such routing. Our coloring now assigns the same color from $\{0, 1, \bot\} \times \{\bot\}^2$ to each of the corresponding vertices on the first and last layers, and propagates this assignment through the middle layers according to the routing scheme; each vertex has a color in $\{0, 1, \bot\}^3$.

It is important to note that the middle layers do not perform any computation - they only copy symbols according to a fixed routing. In particular, if we change one entry in the representation of LOCALENDOFALINE vertex $u$, we can locally compute the difference of the old and new assignments for every middle layer node, even if we don't know the old assignment itself (which may correspond to up to three values routed through that node).

Notice that in order to verify that $u \in V^{\text{LOCAL}}$, we need to check that the active/inactive pattern matches the counter, and the "local conditions". Given a coloring of the Extended Shuffle-$\mathcal{F}$-Exchange graph, checking the local conditions reduces to checking that the color for each first-layer vertex is equal to the color on the respective vertex on the last layer, and that the color of every vertex not in the first layer is computed correctly from its incoming neighbors.

## 6.2 Arithmetization

We now want to represent the verification of $u \in V^{\text{LOCAL}}$ as constraints on polynomials over $\mathcal{G}$ (do not confuse with polynomials over $\mathbb{F}_2$!).

We represent each vertex $u \in V^{\text{LOCAL}}$ as two polynomials: $T_u \colon \mathcal{F}^{t-1} \times \mathcal{E} \to \mathcal{C}$ and $T_{\text{COUNTER}} \colon \mathcal{F}^{t-1} \to \mathcal{C}$, where $\mathcal{C} \subset \mathcal{F}$ is a subset of $\mathcal{F}$ of constant size. For each $(x_1, \ldots, x_{t-1}, \alpha^i) \in \mathcal{F}^{t-1} \times \mathcal{E}$, we set $T_u(x_1, \ldots, x_{t-1}, \alpha^i)$ to be equal to the color assigned to the respective vertex. $T_{\text{COUNTER}}(x_1, \ldots, x_{t-1})$ is assigned 1 if the corresponding line should be active, and 0 otherwise. Additionally, we construct a polynomial $T_{S,P} \colon \mathcal{F}^{t-1} \times \mathcal{E} \to \mathcal{C}'$, for $\mathcal{C} \subset \mathcal{C}' \subset \mathcal{F}$ of size $|\mathcal{C}'| = \Theta\left(\log^{3/2} n\right)$; $T_{S,P}(\cdot)$ represents the instance of ENDOFALINE: for each $(x_1, \ldots, x_{t-1}, \alpha^{5t\ell}) \in \mathcal{F}^{t-1} \times \{\alpha^{5t\ell}\}$, it specifies the gate functionality, whereas for $(x_1, \ldots, x_{t-1}, \alpha^i)$ $(i < 5t\ell)$ it specifies the routing through this vertex. Note that each vertex takes its inputs from at most 3 out of $\Theta\left(\sqrt{\log n}\right)$ incoming edges.

Notice that $T_u, T_{\text{COUNTER}}, T_{S,P}$ are all polynomials of degree at most $|\mathcal{F}|, |\mathcal{E}|$ in their respective variables. The verifier will expect low degree extensions of $T_u, T_{S,P}$ on all of $\mathcal{G}^t$, and analogously for $T_{\text{COUNTER}}$ over $\mathcal{G}^{t-1}$.

In the following, keep in mind that $T_{\text{COUNTER}}$ is constructed implicitly (once we decode all $\log_2 n + O(1)$ bits of the counter), and $T_{S,P}$ is part of the input of the problem, so we only need to locally decode and verify the validity of $T_u$.

First, we construct a polynomial

$$\Psi(\mathbf{x}) \triangleq \psi(T_u(\mathbf{x})) \triangleq \prod_{c \in \mathcal{C}} (T_u(\mathbf{x}) - c), \tag{10}$$

such that $\psi(T_u(\mathbf{x})) = 0$ iff $T_u(\mathbf{x}) \in \mathcal{C}$.

Similarly, we would like to construct a polynomial that verifies that the colors are propagated correctly through the middle layers; for the last layer, it should also verify that the gate functionality is implemented correctly. First, we describe the edges of the Extended Shuffle-$\mathcal{F}$-Exchange graph with the following affine transformations:

$$\begin{aligned}
\rho_{\text{STATIC}}\left(x_1, \ldots, x_{t-1}, \alpha^{i+1}\right) &\triangleq \left(x_1, \ldots, x_{t-1}, \alpha^i\right) \\
\rho_{\text{SHUFFLE}}\left(x_1, \ldots, x_{t-1}, \alpha^{i+1}\right) &\triangleq \left(x_{t-1}, x_1, \ldots, x_{t-2}, \alpha^i\right) \\
\rho_0\left(x_1, \ldots, x_{t-1}, \alpha^{i+1}\right) &\triangleq \left(x_1 + \left[z^0\right], \ldots, x_{t-1}, \alpha^i\right) \\
&\vdots \\
\rho_{\ell-1}\left(x_1, \ldots, x_{t-1}, \alpha^{i+1}\right) &\triangleq \left(x_1 + \left[z^{\ell-1}\right], \ldots, x_{t-1}, \alpha^i\right)
\end{aligned}$$

(So each vertex $\mathbf{x} \in \mathcal{F}^{t-1} \times (\mathcal{E} \setminus \{1\})$ has incoming edges from $\rho_{\text{STATIC}}(\mathbf{x}), \rho_{\text{SHUFFLE}}(\mathbf{x}), \rho_0(\mathbf{x}), \ldots, \rho_{\ell-1}(\mathbf{x})$.)

Now we can define a polynomial

$$\Phi(\mathbf{x}) \triangleq \phi\left(T_u(\mathbf{x}), T_u(\rho_{\text{STATIC}}(\mathbf{x})) T_u(\rho_{\text{SHUFFLE}}(\mathbf{x})), T_u(\rho_0(\mathbf{x})), \ldots, T_u(\rho_{\ell-1}(\mathbf{x})), T_{(S,P)}(\mathbf{x})\right), \tag{11}$$

where for $\mathbf{x} \in \mathcal{F}^{t-1} \times (\mathcal{E} \setminus \{1, \alpha^{5t\ell}\})$, $\phi(\cdots) = 0$ iff the colors are propagated correctly to the corresponding vertex, and for $\mathbf{x} \in \mathcal{F}^t \times \{\alpha^{5t\ell}\}$, $\phi(\cdots)$ further checks that the gate functionality is implemented correctly.

Finally, we add a third polynomial over $\mathcal{F}^{t-1}$,

$$\Xi(\mathbf{x}) \triangleq \xi\left(T_u(x_1, \ldots, x_{t-1}, 1), T_u\left(x_1, \ldots, x_{t-1}, \alpha^{5t\ell}\right), T_{\text{COUNTER}}(x_1, \ldots, x_{t-1})\right), \tag{12}$$

which is zero iff $T_u(x_1, \ldots, x_{t-1}, 1) = T_u(x_1, \ldots, x_{t-1}, \alpha^{5t\ell})$ and the corresponding line is active or inactive as dictated by $T_{\text{COUNTER}}(x_1, \ldots, x_{t-1})$.

We now have that $u \in V^{\text{LOCAL}}$ is equivalent to AND of the following conditions:

- Values-in-domain: $\Psi(\mathbf{x}) = 0$ for all $\mathbf{x} \in \mathcal{F}^{t-1} \times \mathcal{E}$;

- Value-propagation and computation: $\Phi(\mathbf{x}) = 0$ for all $\mathbf{x} \in \mathcal{F}^{t-1} \times (\mathcal{E} \setminus \{1\})$; and

- Active-vs-inactive: $\Xi(\mathbf{x}) = 0$ for all $\mathbf{x} \in \mathcal{F}^{t-1}$.

Notice also that $\psi, \xi$ are constant total degree polynomials, whereas $\phi$ has total degree polylog$n$; thus $\Psi, \Phi, \Xi$ are of degrees $|\mathcal{F}| \cdot$ polylog$n, |\mathcal{E}| \cdot$ polylog$n \leq |\mathcal{G}|/2$ in their respective variables.

## 6.3 PCP

Our verifier should test that $\Psi, \Phi, \Xi$ are zero everywhere on the respective domains ($\mathcal{F}^{t-1} \times \mathcal{E}, \mathcal{F}^{t-1} \times (\mathcal{E} \setminus \{1\}), \mathcal{F}^{t-1}$). We describe the test for $\Psi$; the corresponding tests for $\Phi$ and $\Xi$ follow analogously.

In order to test that $\Psi(\mathbf{x})$ is zero over all of $\mathcal{F}^{t-1} \times \mathcal{E}$, consider the following polynomials:

$$\Psi'\left(x_1, \ldots, x_{t-1}, \alpha^j\right) \triangleq \sum_{\substack{f_{i_{t/2+1}}, \ldots, f_{i_{t-1}} \in \mathcal{F} \\ j' \in \{1, \ldots, 5t\ell\}}} \Psi\left(x_1, \ldots, x_{t/2}, f_{i_{t/2+1}}, \ldots, f_{i_{t-1}}, \alpha^{j'}\right) \prod_{k=t/2+1}^{t-1} x_k^{i_k-1} \cdot \left(\alpha^j\right)^{j'} \tag{13}$$

$$\Psi''\left(x_1, \ldots, x_{t-1}, \alpha^j\right) \triangleq \sum_{f_{i_1}, \ldots, f_{i_{t/2}} \in \mathcal{F}} \Psi'\left(f_{i_1}, \ldots, f_{i_{t/2}}, x_{t/2+1}, \ldots, x_{t-1}, \alpha^j\right) \prod_{k=1}^{t/2} x_k^{i_k-1}. \tag{14}$$

In particular, observe that

$$\Psi''\left(x_1, \ldots, x_{t-1}, \alpha^j\right) = \sum_{\substack{f_{i_1}, \ldots, f_{i_{t-1}} \in \mathcal{F} \\ j' \in \{1, \ldots, 5t\ell\}}} \Psi\left(f_{i_1}, \ldots, f_{i_{t-1}}, \alpha^{j'}\right) \cdot \prod_{k=1}^{t-1} x_k^{i_k-1} \cdot \left(\alpha^j\right)^{j'}.$$

The main point of this construction is that $\Psi''$ is the zero polynomial if and only if $\Psi(\mathbf{x}) = 0$ for every $\mathbf{x} \in \mathcal{F}^{t-1} \times \mathcal{E}$. Now, by definition, $\Psi''$ has degrees at most $|\mathcal{F}|, |\mathcal{E}|$ in each variable, respectively, so it suffices to test that it is indeed zero at one random point from $\mathcal{G}^t$.

The verifier should also ensure that $\Psi'$ and $\Psi''$ are constructed correctly. This is done by picking a uniformly random $\mathbf{x} \in \mathcal{G}^t$ and verifying that (13) and (14) hold. If $\Psi$ is constructed correctly, then $\Psi'$ also has low degrees in each variable (compared to $|\mathcal{G}|$), so it suffices to test on a single $\mathbf{x} \in \mathcal{G}^t$. Similarly, the verifier should also check that $\Psi$ is constructed correctly as in (10); thanks to the low-degree guarantee, a single $\mathbf{x} \in \mathcal{G}^t$ suffices here as well.

Notice that the verifier does not need to receive $\Psi''$ explicitly. (This is crucial later for the proof construction!) Testing that $\Psi''$ is constructed correctly at $\left(x_1, \ldots, x_{t-1}, \alpha^j\right)$ and that it is zero at the same point is equivalent to simply testing that

$$\sum_{f_{i_1}, \ldots, f_{i_{t/2}} \in \mathcal{F}} \Psi'\left(f_{i_1}, \ldots, f_{i_{t/2}}, x_{t/2+1}, \ldots, x_{t-1}, \alpha^j\right) \prod_{k=1}^{t/2} x_k^{i_k-1} = 0. \tag{15}$$

We also need to test that all polynomials are indeed low degree. We do this later, in Subsection 6.7.

Define and test $\Phi', \Xi'$ analogously to $\Psi'$. We conclude that if $T_u, \Psi, \Phi, \Xi, \Psi', \Phi', \Xi'$ are all indeed of low degrees as promised and the verifier accepts with probability greater than $\epsilon$, then $T_u, T_{\text{COUNTER}}$ represent a real vertex $u \in V^{\text{LOCAL}}$. Notice that all of our tests choose $\mathbf{x} \in \mathcal{G}^{t/2}$ with uniform marginal probabilities.

For simplicity of notation, we henceforth treat $\Xi(\cdot)$ as $t$-variate polynomial (with the $t$-th variable being a dummy variable).

## 6.4 The final encoding

Our proof $\Pi(u)$ consists of the following strings.

- $C(u)$ consists of a good encoding of the "counter" of $u$. (Notice that we require a succinct encoding that can be fully decoded, and then the entire $T_{\text{COUNTER}}$ can be computed locally.)

- $E(u)$ consists of the polynomial $T_u \colon \mathcal{G}^t \to \mathcal{G}$.

- $\pi(u)$ consists of the polynomials $\Psi, \Psi', \Phi, \Phi', \Xi, \Xi' \colon \mathcal{G}^t \to \mathcal{G}$.

Although they have different lengths (in particular, $C(u)$ is much shorter), we think of them as having equal "weights", i.e. we allow only a constant fraction of errors in each. Also, we think of symbols in larger alphabets as encoded with a good binary error correcting code.

## 6.5 Local decoding and error correction

When a typical PCP verifier is given a proof that is wrong even at a single bit, it is already allowed to reject the proof. For our purpose, we want a more lenient verifier that, when given a proof that is wrong at a small fraction of the bits, corrects those bits and accepts. Furthermore, we want to ensure that our verifier only uses very little randomness: only $(1/2 + o(1)) \log_2 n$ random bits. A second goal of this subsection is to establish the desideratum that the verifier can adaptively decode an arbitrary $q_{\text{CRITICAL}}$-tuple of entries from the proof.

We describe how to locally decode and correct $\Psi$; $T_u, \Psi', \Phi, \Xi, \Phi', \Xi'$ follow with minor modifications, whereas the counter has a succinct representation and can be fully decoded. Specifically, the verifier needs to decode $\Psi$ in order to test (10), (13). For (10) the verifier only needs to decode $\Psi$ at one uniformly random point, so this is the easy case. For (13), the verifier wants to decode an entire axis-parallel $(t/2)$-dimensional affine subspace, $Q_{(x_1, \ldots, x_{t/2})} \triangleq \{(x_1, \ldots, x_{t/2})\} \times \mathcal{G}^{t/2}$, for a uniformly random choice of $(x_1, \ldots, x_{t/2}) \in \mathcal{G}^{t/2}$.

Let $S \subset \mathcal{G}^{t/2}$ be a $\left(1/\left(t^2 \log |\mathcal{G}|\right)\right)$-biased set of cardinality $|S| = \text{poly}(t, \log |\mathcal{G}|)$ (as guaranteed by Section 2.1). We pick a uniformly random $\mathbf{x} = (x_1, \ldots, x_{t/2}) \in \mathcal{G}^{t/2}$, and a uniformly random $\mathbf{y} = (y_1, \ldots, y_{t/2}) \in S$. We then consider the $(t/2 + 1)$-dimensional affine subspace $R_{\mathbf{x}, \mathbf{y}} \triangleq \bigcup_{\beta \in \mathcal{G}} Q_{(\mathbf{x} + \beta \mathbf{y})}$. If $\widehat{\Psi}$ is $O(\epsilon_{\text{DECODE}})$-close to $\Psi$, we have by Lemma 2.4 that with probability $1 - o(1)$, the restriction of $\widehat{\Psi}$ to $R_{\mathbf{x}, \mathbf{y}}$, denoted $\widehat{\Psi}\,|_{R_{\mathbf{x}, \mathbf{y}}}$, is also $O(\epsilon_{\text{DECODE}})$-close to $\Psi\,|_{R_{\mathbf{x}, \mathbf{y}}}$. Whenever this is the case, the verifier correctly decodes $\Psi$ on $R_{\mathbf{x}, \mathbf{y}}$.

Notice that so far, our verifier queries $|\mathcal{G}|^{t/2+1} = n^{1/2+o(1)}$ symbols, and uses $\log_2 |\mathcal{G}^{t/2}| + \log_2 |S| = (1/2 + o(1)) \log_2 n$ random bits.

The local decoding for $T_u, \Psi', \Phi, \Xi, \Phi', \Xi'$ follows with minor modifications which we now describe. Notice that we can reuse the same random bits from the decoding of $\Psi$ to decode each of the other polynomials. For (14) the verifier wants to decode $\Psi'$ (equivalently for $\Phi', \Xi'$) on a subspace that fixes the second half of the coordinates, $Q'_{(x_{t/2+1}, \ldots, x_t)} \triangleq \mathcal{G}^{t/2} \times \{(x_{t/2+1}, \ldots, x_t)\}$.

For (10) and (12), the verifier also needs to decode one or two entries of $T_u$, but this is no harder than decoding an entire plane. For (11), there is some subtlety: $\Phi(\mathbf{x})$ is a function of the values of $T_u$ on a super-constant number $(\ell + 3 = \Theta(\sqrt{\log n}))$ of points $\mathbf{x}, \rho_{\text{STATIC}}(\mathbf{x}), \rho_{\text{SHUFFLE}}(\mathbf{x}), \rho_0(\mathbf{x}), \ldots, \rho_{\ell-1}(\mathbf{x})$. Observe that $\mathbf{x}, \rho_{\text{STATIC}}(\mathbf{x}), \rho_0(\mathbf{x}), \ldots, \rho_{\ell-1}(\mathbf{x})$ all belong to the same $(t/2)$-dimensional subspace (only the first and last coordinates change); thus we only need to apply subspace-decoding twice.

Notice that our decoding mechanism also suffices to locally correct errors (with probability at least $(1 - o(1))$, and assuming that we start with a function that is close to low-degrees). Furthermore, by Lemma 2.4, the distance $\left|\widehat{\Psi}\,|_{R_{\mathbf{x},\mathbf{y}}} - \Psi\,|_{R_{\mathbf{x},\mathbf{y}}}\right|$ is within $\pm o(1)$ of the global distance $\left|\widehat{\Psi} - \Psi\right|$. Handling the rest of the polynomials as described in the previous paragraphs, we can thus estimate $\left|\widehat{\Pi}\,|_{E,\pi} - \Pi(u)\,|_{E,\pi}\right|$ to within $\pm o(1)$; for the bits that correspond to the counter, we can compute the distance exactly.

### Decoding a particular subspace

Above, we described how to locally decode a random subspace. Suppose instead that we want to locally decode $\Psi\,|_{Q_{(w_1,\ldots,w_{t/2})}}$ for some particular $\mathbf{w} \in \mathcal{G}^{t/2}$ (specifically, when $Q_{(w_1,\ldots,w_{t/2})}$ contains information about one of the critical bits). Since we did not waste any random bits on picking $\mathbf{w}$, we can afford to pick $\mathbf{y}$ uniformly from all of $\mathcal{G}^{t/2}$. On the other hand, since $\mathbf{w}$ is arbitrary, there may still be an $\Theta(\epsilon_{\text{DECODE}})$ chance that $\left|\widehat{\Psi}\,|_{R_{\mathbf{w},\mathbf{y}}} - \Psi\,|_{R_{\mathbf{w},\mathbf{y}}}\right|$ is too large ($> 1/2$) to guarantee correct decoding. Instead, we pick a third vector $\mathbf{z} \in S$, and query $\widehat{\Psi}$ on the $(t/2+2)$-dimensional affine subspace $P_{\mathbf{w},\mathbf{y},\mathbf{z}} \triangleq \bigcup_{\beta,\gamma \in \mathcal{G}} Q_{(\mathbf{w}+\beta(\mathbf{y}+\gamma\mathbf{z}))}$; notice that $P_{\mathbf{w},\mathbf{y},\mathbf{z}} \triangleq \bigcup_{\gamma \in \mathcal{G}} R_{\mathbf{w},\mathbf{y}+\gamma\mathbf{z}}$. In expectation, for each $(\mathbf{y}+\gamma\mathbf{z})$'s, the restriction $\widehat{\Psi}\,|_{R_{\mathbf{w},\mathbf{y}+\gamma\mathbf{z}}}$ is at distance at most $O(\epsilon_{\text{DECODE}})$ from $\Psi\,|_{R_{\mathbf{w},\mathbf{y}+\gamma\mathbf{z}}}$. Therefore, by Lemma 2.4, with probability $1 - o(1)$, we have that $\left|\widehat{\Psi}\,|_{P_{\mathbf{w},\mathbf{y},\mathbf{z}}} - \Psi\,|_{P_{\mathbf{w},\mathbf{y},\mathbf{z}}}\right| = O(\epsilon_{\text{DECODE}})$.

## 6.6 Local proof-construction

We now want to establish the following unusual property of our holographic proof scheme: we can, with high probability, locally construct any subspace $Q_{(x_1,\ldots,x_{t/2})}$ of the proof $\pi(u)$, by reading only a small part of the encoding $E(u)$ (in particular, we need to decode a constant number of $(t/2+1)$-dimensional subspaces from $E(u)$). Here, it is important that $Q_{(x_1,\ldots,x_{t/2})}$ is defined via a restriction of the *first* $t/2$ coordinates, but we assume access to decoded subspaces of $E(u)$ with restriction of any $(t/2)$-tuple of coordinates.

Recall that in order to compute $\Psi(\mathbf{x})$ or $\Xi(\mathbf{x})$, it suffices to know the values of $T_u$ at a constant number of locations ((10) and (12)), so here this property trivially holds. For $\Phi(\mathbf{x})$, we need to know $T_u$ at $\Theta(\sqrt{\log n})$ locations (11); however, as we argue in Subsection 6.5, they belong to two axis-parallel subspaces, so it again suffices to decode $T_u$ on a constant number of subspaces.

In order to compute $\Psi'(\mathbf{x}), \Phi'(\mathbf{x}), \Xi'(\mathbf{x})$, we need to correctly decode the values of $\Psi, \Phi, \Xi$ on (almost) an entire axis-parallel subspace $Q_{(x_1,\ldots,x_{t/2})}$. The values of $\Psi, \Phi, \Xi$ are not given as part of $E(u)$, but (as we argued in the previous paragraph), we can reconstruct them from $T_u$. Specifically, in order to know the values of $\Psi$ or $\Xi$ on $Q_{(x_1,\ldots,x_{t/2})}$, we simply need to decode $T_u$ on $Q_{(x_1,\ldots,x_{t/2})}$. In order to construct $\Phi$ on an entire subspace, observe that for every $\left(x_1,\ldots,x_{t/2}, f_{i_{t/2+1}},\ldots,f_{i_{t-1}}, \alpha^j\right) \in Q_{(x_1,\ldots,x_{t/2})}$, all the vectors $\rho_{\text{STATIC}}\left(x_1,\ldots,x_{t/2}, f_{i_{t/2+1}},\ldots,f_{i_{t-1}}, \alpha^j\right), \rho_0\left(x_1,\ldots,x_{t/2}, f_{i_{t/2+1}},\ldots,f_{i_{t-1}}, \alpha^j\right),\ldots,$ $\rho_{\ell-1}\left(x_1,\ldots,x_{t/2}, f_{i_{t/2+1}},\ldots,f_{i_{t-1}}, \alpha^j\right)$ belong to the same subspace $Q_{(x_1,\ldots,x_{t/2})}$. Furthermore, all the vectors $\rho_{\text{SHUFFLE}}\left(x_1,\ldots,x_{t/2}, f_{i_{t/2+1}},\ldots,f_{i_{t-1}}, \alpha^j\right) =$ $\left(f_{i_{t-1}}, x_1, \ldots, x_{t/2}, f_{i_{t/2+1}}, \ldots, f_{i_{t-2}}, \alpha^j\right)$ belong to the subspace $Q^{\text{Shuffle}}_{(x_1,\ldots,x_{t/2})} \triangleq \mathcal{G} \times (x_1, \ldots, x_{t/2}) \times \mathcal{G}^{t/2-1}$. Therefore the value of $\Phi'(\mathbf{x})$ can also be computed from the values of $T_u$ on a constant number of subspaces.

## 6.7 Local robust testing

So far, we assumed that we are given functions that are low degree (Subsection 6.3), or approximately low degree (Subsection 6.5). However, we must verify that this is indeed the case.

We will use results for locally testing tensor codes [68]. In order to describe them, let us briefly introduce some minimal background and notation. Given linear codes $C_1, C_2$ with respective generator matrices $M_1, M_2$, their *tensor product*, denoted $C_1 \otimes C_2$ is the linear code generated by the tensor product of the matrices $M_1 \otimes M_2$. We say that $C_1 \otimes C_2$ is a *tensor code*. In general, we can talk about the tensor product of $k$ codes, $C_1 \otimes \cdots \otimes C_k$, which is defined recursively. We say that $C_1 \otimes \cdots \otimes C_k$ is the $k$-th power of $C$ (denoted $C^k$), if $C_i = C$ for all $i \in [k]$. In our case, for example, the code $\mathcal{C}$ of all $(|\mathcal{G}|/2)$-individual degrees polynomials $T: \mathcal{G}^t \to \mathcal{G}$ is the $t$-th power of the code of all $(|\mathcal{G}|/2)$-degree polynomials $t: \mathcal{G} \to \mathcal{G}$.

Alternatively, we can think of any product code as a function from some product domain to some range. The (axis-parallel) *hyperplane tester* picks a uniformly random index $i \in [k]$, and a uniformly random value $v_i$ from the $i$-th domain. It reads the assignment of the function for all vectors that have $v_i$ in their $i$-th coordinate, and tests whether they form (or are close to) a codeword from $C_1 \otimes \cdots \otimes C_{i-1} \otimes C_{i+1} \otimes \cdots \otimes C_k$. Going back to our setting, this could mean selecting a random $i \in \{1, \ldots, t\}$, and a random value $v_i \in \mathcal{G}$; reading $T(\mathbf{x})$ for all $\mathbf{x}$ such that $x_i = v_i$, and verifying that the restricted $(t-1)$-variate function is close to a low degrees polynomial.

To optimize query complexity, it will be more convenient for us to think of the code of $(t/4)$-variate $(|\mathcal{G}|/2)$-individual degrees polynomials $T_{t/4}: \mathcal{G}^{t/4} \to \mathcal{G}$, henceforth denoted $\mathcal{C}_{1/4}$. The code $\mathcal{C}$ can now be written $\mathcal{C} = (\mathcal{C}_{1/4})^4$. The hyperplane tester now chooses a random $i \in \{1, 2, 3, 4\}$ and a random value $v_i \in \mathcal{G}^{t/4}$, and reads $T(\mathbf{x})$ for all $\mathbf{x}$ whose restriction to the $i$-th $(t/4)$-tuple of coordinates is equal to $v_i$.

We can now apply the following theorem due to Viderman [68]:

**Theorem 6.2.** *[68, Theorem 4.4] Let $\mathcal{C} = C_1 \otimes \cdots \otimes C_k$ be the tensor product of $k > 2$ linear codes of relative distances $\delta_1, \ldots, \delta_k$. Let $M$ be a string which is $\epsilon$-far from the tensor code $\mathcal{C}$. Then the restriction of $M$ to a random hyperplane drawn from the above distribution is, in expectation, at least $\left(\epsilon \cdot \frac{\prod_{j=1}^k \delta_j}{2k^2}\right)$-far from the restricted code $C_1 \otimes \cdots \otimes C_{i-1} \otimes C_{i+1} \otimes \cdots \otimes C_k$.*

Let $S_{1/4} \subset \mathcal{G}^{t/4}$ be a $\left(1/(t^2 \log |\mathcal{G}|)\right)$-biased set of cardinality $|S_{1/4}| = \mathsf{poly}(t, \log |\mathcal{G}|)$ (for example, take the restriction of each $\mathbf{y} \subset S$ to its first $t/4$ entries). Sample $\mathbf{x} \in \mathcal{G}^{t/4}$ and $\mathbf{y} \in S_{1/4}$ uniformly at random. Run the hyperplane tester for $(\mathcal{C}_{1/4})^4$ on each alleged polynomial $4|\mathcal{G}|$ times, taking all the possibilities for $i \in \{1, 2, 3, 4\}$ and $v_i = (\mathbf{x} + \beta \mathbf{y})$ for all $\beta \in \mathcal{G}$. By Theorem 6.2, if the polynomial is $\epsilon_{\text{Sound}}$-far from $(\mathcal{C}_{1/4})^4$, then the expected distance on each of the $4|\mathcal{G}|$ tests is $\left(\epsilon_{\text{Sound}} \cdot \frac{\prod_{j=1}^k \delta_j}{2k^2}\right) = \epsilon_{\text{Sound}}/512$ (since $k = 4$ and $\delta_j \geq 1/2$). By Lemma 2.4, the average over all the tests is, with probability $1 - o(1)$, within $\pm o(1)$ of this expectation.

So far the tester requires $(1/4 + o(1)) \log_2 n$ random bits to sample a hyperplane, but has prohibitively high query complexity: $n^{3/4+o(1)}$. In order to drive down the query complexity, we can recurse on the test using fresh $(1/4 + o(1)) \log_2 n$ random bits: we re-partition $\left(\mathcal{C}_{1/4}\right)^3$ as a tensor product of three linear codes, and reapply Viderman's theorem. Our query complexity is now $n^{1/2+o(1)}$, the total random-bits complexity is $(1/2 + o(1)) \log_2 n$, and the robustness guarantee is $\left(\epsilon_{\text{SOUND}} \cdot \frac{2^{-4}}{2 \cdot 4^2} \cdot \frac{2^{-3}}{2 \cdot 3^2}\right) - o(1) > \epsilon_{\text{SOUND}}/10^5$.

Finally, our tester accepts iff the average distance across all $(t/2)$-dimensional subspaces from the nearest low-degrees polynomial is less than $2\sqrt{\epsilon_{\text{COMPLETE}}} < \epsilon_{\text{SOUND}}/\left(2 \cdot 10^5\right)$. Thus our tester accepts with high probability whenever $\widehat{\Psi}$ is $\epsilon_{\text{COMPLETE}}$-close to the true $\Psi$, and rejects with high probability whenever $\widehat{\Psi}$ is $\epsilon_{\text{SOUND}}$-far from any low degrees polynomial. (Testing $\widehat{T_u}, \widehat{\Phi}, \widehat{\Xi}, \widehat{\Psi'}, \widehat{\Phi'}, \widehat{\Xi'}$ follows analogously.)

## 6.8 Summary

The verifier, on input $\widehat{\Pi} = \left(\widehat{E}, \widehat{C}, \widehat{\pi}\right)$, does the following.

**Randomness**

In a preprocessing step, the verifier fixes in advance a $\left(1/\left(t^2 \log |\mathcal{G}|\right)\right)$-biased set $S$ of cardinality $|S| = \mathsf{poly}\left(t, \log |\mathcal{G}|\right)$. Let $S_{1/4}$ denote the restriction of $S$ to the first $t/4$ coordinates.

The tester then:

- draws $\mathbf{x}$ uniformly at random from $\mathcal{G}^{t/2}$, and $\mathbf{y}$ uniformly at random from $S$; and

- reusing the same randomness, draws $\mathbf{x}_1, \mathbf{x}_2 \in \mathcal{G}^{t/4}$ and $\mathbf{y}_1, \mathbf{y}_2 \in S_{1/4}$.

**Queries**

The verifier reads $\widehat{T_u}$ on all vectors $\mathbf{z} \in \mathcal{G}^t$ such that one of $\left\{(z_1, \ldots, z_{t/2}), (z_2, \ldots z_{t/2+1})\right\}$ is of the form $(\mathbf{x} + \beta \mathbf{y})$ for some $\beta \in \mathcal{G}$. It also reads $\widehat{\Psi}, \widehat{\Phi}, \widehat{\Xi}, \widehat{\Psi'}, \widehat{\Phi'}, \widehat{\Xi'}$ on all $\mathbf{z} \in \mathcal{G}^t$ such that $(z_1, \ldots z_{t/2}) = (\mathbf{x} + \beta \mathbf{y})$ for some $\beta \in \mathcal{G}$ (denote those vectors $R_{\mathbf{x},\mathbf{y}}$). Similarly, it also reads $\Psi', \Phi'$, and $\Xi'$ on all $\mathbf{z} \in \mathcal{G}^t$ such that $(z_{t/2+1}, \ldots, z_t) = (\mathbf{x} + \beta \mathbf{y})$ (denote those vectors $R'_{\mathbf{x},\mathbf{y}}$). The union of all those subspaces (i.e. $R_{\mathbf{x},\mathbf{y}} \cup R'_{\mathbf{x},\mathbf{y}}$) forms $G_{\text{PCP}}$, while $G_{\text{SAMPLE}} \triangleq R_{\mathbf{x},\mathbf{y}}$.

For each choice of $i_1 \neq i_2 \in \{1, 2, 3, 4\}$ and each $\beta_1, \beta_2 \in \mathcal{G}$, the verifier reads each polynomial on all vectors $\mathbf{z} \in \mathcal{G}^t$ such that the $i_1$-th $(t/4)$-tuple of $\mathbf{z}$ is equal to $\mathbf{x}_1 + \beta_1 \mathbf{y}_1$, and the $i_2$-th $(t/4)$-tuple of $\mathbf{z}$ is equal to $\mathbf{x}_2 + \beta_2 \mathbf{y}_2$. The union of all those subspaces forms $G_{\text{LTC}}$.

Finally, the verifier also reads the entire encoding $C(u)$ of the counter of $u$. Given $C(u)$, the verifier also picks $q_{\text{CRITICAL}}$ vectors $\mathbf{w}^1, \ldots, \mathbf{w}^{q_{\text{CRITICAL}}} \in \mathcal{G}^{t/2}$ that correspond to each of the critical bits, and reads $T_u$ on all vectors $\mathbf{z} \in \mathcal{G}^t$ such that $(z_1, \ldots, z_{t/2}) = \left(\mathbf{w}^i + \beta (\mathbf{x} + \beta \mathbf{y})\right)$ for some $\beta, \gamma \in \mathcal{G}$ and $i \in \{1, \ldots, q_{\text{CRITICAL}}\}$. Those are $G_{\text{CRITICAL}}$.

**Test**

The verifier uses the hyperplane tester from Subsection 6.7 to verify that all the polynomials are indeed close to low-degree polynomials. (If not, it suffices to return an error message and abort.)

**Correct & decode**

For each subspace queried, the verifier finds the unique low-degree polynomials $T_u, \Psi, \Phi, \Xi, \Psi', \Phi', \Xi'$ that are close to the values read.

**Verify**

The verifier checks that the corrected polynomials $T_u, \Psi, \Phi, \Xi, \Psi', \Phi', \Xi'$ satisfy (10), (11), (12), and (13) on all of $R_{\mathbf{x},\mathbf{y}}$. It also checks that (15) (and its analogues for $\Phi', \Xi'$) are satisfied on all of $R'_{\mathbf{x},\mathbf{y}}$.

**Estimate distance**

The verifier also computes the hamming distance between the values of $\widehat{T_u}, \widehat{\Psi}, \widehat{\Phi}, \widehat{\Xi}, \widehat{\Psi'}, \widehat{\Phi'}, \widehat{\Xi'}$ and $T_u, \Psi, \Phi, \Xi, \Psi', \Phi', \Xi'$ on $R_{\mathbf{x},\mathbf{y}}$.

**Extrapolate**

Given $C(u)$, the verifier can locally reconstruct $C(S^{\text{LOCAL}}(u))$. Additionally, after the verifier decoded the values of $T_u$ at the critical bits, it also knows the current and new values of all the entries that change between $u$ and $S^{\text{LOCAL}}(u)$. Since $E(\cdot)$ is a linear encoding, the verifier can also completely construct the difference vector $E(S^{\text{LOCAL}}(u)) - E(u)$. In other words, given the value of $T_u(\mathbf{x})$ for any $\mathbf{x}$, the verifier can locally compute $T_{S^{\text{LOCAL}}(u)}(\mathbf{x})$.

The verifier can use the ideas from Subsection (6.6) to also construct the values of all the polynomials $(\Psi, \Phi, \Xi, \Psi', \Phi', \Xi')$ for $S^{\text{LOCAL}}(u)$, on all of $R_{\mathbf{x},\mathbf{y}}$.

For $P^{\text{LOCAL}}(u)$, the verifier must first check that $u$ is not the special vertex $u_\mathbf{0}$. This is easy to do since $T_u$ and the respective $T_{u_\mathbf{0}}$ are both low degree polynomials and must differ (if $u \neq u_\mathbf{0}$) on a constant fraction of their entries. If indeed $u \neq u_\mathbf{0}$, recovering information about $P^{\text{LOCAL}}(u)$ is completely analogous to the procedure for $S^{\text{LOCAL}}(u)$. Otherwise, we simply have $P^{\text{LOCAL}}(u_\mathbf{0}) = u_\mathbf{0}$.

# 7 Polymatrix WeakNash

In this section we prove our main technical result:

**Theorem 7.1.** *There exists a constant $\epsilon_{\text{NASH}} > 0$, such that there is a $2^{n^{1/2+o(1)}}$-time reduction from* ENDOFALINE *to finding an $(\epsilon_{\text{NASH}}, \epsilon_{\text{NASH}})$-*WELL-SUPPORTED-WEAKNASH *of a complete bipartite polymatrix game between $n^{1/2+o(1)}$ players with $2^{n^{1/2+o(1)}}$ actions each; the payoffs in each bimatrix subgame are in $[0, 1/n_B], [0, 1/n_A]$, where $n_A, n_B$ denote the number of players on each side of the bipartite graph.*

We construct a subexponential polymatrix game that implements a variant of the hard Brouwer function constructed in Section 3. The new Brouwer function is very similar, but we make the following modifications:

- We use the vertices of a LOCALENDOFALINE instance instead of ENDOFALINE.

- Instead of encoding each LOCALENDOFALINE vertex $u$ with an arbitrary error correcting code, the Brouwer vertices correspond to the holographic proofs $\Pi(u)$ from Section 6.

- For convenience of notation, instead of maintaining $m$ coordinates (which we expect to be identical anyway), for each of the auxiliary compute-vs-copy bit and the special direction, we keep just one coordinate for each, but give each a constant relative weight.

- In particular, the "first" Brouwer line segment goes from $\mathbf{z}_2 \triangleq (\Pi(u_\mathbf{0}), \Pi(u_\mathbf{0}), 0, 2)$ to $(\Pi(u_\mathbf{0}), \Pi(u_\mathbf{0}), 0, 0)$.

The rest proof proceeds as follows: In Subsection 7.1, we define the strategy space for the players and relate it to the holographic proof system from Section 6. In Subsection 7.2 we introduce some important notation used throughout this section. In the next two subsections (7.3 and 7.4) we define the players' payoffs, and implement an imitation gadget, assuming that the Brouwer function $f$ can be locally computed. In Subsection 7.5, we translate the PCP guarantees from Proposition 6.1 to guarantees about approximate equilibria of our polymatrix game. This completes the setup of the argument. The rest of this section (Subsections 7.6-7.9) shows that we can indeed locally compute $f$ in the different regions where it is defined (outside the picture, close to a Brouwer line segment, etc.).

## 7.1 Players and strategies

Let $\mathcal{R}_{\mathrm{PCP}}$ denote the set of random strings used by the PCP verifier in Proposition 6.1. For each $r \in \mathcal{R}_{\mathrm{PCP}}$ we construct a player $(A, r)$ on one side of the bipartite game. On the opposite side, we construct a player $(B, \mathbf{q})$ for every $\mathbf{q} \in \mathcal{G}^{t/2}$. We refer to the respective sides of the bipartite game as Alice's players and Bob's players. (In Section 8 we will construct a bimatrix game between Alice and Bob where each player "controls" the vertices on her or his side of the polymatrix game.)

Each player has $(3/\epsilon_{\mathrm{PRECISION}} + 1)^{n^{1/2+o(1)}+2}$ actions. We think of each action as an ordered tuple of:

- two vectors in $[-1, 2]^{n^{1/2+o(1)}}$, where the interval $[-1, 2]$ is discretized into $\{-1, -1 + \epsilon_{\mathrm{PRECISION}}, \ldots, 2 - \epsilon_{\mathrm{PRECISION}}, 2\}$; and

- two more variables in $[-1, 2]$.

Recall that each Brouwer vertex corresponds to a pair $(u, v)$ of either identical or consecutive vertices from the LOCALENDOFALINE instance. For player $(A, r)$, the two vectors allegedly correspond to bits read from two holographic proofs (a-la Section 6) $(\Pi(u), \Pi(v))$ by the PCP verifier with random string $r$. For player $(B, \mathbf{q})$ the vectors represent (a binary encoding of) the counters $C(u), C(v)$, and the entries of $E(u), \pi(u), E(v), \pi(v)$ on a $(t/2)$-dimensional subspace $G^{(B,\mathbf{q})} \subset \mathcal{G}^t$ to be specified below.

The additional two variables represent the compute-vs-copy bit and the special direction.

In addition to the assignments described above, each player on Alice's side has an additional choice among $(3/\epsilon_{\mathrm{PRECISION}} + 1)^{n^{1/2+o(1)}}$ actions. These additional actions allegedly correspond to $2q_{\mathrm{CRITICAL}}$ (affine, $(t/2 + 2)$-dimensional) subspaces from which we can decode the $q_{\mathrm{CRITICAL}}$ critical bits of each of the vertices. (We know which entries are represented by reading the "counter" part of the proof.)

**Bob's subspace $G^{(B,\mathbf{q})}$**

Let $G_{\mathrm{ALL}}^{(A,r)} \subset \mathcal{G}^t$ denote the set of entries of $E(\cdot), \pi(\cdot)$ read by the PCP verifier with random string $r$. Recall that the PCP verifier picks a set of $(t/2)$-dimensional axis-parallel affine subspaces; each subspace is an affine shift of a linear subspace, spanned by one of a constant number of $(t/2)$-tuples of standard basis vectors. We pick $G^{(B,\mathbf{q})}$ so that it is spanned by vectors that are linearly independent of all those tuples. In particular, $G^{(B,\mathbf{q})}$ intersects each of $(A,r)$'s subspaces at exactly one point.

For any $(t/2)$-tuple of standard basis vectors of $\mathcal{G}^t$, a uniformly random $(t/2)$-tuple of new vectors in $\mathcal{G}^t$ completes it to a basis with high probability. To see this, consider adding the new vectors one by one; at each step, there is at least a $(1 - 1/|\mathcal{G}|)$-probability that it is linearly independent of the previously chosen vectors; the claim follows by union bound and $t = o(|\mathcal{G}|)$. Therefore, a random $(t/2)$-tuple of vectors will also, with high probability, be simultaneously linearly independent of each of the relevant tuples of standard basis vectors. Fix any such a $(t/2)$-tuple (there are also efficient ways to deterministically find such a tuple, but in quasi-polynomial time we can simply brute-force enumerate through all tuples).

Finally, we let $G^{(B,\mathbf{0}_{t/2})}$ be the linear subspace spanned by those vectors; for general $\mathbf{q} \in \mathcal{G}^{t/2}$, we let $G^{(B,\mathbf{q})}$ be the same subspace shifted by $(\mathbf{q}, \mathbf{0}_{t/2})$.

## 7.2 Notation

We introduce some necessary additional notation for formally handling vectors in this section. The notation for $(A,r)$'s assignment to the critical bits is analogous in spirit but more subtle to formalize; we come back to it to Subsection 7.3.

**Definition 7.2.** We say that a player is *happy* if every strategy on her support is $\epsilon_{\text{NASH}}$-optimal.

**Rounding** For $x \in \mathbb{R}$, let $\nu(x) \triangleq \begin{cases} 1 & x > 1/2 \\ 0 & \text{otherwise} \end{cases}$ denote the rounding of $x$ to $\{0,1\}$; for $\mathbf{x} \in \mathbb{R}^n$, we compute $\nu(\mathbf{x})$ coordinate-wise.

**Coordinates** Let $M \triangleq \{1, \ldots, 2m+2\}$ denote the set of all coordinates. We partition $M$ into: $M_1 \triangleq \{1, \ldots, m\}$ (the first proof), $M_2 \triangleq \{m+1, \ldots, 2m\}$ (the second proof), $M_3 \triangleq \{2m+1\}$ (the Compute-vs-Copy bit), and $M_4 \triangleq \{2m+2\}$ (the special direction). We will also use the abbreviations $M_{1,2} \triangleq M_1 \cup M_2$ and $M_{3,4} \triangleq M_3 \cup M_4$.

Fix some player $(A,r)$ on Alice's side; we further detail $(A,r)$'s strategy space as follows. Let $G^{(A,r)}_{\text{SAMPLE}}, G^{(A,r)}_{\text{PCP}}, G^{(A,r)}_{\text{LTC}} \subset \mathcal{G}^t$ denote the sets of entries of $E(\cdot), \pi(\cdot)$ read by the PCP verifier with random string $r$ from the alleged proofs. Let $I^{(A,r)}_{\text{SAMPLE}}, I^{(A,r)}_{\text{PCP}}, I^{(A,r)}_{\text{LTC}} \subset M_{1,2}$ denote the indices of the corresponding bits, respectively. For player $(B, \mathbf{q})$ on Bob's side, let $I^{(B,\mathbf{q})} \subset M_{1,2}$ denote the indices of the bits corresponding to $G^{(B,\mathbf{q})}$. We also let $K \subset M_{1,2}$ ($K$ for Kounter) denote the set of indices that correspond to the encodings of counters in both proofs. (Notice that $K$ is the same for every player on both sides.)

Finally we will be particularly interested in the following sets:

$$\begin{aligned} J^{(A,r)} &\triangleq K \cup I^{(A,r)}_{\text{SAMPLE}} \cup M_{3,4} \\ J^{(B,\mathbf{q})} &\triangleq K \cup I^{(B,\mathbf{q})} \cup M_{3,4}. \end{aligned}$$

**Partial vectors** Player $(A,r)$'s strategy $\mathbf{a}$ naturally corresponds to a *partial vector* $\mathbf{x}^{(\mathbf{a})}_{\text{ALL}} \in ([-1,2] \cup \{\bot\})^M$. For $i \in K \cup I^{(A,r)}_{\text{PCP}} \cup I^{(A,r)}_{\text{LTC}} \cup M_{3,4}$, we set the $i$-th coordinate of $\mathbf{x}^{(\mathbf{a})}_{\text{ALL}}$ to the value that $\mathbf{a}$ assigns to the corresponding coordinate; for all other $i$'s, we let $\mathbf{x}^{(\mathbf{a})}_{\text{ALL}}|_i \triangleq \bot$.

In particular we are interested in the partial vector $\mathbf{x}^{(\mathbf{a})} \in ([-1,2] \cup \{\bot\})^M$ that describes $(A,r)$'s partial assignment to $J^{(A,r)}$. For $i \in J^{(A,r)}$, we set the $i$-th coordinate of $\mathbf{x}^{(\mathbf{a})}$ to the value that $\mathbf{a}$ assigns to the corresponding coordinate; for all other $i$'s, we let $\mathbf{x}^{(\mathbf{a})}|_i \triangleq \bot$.

For action $\mathbf{b}$ taken by player $(B,\mathbf{q})$, we can simply define one partial vector $\mathbf{x}^{(\mathbf{b})} \in ([-1,2] \cup \{\bot\})^M$, where $\mathbf{x}^{(\mathbf{b})}|_i \triangleq \bot$ for $i \notin J^{(B,\mathbf{q})}$.

**Partial norms** All norms in this section are 2-norm: we henceforth drop the 2 subscript from $\|\cdot\|_2$. Instead, we will use the subscript to denote the subset of coordinates over which we want to compute the norm. For example,

$$\|\mathbf{x}\|_K^2 \triangleq \mathsf{E}_{i \in K}\left[x_i^2\right].$$

In fact, we are often interested in expectation with respect to a non-uniform distribution; for example:

$$\|\mathbf{x}\|_{J^{(A,r)}}^2 \triangleq \frac{1}{4}\mathsf{E}_{i \in I_{\text{SAMPLE}}^{(A,r)}}\left[x_i^2\right] + \frac{1}{4}\mathsf{E}_{i \in K}\left[x_i^2\right] + \frac{1}{2}\mathsf{E}_{i \in M_{3,4}}\left[x_i^2\right]$$

$$\|\mathbf{x}\|_{J^{(B,\mathbf{q})}}^2 \triangleq \frac{1}{4}\mathsf{E}_{i \in I^{(B,\mathbf{q})}}\left[x_i^2\right] + \frac{1}{4}\mathsf{E}_{i \in K}\left[x_i^2\right] + \frac{1}{2}\mathsf{E}_{i \in M_{3,4}}\left[x_i^2\right].$$

The distribution is set such that each part (e.g. $K$, $I_{\text{PCP}}^{(A,r)}$, $I^{(B,\mathbf{q})}$, or $M_1, \ldots, M_4$) receives an equal weight (up to constant factors). The distribution will henceforth be implicit, and only the subset of coordinates will be explicit.

This notation also allows us to talk about distances between partial vectors, e.g.

$$\left\|\mathbf{x}^{(\mathbf{a})} - \mathbf{x}^{(\mathbf{b})}\right\|_{J^{(A,r)} \cap J^{(B,\mathbf{q})}}^2 \triangleq \frac{1}{4}\left\|\mathbf{x}^{(\mathbf{a})} - \mathbf{x}^{(\mathbf{b})}\right\|_K^2 + \frac{1}{2}\left\|\mathbf{x}^{(\mathbf{a})} - \mathbf{x}^{(\mathbf{b})}\right\|_{M_{3,4}}^2 + \frac{1}{4}\left\|\mathbf{x}^{(\mathbf{a})} - \mathbf{x}^{(\mathbf{b})}\right\|_{I_{\text{SAMPLE}}^{(A,r)} \cap I^{(B,\mathbf{q})}}^2.$$

**Aggregate vectors** Let $\mathcal{A}$ denote the joint (product) distribution over all Alice's players' mixed strategies, and define $\mathcal{B}$ analogously for Bob's players. We define $\mathbf{x}^{(\mathcal{A})}, \mathbf{x}^{(\mathcal{B})} \in [-1,2]^M$ to be the weighted entry-wise average of $\mathbf{x}^{(\mathbf{a})}, \mathbf{x}^{(\mathbf{b})}$, respectively, where for each coordinate we only take the expectation over vectors for which it is assigned a real (non-$\bot$) value.

### 7.3 Alice's imitation gadget

Our first goal is to incentivize player $(A,r)$ to imitate $\mathbf{x}^{(\mathcal{B})}$ on her assigned coordinates. We break up her utility on each bimatrix subgame as follows[9]:

$$U^{(A,r)} \triangleq U_{\text{SAMPLE}}^{(A,r)} + U_{\text{PCP}}^{(A,r)} + U_{\text{LTC}}^{(A,r)} + U_{\text{COUNTER}}^{(A,r)} + U_{M_{3,4}}^{(A,r)} + U_{\text{CRITICAL}}^{(A,r)}.$$

For each player $(B,\mathbf{q})$ on Bob's side, we simply define the first five parts to incentivize copying $(B,\mathbf{q})$'s strategy on each shared coordinate. Let $\mathbf{a}, \mathbf{b}$ denote $(A,r)$ and $(B,\mathbf{q})$ respective strategies; then

---

[9]For convenience of notation, we define bimatrix subgames with payoffs in $[-54/n_B, 0], [-36/n_A, 0]$. Payoffs $[0, 1/n_B], [0, 1/n_A]$ can be obtained by scaling and shifting.

$$U_{\text{SAMPLE}}^{(A,r)} \triangleq -\left\|\mathbf{x}_{\text{ALL}}^{(\mathbf{a})} - \mathbf{x}^{(\mathbf{b})}\right\|^2_{I_{\text{SAMPLE}}^{(A,r)} \cap I^{(B,\mathbf{q})}}$$

$$U_{\text{PCP}}^{(A,r)} \triangleq -\left\|\mathbf{x}_{\text{ALL}}^{(\mathbf{a})} - \mathbf{x}^{(\mathbf{b})}\right\|^2_{I_{\text{PCP}}^{(A,r)} \cap I^{(B,\mathbf{q})}}$$

$$U_{\text{LTC}}^{(A,r)} \triangleq -\left\|\mathbf{x}_{\text{ALL}}^{(\mathbf{a})} - \mathbf{x}^{(\mathbf{b})}\right\|^2_{I_{\text{LTC}}^{(A,r)} \cap I^{(B,\mathbf{q})}}$$

$$U_{\text{COUNTER}}^{(A,r)} \triangleq -\left\|\mathbf{x}_{\text{ALL}}^{(\mathbf{a})} - \mathbf{x}^{(\mathbf{b})}\right\|^2_{K}$$

$$U_{M_{3,4}}^{(A,r)} \triangleq -\left\|\mathbf{x}_{\text{ALL}}^{(\mathbf{a})} - \mathbf{x}^{(\mathbf{b})}\right\|^2_{M_{3,4}}.$$

Notice that

$$\frac{1}{8}U_{\text{PCP}}^{(A,r)} + \frac{1}{8}U_{\text{LTC}}^{(A,r)} + \frac{1}{4}U_{\text{COUNTER}}^{(A,r)} + \frac{1}{2}U_{M_{3,4}}^{(A,r)} = -\left\|\mathbf{x}_{\text{ALL}}^{(\mathbf{a})} - \mathbf{x}^{(\mathbf{b})}\right\|^2_{\left(K \cup I_{\text{PCP}}^{(A,r)} \cup I_{\text{LTC}}^{(A,r)} \cup M_{3,4}\right) \cap J^{(B,\mathbf{q})}},$$

whereas

$$\frac{1}{4}U_{\text{SAMPLE}}^{(A,r)} + \frac{1}{4}U_{\text{COUNTER}}^{(A,r)} + \frac{1}{2}U_{M_{3,4}}^{(A,r)} = -\left\|\mathbf{x}^{(\mathbf{a})} - \mathbf{x}^{(\mathbf{b})}\right\|^2_{J^{(A,r)} \cap J^{(B,\mathbf{q})}}.$$

As we mentioned earlier, we also want $(A, r)$ to encode the critical bits for each LOCAL-ENDOFALINE vertex. The indices of this encoding depend on both:

- the indices of the critical bits - which depends on the "counter" of the two LOCALEND-OFALINE vertices, allegedly encoded in both $\mathbf{a}$ and $\mathbf{b}$; and

- the directions in $\mathcal{G}^{t/2}$ that the PCP verifier uses to locally decode those bits - which depend on the random string $r$.

We let $G_{\text{CRITICAL}}^{(A,r)}(\mathbf{a}) \subset \mathcal{G}^t$ denote the set of additional entries that encode the critical bits - as induced by $(A, r)$'s strategy $\mathbf{x}^{(\mathbf{a})}|_K$ and random string $r$; let $I_{\text{CRITICAL}}^{(A,r)}(\mathbf{a}) \subset M_{1,2}$ denote the corresponding bits' indices. In order to define $(A, r)$'s utilities, it is actually more convenient to use the critical bits determined by $(B, \mathbf{q})$'s $\mathbf{b}$, henceforth denoted $I_{\text{CRITICAL}}^{(A,r)}(\mathbf{b})$. (The superscript remains $(A, r)$ because we still decode those bits according to random string $r$.) This added level of indirection prevents $(A, r)$ from manipulating her encoding of the counter to increase her utility from the critical bits. In other words, $\mathbf{a}$ includes an assignment to $q_{\text{CRITICAL}}$ affine $(t/2 + 2)$-dimensional subspaces of $\mathcal{G}^t$. On each subgame, we subjectively interpret $\mathbf{a}$'s assignment as a partial vector $\mathbf{x}_{\text{CRITICAL}}^{(\mathbf{a})}$ with values in $[-1, 2]$ on $I_{\text{CRITICAL}}^{(A,r)}(\mathbf{b})$. In particular, we define

$$U_{\text{CRITICAL}}^{(A,r)} \triangleq -\left\|\mathbf{x}_{\text{CRITICAL}}^{(\mathbf{a})} - \mathbf{x}^{(\mathbf{b})}\right\|^2_{I_{\text{CRITICAL}}^{(A,r)}(\mathbf{b}) \cap I^{(B,\mathbf{q})}}.$$

For some choices of $\mathbf{b}$, the induced assignment to the counter may be ambiguous or just far from every codeword. However, in that case we are probably far from every Brouwer line segment, so we don't care what $(A, r)$ assigns to the critical bits.

*Claim* 7.3. For every happy player $(A, r)$ on Alice's side and any action $\mathbf{a}$ in her support, for each $I \in \left\{I_{\text{SAMPLE}}^{(A,r)}, I_{\text{PCP}}^{(A,r)}, I_{\text{LTC}}^{(A,r)}\right\}$

$$\left\|\mathbf{x}_{\text{ALL}}^{(\mathbf{a})} - \mathbf{x}^{(\mathcal{B})}\right\|^2_I = O\left(\epsilon_{\text{NASH}}\right).$$

*Proof.* Recall that $(A, r)$'s total utility from her assignment to $I_{\text{LTC}}^{(A,r)}$ is given by

$$U_{\text{LTC}}^{(A,r)} = -\mathsf{E}_{\mathbf{q} \in \mathcal{G}^{t/2}} \mathsf{E}_{\mathbf{b} \sim \mathcal{B}[(B,\mathbf{q})]} \left[ \left\| \mathbf{x}_{\text{ALL}}^{(\mathbf{a})} - \mathbf{x}^{(\mathbf{b})} \right\|_{I_{\text{LTC}}^{(A,r)} \cap I^{(B,\mathbf{q})}}^2 \right]$$

$$= -\mathsf{E}_{i \in I_{\text{LTC}}^{(A,r)}} \left[ \left( \mathbf{x}_{\text{ALL}}^{(\mathbf{a})} |_i - \mathsf{E}_{\mathbf{b} \sim \mathcal{B}[(B,\mathbf{q}(i))]} \left[ \mathbf{x}^{(\mathbf{b})} |_i \right] \right)^2 \right] - \mathsf{E}_{i \in I_{\text{LTC}}^{(A,r)}} \left[ \text{Var}_{\mathbf{b} \sim \mathcal{B}[(B,\mathbf{q}(i))]} \left[ \mathbf{x}^{(\mathbf{b})} |_i \right] \right], \quad (16)$$

where $(B, \mathbf{q}(i))$ is the player on Bob's side for which $i \in I^{(B,\mathbf{q}(i))}$ (i.e. $\mathbf{q}(i)$ denotes the value $\mathbf{q} \in \mathcal{G}^{t/2}$ such that bit $i$ is part of the representation of some $\mathbf{g} \in G^{(B,\mathbf{q})}$).

Notice that the second term of (16) does not depend on $(A, r)$'s strategy, so she simply wants to maximize the first term. For every $i$, she can approximate $\mathsf{E}_{\mathbf{b} \sim \mathcal{B}[(B,\mathbf{q}(i))]} \left[ \mathbf{x}^{(\mathbf{b})} |_i \right]$ to within $\pm \epsilon_{\text{PRECISION}}$. Therefore, if $\mathbf{a}$ is an $\epsilon_{\text{NASH}}$-optimal strategy, we have

$$U_{\text{PCP}}^{(A,r)} \geq -\mathsf{E}_{i \in I_{\text{PCP}}^{(A,r)}} \left[ \text{Var}_{\mathbf{b} \sim \mathcal{B}[(B,\mathbf{q}(i))]} \left[ \mathbf{x}^{(\mathbf{b})} |_i \right] \right] - \underbrace{\epsilon_{\text{NASH}} - \epsilon_{\text{PRECISION}}^2}_{=O(\epsilon_{\text{NASH}})}.$$

An analogous argument works for $I_{\text{SAMPLE}}^{(A,r)}, I_{\text{PCP}}^{(A,r)}$. □

The following claims follow along the same lines:

*Claim* 7.4. For every happy player $(A, r)$ on Alice's side and any action $\mathbf{a}$ in her support,

$$\left\| \mathbf{x}^{(\mathbf{a})} - \mathbf{x}^{(\mathcal{B})} \right\|_K^2 = O(\epsilon_{\text{NASH}}).$$

*Claim* 7.5. For every happy player $(A, r)$ on Alice's side and any action $\mathbf{a}$ in her support,

$$\left\| \mathbf{x}^{(\mathbf{a})} - \mathbf{x}^{(\mathcal{B})} \right\|_{J^{(A,r)}}^2 = O(\epsilon_{\text{NASH}}).$$

**Corollary 7.6.** *Let $(\mathcal{A}, \mathcal{B})$ be an $(\epsilon_{\text{NASH}}, \epsilon_{\text{NASH}})$-Well-Supported-WeakNash, then*

$$\left\| \mathbf{x}^{(\mathcal{A})} - \mathbf{x}^{(\mathcal{B})} \right\|_M^2 = O(\epsilon_{\text{NASH}}).$$

### 7.4 Bob's imitation gadget

We would like to also have the players on Bob's side minimize $\left\| f\left(\mathbf{x}^{(\mathcal{A})}\right) - \mathbf{x}^{(\mathbf{b})} \right\|_{J^{(B,\mathbf{q})}}^2$. In order to implement this in a polymatrix game, we want to locally compute $f(\cdot)$ given access only to the partial information provided by player $(A, r)$'s strategy $\mathbf{a}$. In Section 3.3 we argued that we can locally compute $f_i(\mathbf{x})$ using only approximate, partial information about $\mathbf{x}$. The main goal of this section is to argue that for most $i \in J^{(A,r)}$, we can compute this partial information from $\mathbf{a}$, approximately and with high probability over $r$. Before we do that, however, let us assume that that we have some estimate $f^{(\mathbf{a})} |_i$ and derive the utility of player $(B, \mathbf{q})$.

For any action $\mathbf{a}$ that $(A, r)$ plays, we define a partial target vector $f^{(\mathbf{a})} \in ([-1, 2] \cup \{\bot\})^M$ that we would like $(B, \mathbf{q})$ to imitate. In particular, $(B, \mathbf{q})$ is incentivized to play a strategy that is close to $f^{(\mathbf{a})}$ on the coordinates where both $f^{(\mathbf{a})}$ and $\mathbf{x}^{(\mathbf{b})}$ are defined. We define,

$$U^{(B,\mathbf{q})} \triangleq - \left\| f^{(\mathbf{a})} - \mathbf{x}^{(\mathbf{b})} \right\|_{J^{(A,r)} \cap J^{(B,\mathbf{q})}}^2.$$

Similarly to $\mathbf{x}^{(\mathcal{A})}$ and $\mathbf{x}^{(\mathcal{B})}$, let $f^{(\mathcal{A})} \in [-1, 2]^M$ denote the weighted entry-wise average of $f^{(\mathbf{a})}$, where for each $i$ we take an average over $f^{(\mathbf{a})} |_i$ for all $\mathbf{a}$ in the support of $(A, r)$, for $r$ such that $i \in J^{(A,r)}$.

*Claim* 7.7. For every happy player $(B, \mathbf{q})$, on Bob's side and any action $\mathbf{b}$ in his support,

$$\left\| \mathbf{x}^{(\mathbf{b})} - f^{(\mathcal{A})} \right\|^2_{J^{(B,\mathbf{q})}} = O\left(\epsilon_{\text{NASH}}\right).$$

*Proof.* Analogous to Claim 7.3. □

**Corollary 7.8.** *Let $(\mathcal{A}, \mathcal{B})$ be an $(\epsilon_{\text{NASH}}, \epsilon_{\text{NASH}})$-Well-Supported-WeakNash, then*

$$\left\| \mathbf{x}^{(\mathcal{B})} - f^{(\mathcal{A})} \right\|^2_M = O\left(\epsilon_{\text{NASH}}\right).$$

For the rest of this section, our goal is to prove that $f^{(\mathcal{A})}$ is also close to $f\left(\mathbf{x}^{(\mathcal{B})}\right)$ - this would prove that $\mathbf{x}^{(\mathcal{B})}$ is an approximate fixed point of $f$.

### 7.5 Reading the holographic proofs

In this subsection we translate the desiderata from Proposition 6.1 to guarantees about approximate equilibria in our game. Our first task in approximating $f\left(\mathbf{x}^{(\mathcal{B})}\right)$ is deciding whether $\mathbf{x}^{(\mathcal{B})}$ is close to a Brouwer line segment, close to a Brouwer vertex, or far from both. Close to any Brouwer vertex, both restrictions $\mathbf{x}^{(\mathcal{B})}|_{M_1}$ and $\mathbf{x}^{(\mathcal{B})}|_{M_2}$ are close to some $\Pi(u), \Pi(v)$, for some $u, v \in V^{\text{LOCAL}}$; close to a Brouwer line, at least one of $\mathbf{x}^{(\mathcal{B})}|_{M_1}$ and $\mathbf{x}^{(\mathcal{B})}|_{M_2}$ is close to some $\Pi(u)$.

The first step of this first task is to decide whether $\mathbf{x}^{(\mathcal{B})}|_{M_1}$ is close to some $\Pi(u)$. Formally, we have a *game verifier* that encapsulates rounding assignments in $[-1, 2]^M$ to $\{0, 1\}^M$ and feeding them to the PCP verifier. Given that player $(A, r)$ chooses strategy $\mathbf{a}$, we say that the game verifier *accepts* $\mathbf{a}|_{M_1}$ if:

- $\mathbf{x}^{(\mathbf{a})}|_{M_1}$ is close to binary, i.e.

$$\left\| \mathbf{x}^{(\mathbf{a})} - \nu\left(\mathbf{x}^{(\mathbf{a})}|_{M_1}\right) \right\|^2_{J^{(A,r)}} < \sqrt{\epsilon_{\text{COMPLETE}}};$$

- and the PCP verifier accepts the rounded bits it reads from the first proof, $\nu\left(\mathbf{x}^{(\mathbf{a})}_{\text{ALL}}|_{M_1}\right)$.

If either of those does not hold, we say that the game verifier *rejects* $\mathbf{a}|_{M_1}$.

The main idea in all the lemmata below is that by the Good Sample desideratum in Proposition 6.1, $G^{(A,r)}_{\text{SAMPLE}}$, $G^{(A,r)}_{\text{PCP}}$, $G^{(A,r)}_{\text{LTC}}$ and $G^{(A,r)}_{\text{CRITICAL}}(\mathbf{a})$ are all representing samples of $\mathcal{G}^t$. Therefore, if $\mathbf{x}^{(\mathcal{B})}|_{M_1}$ is close to $\{0,1\}^{M_1}$, then we expect that $\mathbf{x}^{(\mathcal{B})}|_{M_1 \cap I^{(A,r)}_{\text{PCP}}}$ is also close to $\{0,1\}^{M_1 \cap I^{(A,r)}_{\text{PCP}}}$. By Claim 7.3, we can then also expect that $\mathbf{x}^{(\mathbf{a})}_{\text{ALL}}|_{M_1 \cap I^{(A,r)}_{\text{PCP}}}$ is close to the *same* vector in $\{0,1\}^{M_1 \cap I^{(A,r)}_{\text{PCP}}}$. Furthermore, if $\mathbf{x}^{(\mathbf{a})}_{\text{ALL}}|_{M_1 \cap I^{(A,r)}_{\text{PCP}}}$ is $\epsilon$-close to some binary vector $\nu\left(\mathbf{x}^{(\mathcal{B})}|_{M_1 \cap I^{(A,r)}_{\text{PCP}}}\right)$, then its rounding, $\nu\left(\mathbf{x}^{(\mathbf{a})}_{\text{ALL}}|_{M_1 \cap I^{(A,r)}_{\text{PCP}}}\right)$, is also $O(\epsilon)$-close to $\nu\left(\mathbf{x}^{(\mathcal{B})}|_{M_1 \cap I^{(A,r)}_{\text{PCP}}}\right)$: On each coordinate $i$, if the two binary vectors disagree (if $\left|\nu\left(\mathbf{x}^{(\mathcal{B})}|_i\right) - \nu\left(\mathbf{x}^{(\mathbf{a})}_{\text{ALL}}|_i\right)\right| = 1$), then we must have $\left|\nu\left(\mathbf{x}^{(\mathcal{B})}|_i\right) - \mathbf{x}^{(\mathcal{B})}|_i\right| + \left|\mathbf{x}^{(\mathcal{B})}|_i - \mathbf{x}^{(\mathbf{a})}_{\text{ALL}}|_i\right| \geq 1/2$. Finally, if $\nu\left(\mathbf{x}^{(\mathbf{a})}_{\text{ALL}}|_{M_1 \cap I^{(A,r)}_{\text{PCP}}}\right)$ is close to $\nu\left(\mathbf{x}^{(\mathcal{B})}|_{M_1 \cap I^{(A,r)}_{\text{PCP}}}\right)$ for most $\mathbf{a}$, then we can argue that reading $\nu\left(\mathbf{x}^{(\mathbf{a})}_{\text{ALL}}|_{M_1 \cap I^{(A,r)}_{\text{PCP}}}\right)$ is almost like sampling $\nu\left(\mathbf{x}^{(\mathcal{B})}|_{M_1}\right)$, and then letting an adaptive adversary corrupt a small fraction of the entries. Therefore, we can use the guarantees of the PCP verifier with robust soundness / decoding / completeness.

### 7.5.1 Soundness

**Lemma 7.9.** *Let $(\mathcal{A}, \mathcal{B})$ be an $(\epsilon_{\text{NASH}}, \epsilon_{\text{NASH}})$-Well-Supported-WeakNash. If $\left\|\mathbf{x}^{(\mathcal{B})} - \Pi(u)\right\|^2_{M_1} > 4\epsilon_{\text{SOUND}}$ for every $u \in V^{\text{LOCAL}}$, then for a $(1 - O(\epsilon_{\text{NASH}}))$-fraction of $r$'s, the game verifier rejects $\mathbf{a}\,|_{M_1}$ for every strategy $\mathbf{a}$ in the support of $(A, r)$.*

*Proof.* Suppose that $\left\|\mathbf{x}^{(\mathcal{B})} - \Pi(u)\right\|^2_{M_1} > 4\epsilon_{\text{SOUND}}$. Then by triangle inequality we have that at least one of the following holds:

$$\left\|\mathbf{x}^{(\mathcal{B})} - \nu\left(\mathbf{x}^{(\mathcal{B})}\right)\right\|^2_{M_1} > \epsilon_{\text{SOUND}}^3 \tag{17}$$

$$\left\|\nu\left(\mathbf{x}^{(\mathcal{B})}\right) - \Pi(u)\right\|^2_{M_1} > \epsilon_{\text{SOUND}} \tag{18}$$

**Far from $\{0,1\}^{M_1}$**

Assume first that (17) holds. We can further break (17) into its $K$-component and $(E, \pi)$-component:

$$\left\|\mathbf{x}^{(\mathcal{B})} - \nu\left(\mathbf{x}^{(\mathcal{B})}\right)\right\|^2_{M_1} = \frac{1}{2}\left\|\mathbf{x}^{(\mathcal{B})} - \nu\left(\mathbf{x}^{(\mathcal{B})}\right)\right\|^2_{M_1 \cap K} + \frac{1}{2}\left\|\mathbf{x}^{(\mathcal{B})} - \nu\left(\mathbf{x}^{(\mathcal{B})}\right)\right\|^2_{M_1 \setminus K} \tag{19}$$

When restricted to a random $r \in \mathcal{R}_{\text{PCP}}$, we can learn the first term exactly, and estimate the second term via the proxy:

$$\left\|\mathbf{x}^{(\mathcal{B})} - \nu\left(\mathbf{x}^{(\mathcal{B})}\right)\right\|^2_{M_1 \cap I_{\text{SAMPLE}}^{(A,r)}}.$$

By the Good Sample guarantee from Proposition 6.1, the $\left(M_1 \cap I_{\text{SAMPLE}}^{(A,r)}\right)$-restricted distance between $\mathbf{x}^{(\mathcal{B})}$ and $\nu\left(\mathbf{x}^{(\mathcal{B})}\right)$ concentrates (to within $\pm o(1)$, with high probability) around its expectation (the last term of (19)). Therefore, with probability $(1 - o(1))$ over the choice of $r \in \mathcal{R}_{\text{PCP}}$,

$$\left\|\mathbf{x}^{(\mathcal{B})} - \nu\left(\mathbf{x}^{(\mathcal{B})}\right)\right\|^2_{J^{(A,r)}} \geq \frac{1}{4}\left\|\mathbf{x}^{(\mathcal{B})} - \nu\left(\mathbf{x}^{(\mathcal{B})}\right)\right\|^2_{\underbrace{M_1 \cap J^{(A,r)}}_{=M_1 \cap \left(I_{\text{SAMPLE}}^{(A,r)} \cup K\right)}} = \Omega\left(\epsilon_{\text{SOUND}}^3\right).$$

For any such $r$, if player $(A, r)$ is also happy, we have by Claim 7.5 that for every strategy $\mathbf{a}$ in her support,

$$\left\|\mathbf{x}^{(\mathbf{a})} - \nu\left(\mathbf{x}^{(\mathbf{a})}\right)\right\|^2_{J^{(A,r)}} = \Omega\left(\epsilon_{\text{SOUND}}^3\right) \gg \sqrt{\epsilon_{\text{COMPLETE}}}. \tag{20}$$

**Close to $\{0,1\}^{M_1}$, but far from every $\Pi(u)$**

Alternatively, assume that $\mathbf{x}^{(\mathcal{B})}\,|_{M_1}$ is close to $\{0,1\}^{M_1}$, but (18) holds: we have a binary vector $\nu\left(\mathbf{x}^{(\mathcal{B})}\,|_{M_1}\right)$ which is $\epsilon_{\text{SOUND}}$-far from a valid proof.

We assume wlog that (17) is false, namely

$$\left\|\mathbf{x}^{(\mathcal{B})} - \nu\left(\mathbf{x}^{(\mathcal{B})}\right)\right\|^2_{M_1} \leq \epsilon_{\text{SOUND}}^3.$$

By the Good Sample guarantee in Proposition (6.1), we have that for $(1 - o(1))$-fraction of $r$'s, also

$$\left\|\mathbf{x}^{(\mathcal{B})} - \nu\left(\mathbf{x}^{(\mathcal{B})}\right)\right\|^2_{M_1 \cap I^{(A,r)}_{\text{PCP}}} = O\left(\epsilon^3_{\text{SOUND}}\right). \qquad (21)$$

By Claim 7.3, for every happy $(A, r)$ and every strategy $\mathbf{a}$ in her support,

$$\left\|\mathbf{x}^{(\mathbf{a})}_{\text{ALL}} - \mathbf{x}^{(\mathcal{B})}\right\|^2_{M_1 \cap I^{(A,r)}_{\text{PCP}}} = O\left(\epsilon^3_{\text{SOUND}}\right).$$

Therefore, combining with (21), we have:

$$\left\|\nu\left(\mathbf{x}^{(\mathbf{a})}_{\text{ALL}}\right) - \nu\left(\mathbf{x}^{(\mathcal{B})}\right)\right\|^2_{M_1 \cap I^{(A,r)}_{\text{PCP}}} = O\left(\epsilon^3_{\text{SOUND}}\right),$$

and analogous arguments hold for $M_1 \cap K$ and $M_1 \cap I^{(A,r)}_{\text{LTC}}$. Therefore, by the Robust soundness guarantee in Proposition 6.1, the game verifier rejects every $\mathbf{a}\,|_{M_1}$ for all but an $o(1)$-fraction of happy $(A, r)$'s. $\square$

### 7.5.2 Completeness

**Lemma 7.10.** *Let $(\mathcal{A}, \mathcal{B})$ be an $(\epsilon_{\text{NASH}}, \epsilon_{\text{NASH}})$-Well-Supported-WeakNash. If $\left\|\mathbf{x}^{(\mathcal{B})} - \Pi(u)\right\|^2_{M_1} \leq \frac{1}{4}\epsilon_{\text{COMPLETE}}$ for some $u \in V^{\text{LOCAL}}$, then for a $(1 - O(\epsilon_{\text{NASH}}))$-fraction of $r$'s, and every strategy $\mathbf{a}$ in the support of $(A, r)$, the game verifier accepts $\mathbf{a}\,|_{M_1}$.*

*Proof.* First, notice that the premise implies that,

$$\left\|\nu\left(\mathbf{x}^{(\mathcal{B})}\right) - \Pi(u)\right\|^2_{M_1} \leq \epsilon_{\text{COMPLETE}}. \qquad (22)$$

(Since for any $i \in M_1$, $\left\|\nu\left(\mathbf{x}^{(\mathcal{B})}\right) - \Pi(u)\right\|^2_i \leq 4\left\|\mathbf{x}^{(\mathcal{B})} - \Pi(u)\right\|^2_i$ .)

Furthermore, we also have that

$$\left\|\mathbf{x}^{(\mathcal{B})} - \nu\left(\mathbf{x}^{(\mathcal{B})}\right)\right\|^2_{M_1} \leq \left\|\mathbf{x}^{(\mathcal{B})} - \Pi(u)\right\|^2_{M_1} \leq \frac{1}{4}\epsilon_{\text{COMPLETE}}.$$

Therefore, by the Good Sample guarantee from Proposition 6.1, we also have that for a $(1 - o(1))$-fraction of $r$'s,

$$\left\|\mathbf{x}^{(\mathcal{B})} - \nu\left(\mathbf{x}^{(\mathcal{B})}\right)\right\|^2_{M_1 \cap I^{(A,r)}_{\text{PCP}}} = O(\epsilon_{\text{COMPLETE}}).$$

Combining with Claim 7.3, we get that whenever $(A, r)$ is also happy,

$$\left\|\mathbf{x}^{(\mathbf{a})}_{\text{ALL}} - \nu\left(\mathbf{x}^{(\mathcal{B})}\right)\right\|^2_{M_1 \cap I^{(A,r)}_{\text{PCP}}} = O(\epsilon_{\text{COMPLETE}}).$$

Thus, as in (22), we have

$$\left\|\nu\left(\mathbf{x}^{(\mathbf{a})}_{\text{ALL}}\right) - \nu\left(\mathbf{x}^{(\mathcal{B})}\right)\right\|^2_{M_1 \cap I^{(A,r)}_{\text{PCP}}} = O(\epsilon_{\text{COMPLETE}}).$$

Analogous arguments hold for $M_1 \cap K$ and $M_1 \cap I^{(A,r)}_{\text{LTC}}$. Therefore, for a $(1 - O(\epsilon_{\text{NASH}}))$-fraction of $(A, r)$'s, sampling $\nu\left(\mathbf{x}^{(\mathbf{a})}_{\text{ALL}}\,|_{M_1}\right)$ satisfies the distance criteria for the Completeness and Robust completeness in Proposition 6.1. $\square$

### 7.5.3 Decoding

**Lemma 7.11.** *Let* $(\mathcal{A}, \mathcal{B})$ *be an* $(\epsilon_{\text{NASH}}, \epsilon_{\text{NASH}})$-*Well-Supported-WeakNash. If* $\left\|\mathbf{x}^{(\mathcal{B})} - \Pi(u)\right\|_{M_1}^2 \leq \frac{1}{4}\epsilon_{\text{DECODE}}$ *for some* $u \in V^{\text{LOCAL}}$, *then for a* $(1 - O(\epsilon_{\text{NASH}}))$-*fraction of* $r$'s, *and every strategy* $\mathbf{a}$ *in the support of* $(A, r)$:

- **(Error correction)** *The game verifier can compute the restriction of* $\Pi(u)$ *to* $M_1 \cap J^{(A,r)}$ *correctly.*

- **(Critical bits)** *The induced assignment on the critical bits is approximately correct:*
$$\left\|\mathbf{x}^{(\mathbf{a})}_{\text{CRITICAL}} - \Pi(u)\right\|_{M_1 \cap I^{(A,r)}_{\text{CRITICAL}}(\mathbf{a})}^2 = O(\epsilon_{\text{DECODE}}).$$

- **(Extrapolation)** *The game verifier can also compute the restrictions of* $\Pi(S^{\text{LOCAL}}(u))$ *and* $\Pi(P^{\text{LOCAL}}(u))$ *to* $M_1 \cap J^{(A,r)}$ *correctly.*

*Proof.* By analogous argument to Lemma 7.10 (replace $\epsilon_{\text{COMPLETE}}$ with $\epsilon_{\text{DECODE}}$), we have that, for a $(1 - O(\epsilon_{\text{NASH}}))$-fraction of $(A, r)$'s, sampling $\nu\left(\mathbf{x}^{(\mathbf{a})}_{\text{ALL}} \mid_{M_1}\right)$ satisfies the distance criteria for the Decoding and Robust decoding in Proposition 6.1. Thus the decoding of $\Pi(u)$ on $M_1 \cap J^{(A,r)}$ follows from the "Error-correction on a sample" desideratum of Proposition 6.1.

**Critical bits**

By the premise, we have that
$$\left\|\mathbf{x}^{(\mathcal{B})} - C(u)\right\|_{M_1 \cap K}^2 = O(\epsilon_{\text{DECODE}}). \tag{23}$$

Thus, by Corollary 7.8, also
$$\left\|f^{(\mathcal{A})} - C(u)\right\|_{M_1 \cap K}^2 = O(\epsilon_{\text{DECODE}}).$$

By Claims 7.4 and 7.7, for every $\epsilon_{\text{NASH}}$-optimal $\mathbf{a}$ and $\mathbf{b}$ we also have that
$$\left\|\mathbf{x}^{(\mathbf{a})} - C(u)\right\|_{M_1 \cap K}^2 = O(\epsilon_{\text{DECODE}}). \tag{24}$$
$$\left\|\mathbf{x}^{(\mathbf{b})} - C(u)\right\|_{M_1 \cap K}^2 = O(\epsilon_{\text{DECODE}}). \tag{25}$$

Therefore, we have that for every $\epsilon_{\text{NASH}}$-optimal $\mathbf{a}$ and $\mathbf{b}$, the counter's encoding $C(u)$ is decoded correctly; and in particular,
$$M_1 \cap I^{(A,r)}_{\text{CRITICAL}}(\mathbf{a}) = M_1 \cap I^{(A,r)}_{\text{CRITICAL}}(\mathbf{b}). \tag{26}$$

Let $U^{(A,r)}_{\text{CRITICAL}} \mid_{M_1}$ denote the portion of $U^{(A,r)}_{\text{CRITICAL}}$ that is derived from assignments to the critical bits in the first proof. For every happy player $(A, r)$, and every strategy $\mathbf{a}$ in her support, we have

$$U^{(A,r)}_{\text{CRITICAL}} \mid_{M_1} = -\mathsf{E}_{\mathbf{q} \in \mathcal{G}^{t/2}} \mathsf{E}_{\mathbf{b} \sim \mathcal{B}[(B,\mathbf{q})]} \left[\left\|\mathbf{x}^{(\mathbf{a})}_{\text{CRITICAL}} - \mathbf{x}^{(\mathbf{b})}\right\|^2_{M_1 \cap I^{(A,r)}_{\text{CRITICAL}}(\mathbf{b}) \cap I^{(B,\mathbf{q})}}\right]$$

$$= -\mathsf{E}_{\mathbf{q} \in \mathcal{G}^{t/2}} \mathsf{E}_{\mathbf{b} \sim \mathcal{B}[(B,\mathbf{q})]} \left[\left\|\mathbf{x}^{(\mathbf{a})}_{\text{CRITICAL}} - \mathbf{x}^{(\mathbf{b})}\right\|^2_{M_1 \cap I^{(A,r)}_{\text{CRITICAL}}(\mathbf{a}) \cap I^{(B,\mathbf{q})}}\right] \pm O(\epsilon_{\text{NASH}})$$

$$= -\left\|\mathbf{x}^{(\mathbf{a})}_{\text{CRITICAL}} - \mathbf{x}^{(\mathcal{B})}\right\|^2_{M_1 \cap I^{(A,r)}_{\text{CRITICAL}}(\mathbf{a})} - \mathsf{E}_{i \in M_1 \cap I^{(A,r)}_{\text{CRITICAL}}(\mathbf{a})} \left[\text{Var}_{\mathbf{b} \sim \mathcal{B}[(B,\mathbf{q}(i))]} \left[\mathbf{x}^{(\mathbf{b})} \mid_i\right]\right] \pm O(\epsilon_{\text{NASH}}),$$

where the second equality follows from (26). Therefore, for every $\epsilon_{\text{Nash}}$-optimal strategy $\mathbf{a}$,

$$\left\|\mathbf{x}_{\text{CRITICAL}}^{(\mathbf{a})} - \mathbf{x}^{(\mathcal{B})}\right\|_{M_1 \cap I_{\text{CRITICAL}}^{(A,r)}(\mathbf{a})}^2 = O\left(\epsilon_{\text{Precision}}^2 + \epsilon_{\text{Nash}}\right) = O\left(\epsilon_{\text{Nash}}\right). \tag{27}$$

By the Good Sample guarantee from Proposition 6.1, we have that for a $(1 - o(1))$-fraction of $r$'s,

$$\left\|\mathbf{x}^{(\mathcal{B})} - \Pi(u)\right\|_{M_1 \cap I_{\text{CRITICAL}}^{(A,r)}(\mathbf{a})}^2 = O\left(\epsilon_{\text{Decode}}\right), \tag{28}$$

Combining with (27), we have that for a $(1 - O(\epsilon_{\text{Nash}}))$-fraction of $r$'s,

$$\left\|\mathbf{x}_{\text{CRITICAL}}^{(\mathbf{a})} - \Pi(u)\right\|_{M_1 \cap I_{\text{CRITICAL}}^{(A,r)}(\mathbf{a})}^2 = O\left(\epsilon_{\text{Decode}}\right).$$

**Extrapolation**

Notice that (28) in particular implies

$$\left\|\mathbf{x}^{(\mathcal{B})} - \nu\left(\mathbf{x}^{(\mathcal{B})}\right)\right\|_{M_1 \cap I_{\text{CRITICAL}}^{(A,r)}(\mathbf{a})}^2 = O\left(\epsilon_{\text{Decode}}\right).$$

Combining with (27), we have that

$$\left\|\mathbf{x}_{\text{CRITICAL}}^{(\mathbf{a})} - \nu\left(\mathbf{x}^{(\mathcal{B})}\right)\right\|_{M_1 \cap I_{\text{CRITICAL}}^{(A,r)}(\mathbf{a})}^2 = O\left(\epsilon_{\text{Decode}}\right).$$

Therefore, for a $(1 - O(\epsilon_{\text{Nash}}))$-fraction of $(A, r)$'s, sampling $\nu\left(\mathbf{x}_{\text{CRITICAL}}^{(\mathbf{a})} |_{M_1}\right)$ satisfies the distance criteria for the Decoding and Robust decoding in Proposition 6.1. Therefore, by the "Error-correction on the critical bits" desideratum, for a $(1 - O(\epsilon_{\text{Nash}}))$-fraction of $(A, r)$'s, the PCP verifier correctly decodes the assignment to all the critical bits.

From the assignment to the critical bits, the PCP verifier can reconstruct the complete difference vector $E(u) - E(S^{\text{LOCAL}}(u))$. So now, accessing $E(S^{\text{LOCAL}}(u))$ is equivalent to accessing $E(u)$ (with the same error guarantees). Using the local proof construction guarantee from Proposition 6.1, the PCP verifier can correctly compute $\pi(S^{\text{LOCAL}}(u))$ on $I_{\text{SAMPLE}}^{(A,r)}$. Finally, recall that once the PCP verifier successfully decoded $C(u)$, it can also locally compute $C(S^{\text{LOCAL}}(u))$. □

### 7.6 Default displacement

The simplest case is when the restrictions of $\mathbf{x}^{(\mathcal{B})}$ to both proofs are far from any $\Pi(u), \Pi(v)$. In this case $\mathbf{x}^{(\mathcal{B})}$ is far from all Brouwer line segments, and $f^{(\mathcal{A})}$ needs to apply the default displacement.

**Lemma 7.12.** *If $(\mathcal{A}, \mathcal{B})$ is an $(\epsilon_{\text{Nash}}, \epsilon_{\text{Nash}})$-Well-Supported-WeakNash, and both $\left\|\mathbf{x}^{(\mathcal{B})} - \Pi(u)\right\|_{M_1}^2 > 4\epsilon_{\text{Sound}}$ and $\left\|\mathbf{x}^{(\mathcal{B})} - \Pi(v)\right\|_{M_2}^2 > 4\epsilon_{\text{Sound}}$ for every $u, v \in V^{\text{LOCAL}}$, then $\left\|f^{(\mathcal{A})} - f\left(\mathbf{x}^{(\mathcal{B})}\right)\right\|_M^2 = O(\epsilon_{\text{Nash}})$.*

*Proof.* By Lemma 7.9, for a $(1 - O(\epsilon_{\text{Nash}}))$-fraction of $r$'s and every strategy $\mathbf{a}$ in their support, the game verifier rejects both proofs. In this case, for all those $\mathbf{a}$'s, $f^{(\mathbf{a})}$ implements the default displacement. If **a** is also an $\epsilon_{\text{NASH}}$-optimal strategy, then $\left\|\mathbf{x}^{(\mathbf{a})} - \mathbf{x}^{(\mathcal{B})}\right\|^2_{J^{(A,r)}} = O(\epsilon_{\text{NASH}})$, in which case also

$$\left\|f^{(\mathbf{a})} - f\left(\mathbf{x}^{(\mathcal{B})}\right)\right\|^2_{J^{(A,r)}} = O(\epsilon_{\text{NASH}}).$$

Since the rest of Alice's players can have at most an $O(\epsilon_{\text{NASH}})$ effect, we have that

$$\left\|f^{(\mathcal{A})} - f\left(\mathbf{x}^{(\mathcal{B})}\right)\right\|^2_M = O(\epsilon_{\text{NASH}}).$$

□

The following cases are handled with minor modifications:

- $\mathbf{x}^{(\mathcal{B})}$ is outside the picture and far from the first line ($\mathbf{z}_2 \to (\Pi(u_0), \Pi(u_0), 0, 0)$).

- $\mathbf{x}^{(\mathcal{B})}$ is inside the picture, both $\mathbf{x}^{(\mathcal{B})}|_{M_1}$ and $\mathbf{x}^{(\mathcal{B})}|_{M_2}$ are close to valid proofs, but $\mathbf{x}^{(\mathcal{B})}|_{M_{3,4}}$ doesn't match any Brouwer line segment or vertex.

- $\mathbf{x}^{(\mathcal{B})}$ is inside the picture, both $\mathbf{x}^{(\mathcal{B})}|_{M_1}$ and $\mathbf{x}^{(\mathcal{B})}|_{M_2}$ are close to valid proofs $\Pi(u)$ and $\Pi(v)$, but $u$ and $v$ are not consecutive vertices in the LOCALENDOFALINE graph.

## 7.7 Close to a line (1)

Now suppose that $\mathbf{x}^{(\mathcal{B})}$ is close to some Brouwer line segment $(\mathbf{s} \to \mathbf{t})$ from $\mathbf{s} = (\Pi(u), \Pi(u), 0, 0)$ to $\mathbf{t} = (\Pi(u), \Pi(v), 0, 0)$, for $u, v \in V^{\text{LOCAL}}$ such that $v$ is the LOCALENDOFALINE-successor of $u$. (The case of a line from $(\Pi(u), \Pi(v), 1, 0)$ to $(\Pi(v), \Pi(v), 0, 1)$ follows with minor modifications.) Now, the line $(\mathbf{s} \to \mathbf{t})$ consists of points of the form $\beta \mathbf{s} + (1-\beta)\mathbf{t}$ (for $\beta \in [0, 1]$). From player $(A, r)$'s assignment to $M_1$, we can locally decode and verify $\Pi(u)$ on $I^{(A,r)}_{\text{PCP}} \cup I^{(A,r)}_{\text{LTC}} \cup K$. Furthermore, we can reconstruct both $\Pi(u)$ and $\Pi(v)$ on $I^{(A,r)}_{\text{SAMPLE}} \cup K$. Let $\mathbf{s}^{(\mathbf{a})}, \mathbf{t}^{(\mathbf{a})}$ denote the locally reconstructed restrictions of $\mathbf{s}, \mathbf{t}$, respectively, to $J^{(A,r)}$.

If all the tests passed, then we want to locally apply the displacement close to a line, as defined in (1). The assignments of $\mathbf{x}^{(\mathbf{a})}, \mathbf{s}^{(\mathbf{a})}, \mathbf{t}^{(\mathbf{a})}$ also induce a partial vector $\mathbf{z}^{(\mathbf{a})}$, which is the point closest to $\mathbf{x}^{(\mathbf{a})}$ on the line segment $(\mathbf{s}^{(\mathbf{a})} \to \mathbf{t}^{(\mathbf{a})})$. We can also use $\left\|\mathbf{x}^{(\mathbf{a})} - \mathbf{z}^{(\mathbf{a})}\right\|_{J^{(A,r)}}, \left\|\mathbf{t}^{(\mathbf{a})} - \mathbf{s}^{(\mathbf{a})}\right\|_{J^{(A,r)}}$ as estimates of $\|\mathbf{x} - \mathbf{z}\|_M, \|\mathbf{t} - \mathbf{s}\|_M$.

**Lemma 7.13.** *Let $(\mathcal{A}, \mathcal{B})$ be an $(\epsilon_{\text{NASH}}, \epsilon_{\text{NASH}})$-Well-Supported-WeakNash. Suppose that $\mathbf{x}^{(\mathcal{B})}$ is somewhat close to some Brouwer line segment,*

$$\epsilon^2_{\text{COMPLETE}} < \min_\beta \left\|\mathbf{x}^{(\mathcal{B})} - (\Pi(u), \beta \Pi(u) + (1-\beta)\Pi(v), 0, 0)\right\|^2_M < \epsilon_{\text{DECODE}}, \quad (29)$$

*but far from its endpoints,*

$$\left\|\mathbf{x}^{(\mathcal{B})} - (\Pi(u), \Pi(u), 0, 0)\right\|_M > 2\sqrt{h}$$
$$\left\|\mathbf{x}^{(\mathcal{B})} - (\Pi(u), \Pi(v), 0, 0)\right\|_M > 2\sqrt{h}, \quad (30)$$

*where $u, v \in V^{\text{LOCAL}}$ satisfy $v = S^{\text{LOCAL}}(u)$ and $u = P^{\text{LOCAL}}(v)$.*
*Then $\left\|f^{(\mathcal{A})} - f\left(\mathbf{x}^{(\mathcal{B})}\right)\right\|^2_M = O(\epsilon_{\text{NASH}})$.*

*Proof.* By Lemma 7.11, for a $(1 - O(\epsilon_{\text{NASH}}))$-fraction of $r$'s, and every strategy $\mathbf{a}$ in the support of $(A, r)$, the game verifier can compute $\Pi(u)|_{I_{\text{SAMPLE}}^{(A,r)} \cup K}$ and $\Pi(v)|_{I_{\text{SAMPLE}}^{(A,r)} \cup K}$ correctly; let $\widehat{\Pi(u)}$ and $\widehat{\Pi(v)}$ denote the respective results of those computations.

Consider

$$\beta_{(\mathbf{s} \to \mathbf{t})}^{(\mathbf{a})} \triangleq \arg\min_{\beta} \left\| \mathbf{x}^{(\mathbf{a})} - \left( \beta \widehat{\Pi(u)} + (1-\beta) \widehat{\Pi(v)} \right) \right\|_{M_2 \cap \left( I_{\text{SAMPLE}}^{(A,r)} \cup K \right)}^2. \tag{31}$$

and

$$\mathbf{z}^{\left(\beta_{(\mathbf{s} \to \mathbf{t})}^{(\mathbf{a})}\right)} \triangleq \left( \Pi(u), \beta_{(\mathbf{s} \to \mathbf{t})}^{(\mathbf{a})} \Pi(u) + \left(1 - \beta_{(\mathbf{s} \to \mathbf{t})}^{(\mathbf{a})}\right) \Pi(v), 0, 0 \right) \in [-1, 2]^M;$$

By (29), we have

$$\left\| \mathbf{x}^{(\mathcal{B})} - \mathbf{z}^{\left(\beta_{(\mathbf{s} \to \mathbf{t})}^{(\mathbf{a})}\right)} \right\|_M^2 > \epsilon_{\text{COMPLETE}}^2. \tag{32}$$

By the Good Sample guarantee from Proposition 6.1, whenever $\Pi(u)|_{I_{\text{SAMPLE}}^{(A,r)} \cup K}$ and $\Pi(v)|_{I_{\text{SAMPLE}}^{(A,r)} \cup K}$ are indeed decoded correctly,

$$\left\| \mathbf{x}^{(\mathbf{a})} - \mathbf{z}^{(\mathbf{a})} \right\|_{J^{(A,r)}}^2 > \epsilon_{\text{COMPLETE}}^2 - O(\epsilon_{\text{NASH}}) \gg (3h)^2.$$

Therefore, for all those $\mathbf{a}$'s, we have that $f^{(\mathbf{a})}$ correctly applies the default displacement. □

**Lemma 7.14.** *Let $(\mathcal{A}, \mathcal{B})$ be an $(\epsilon_{\text{NASH}}, \epsilon_{\text{NASH}})$-Well-Supported-WeakNash. Suppose that $\mathbf{x}^{(\mathcal{B})}$ is close to some Brouwer line segment,*

$$\min_{\beta} \left\| \mathbf{x}^{(\mathcal{B})} - (\Pi(u), \beta \Pi(u) + (1-\beta) \Pi(v), 0, 0) \right\|_M^2 \leq \epsilon_{\text{COMPLETE}}^2, \tag{33}$$

*but far from its endpoints,*

$$\begin{aligned}
\left\| \mathbf{x}^{(\mathcal{B})} - (\Pi(u), \Pi(u), 0, 0) \right\|_M &> 2\sqrt{h} \\
\left\| \mathbf{x}^{(\mathcal{B})} - (\Pi(u), \Pi(v), 0, 0) \right\|_M &> 2\sqrt{h},
\end{aligned} \tag{34}$$

*where $u, v \in V^{\text{LOCAL}}$ satisfy $v = S^{\text{LOCAL}}(u)$ and $u = P^{\text{LOCAL}}(v)$.*

*Then $\left\| f^{(\mathcal{A})} - f\left(\mathbf{x}^{(\mathcal{B})}\right) \right\|_M^2 = O(\epsilon_{\text{NASH}})$.*

*Proof.* We show that $f^{(\mathbf{a})}$ correctly implements the displacement close to a line (Subsection 7.7) for $\mathbf{s} = (\Pi(u), \Pi(u), 0, 0)$ and $\mathbf{t} = (\Pi(u), \Pi(v), 0, 0)$.

By Lemma 7.10, for a $(1 - O(\epsilon_{\text{NASH}}))$-fraction of $r$'s, and every strategy $\mathbf{a}$ in the support of $(A, r)$, the game verifier accepts $\mathbf{a}|_{\{1,\ldots,m\}}$. Furthermore, by Lemma 7.11, it can compute $\Pi(u)|_{I_{\text{SAMPLE}}^{(A,r)} \cup K}$ and $\Pi(v)|_{I_{\text{SAMPLE}}^{(A,r)} \cup K}$ correctly; we again denote the partial proofs locally computed by the game verifier $\widehat{\Pi(u)}$ and $\widehat{\Pi(v)}$. Denote the induced partial vectors $\mathbf{s}^{(\mathbf{a})}$ and $\mathbf{t}^{(\mathbf{a})}$. Recall that for every $\epsilon$-optimal $\mathbf{a}$, $\left\| \mathbf{x}^{(\mathbf{a})} - \mathbf{x}^{(\mathcal{B})} \right\|_{J^{(A,r)}}^2 = O(\epsilon_{\text{NASH}})$, and also, $\left| \left\| \mathbf{s}^{(\mathbf{a})} - \mathbf{t}^{(\mathbf{a})} \right\|_{J^{(A,r)}} - \left\| \mathbf{s} - \mathbf{t} \right\|_M \right|^2 = o(1)$.

Let

$$\beta_{(\mathbf{s} \to \mathbf{t})}^{(\mathbf{a})} \triangleq \frac{\left(\mathbf{t}^{(\mathbf{a})} - \mathbf{s}^{(\mathbf{a})}\right)}{\left\| \mathbf{s}^{(\mathbf{a})} - \mathbf{t}^{(\mathbf{a})} \right\|_{J^{(A,r)}}} \cdot \left( \mathbf{x}^{(\mathbf{a})} - \mathbf{s}^{(\mathbf{a})} \right),$$

where the $(\cdot)$ denotes a $J^{(A,r)}$-restricted dot-product.

For every $\epsilon$-optimal $\mathbf{a}$, $\left\|\mathbf{x}^{(\mathbf{a})} - \mathbf{x}^{(\mathcal{B})}\right\|^2_{J(A,r)} = O(\epsilon_{\mathrm{NASH}})$; for most of them $\Pi(u)\,|_{I^{(A,r)}_{\mathrm{SAMPLE}} \cup K}$ and $\Pi(v)\,|_{I^{(A,r)}_{\mathrm{SAMPLE}} \cup K}$ are also computed correctly, so $\mathbf{s}^{(\mathbf{a})}, \mathbf{t}^{(\mathbf{a})}$ are the correct restrictions of $\mathbf{s}, \mathbf{t}$ to $J^{(A,r)}$. Then, by the Good Sample guarantee from Proposition 6.1, we have that with probability $(1 - o(1))$, the $J^{(A,r)}$-restricted dot-product is a good approximation. Namely,

$$\left|\left(\mathbf{t}^{(\mathbf{a})} - \mathbf{s}^{(\mathbf{a})}\right) \cdot \left(\mathbf{x}^{(\mathcal{B})}\,|_{J(A,r)} - \mathbf{s}^{(\mathbf{a})}\right) - (\mathbf{t} - \mathbf{s}) \cdot \left(\mathbf{x}^{(\mathcal{B})} - \mathbf{s}\right)\right| = o(1)$$

and

$$\left|\left(\mathbf{t}^{(\mathbf{a})} - \mathbf{s}^{(\mathbf{a})}\right) \cdot \left(\mathbf{x}^{(\mathbf{a})} - \mathbf{x}^{(\mathcal{B})}\,|_{J(A,r)}\right)\right| = O\left(\left\|\mathbf{x}^{(\mathbf{a})} - \mathbf{x}^{(\mathcal{B})}\right\|_{J(A,r)}\right) = O(\sqrt{\epsilon_{\mathrm{NASH}}}),$$

and thus also

$$\left|\left(\mathbf{t}^{(\mathbf{a})} - \mathbf{s}^{(\mathbf{a})}\right) \cdot \left(\mathbf{x}^{(\mathbf{a})} - \mathbf{s}^{(\mathbf{a})}\right) - (\mathbf{t} - \mathbf{s}) \cdot \left(\mathbf{x}^{(\mathcal{B})} - \mathbf{s}\right)\right| = O(\sqrt{\epsilon_{\mathrm{NASH}}}).$$

Because $\Pi(\cdot)$ is a constant relative distance error correcting code, $\|\mathbf{s} - \mathbf{t}\|_M = \Theta(1)$. Therefore,

$$\left|\beta^{(\mathbf{a})}_{(\mathbf{s} \to \mathbf{t})} - \beta_{(\mathbf{s} \to \mathbf{t})}\left(\mathbf{x}^{(\mathcal{B})}\right)\right|^2 = O(\epsilon_{\mathrm{NASH}}).$$

In particular, let

$$\mathbf{z}^{(\mathbf{a})} \triangleq \beta^{(\mathbf{a})}_{(\mathbf{s} \to \mathbf{t})} \mathbf{s}^{(\mathbf{a})} + \left(1 - \beta^{(\mathbf{a})}_{(\mathbf{s} \to \mathbf{t})}\right) \mathbf{t}^{(\mathbf{a})}$$

$$\mathbf{z} \triangleq \beta_{(\mathbf{s} \to \mathbf{t})}\left(\mathbf{x}^{(\mathcal{B})}\right) \mathbf{s} + \left(1 - \beta_{(\mathbf{s} \to \mathbf{t})}\left(\mathbf{x}^{(\mathcal{B})}\right)\right) \mathbf{t};$$

then for a $(1 - O(\epsilon_{\mathrm{NASH}}))$-fraction of $r$'s, we have

$$\left\|\mathbf{z}^{(\mathbf{a})} - \mathbf{z}\right\|^2_{J(A,r)} = O(\epsilon_{\mathrm{NASH}}).$$

By the triangle inequality, also:

$$\left|\left\|\mathbf{x}^{(\mathbf{a})} - \mathbf{z}^{(\mathbf{a})}\right\|_{J(A,r)} - \left\|\mathbf{x}^{(\mathcal{B})} - \mathbf{z}\right\|_{J(A,r)}\right|^2 = O(\epsilon_{\mathrm{NASH}}),$$

and thus by the Good Sample guarantee also

$$\left|\left\|\mathbf{x}^{(\mathbf{a})} - \mathbf{z}^{(\mathbf{a})}\right\|_{J(A,r)} - \left\|\mathbf{x}^{(\mathcal{B})} - \mathbf{z}\right\|_M\right|^2 = O(\epsilon_{\mathrm{NASH}}).$$

Finally, whenever $\mathbf{a}$ is $\epsilon_{\mathrm{NASH}}$-optimal and satisfies all the above, we have by Lipschitz continuity that:

$$\left\|f^{(\mathbf{a})} - f\left(\mathbf{x}^{(\mathcal{B})}\right)\right\|^2_{J(A,r)} = O(\epsilon_{\mathrm{NASH}}).$$

$\square$

## 7.8 Close to a line (2)

There are a few different scenarios where we expect the game verifier to accept both proofs:

- $\mathbf{x}^{(\mathcal{B})}$ may still be far from every Brouwer line segment; e.g. because the corresponding vertices are not neighbors in the LOCALENDOFALINE graph, or the values on $M_{3,4}$ don't match.

- $\mathbf{x}^{(\mathcal{B})}$ may be close to the "first" Brouwer line segment, where only the special direction ($M_4$) changes.

- $\mathbf{x}^{(\mathcal{B})}$ may be close to a Brouwer line segment where only the Compute-vs-Copy bit ($M_3$) changes.

- or $\mathbf{x}^{(\mathcal{B})}$ may be close to a Brouwer vertex.

In the first case, simply apply the default displacement. In this subsection we briefly describe the two cases corresponding to $\mathbf{x}^{(\mathcal{B})}$ close to a single Brouwer line segment. Finally, the case where $\mathbf{x}^{(\mathcal{B})}$ is close to a Brouwer vertex is deferred to Subsection 7.9.

**Close to the first line**

Near the "first" Brouwer line segment, locally computing the displacement is relatively simple. First observe that this case is easy to recognize by local access: by the Good Sample guarantee from Proposition 6.1, we can estimate the hamming distance of the first $2m + 1$ coordinates to $(\Pi(u_\mathbf{0}), \Pi(u_\mathbf{0}), 0)$. Furthermore we know $\mathbf{s} = \mathbf{z}_2$ and $\mathbf{t} = (\Pi(u_\mathbf{0}), \Pi(u_\mathbf{0}), 0, 0)$ exactly, and therefore also $\|\mathbf{t} - \mathbf{s}\|_M = 1$. Finally, let $\mathbf{z}^{(\mathbf{a})}$ be equal to $(\Pi(u_\mathbf{0}), \Pi(u_\mathbf{0}), 0)$ on its first $2m + 1$ coordinates, and $\mathbf{x}^{(\mathbf{a})} |_{M_4}$ on the last one.

**Close to a line that updates the auxiliary compute-vs-copy bit**

Suppose that $\mathbf{x}^{(\mathcal{B})}$ is close to a Brouwer line segment from $\mathbf{s} = (\Pi(u), \Pi(v), 0, 0)$ to $\mathbf{t} = (\Pi(u), \Pi(v), 1, 0)$, for $u, v \in V^{\text{LOCAL}}$ such that $v$ is the LOCALENDOFALINE-successor of $u$. (The case of a Brouwer line segment from $(\Pi(v), \Pi(v), 1, 0)$ to $(\Pi(v), \Pi(v), 0, 0)$ follows with minor modifications.)

By Lemma 7.11, we can locally decode both $\Pi(u)$ and $\Pi(v)$ (and verify that they are valid proofs of consecutive vertices in the LOCALENDOFALINE graph). We can therefore construct partial vectors $\mathbf{s}^{(\mathbf{a})}, \mathbf{t}^{(\mathbf{a})}$ which are equal, for a $(1 - O(\epsilon_{\text{NASH}}))$-fraction of $(A, r)$'s, to the restrictions of the true $\mathbf{s}, \mathbf{t}$ to $J^{(A,r)}$. Finally, let

$$\mathbf{z}^{(\mathbf{a})} \triangleq \left(\mathbf{x}^{(\mathbf{a})} |_{M_3}\right) \mathbf{t}^{(\mathbf{a})} + \left(1 - \mathbf{x}^{(\mathbf{a})} |_{M_3}\right) \mathbf{s}^{(\mathbf{a})}.$$

## 7.9 Close to a vertex

In this subsection we consider the case where $\mathbf{x}^{(\mathcal{B})}$ is close to a Brouwer vertex representing two different LOCALENDOFALINE vertices. In particular, we assume that it is close to a Brouwer vertex of the form $\mathbf{y} = (\Pi(u), \Pi(v), 1, 0)$. The case of $(\Pi(u), \Pi(v), 0, 0)$ follows with minor modifications; we will return to the cases of $(\Pi(u), \Pi(u), 0, 0)$ and $(\Pi(v), \Pi(v), 1, 0)$ in a couple of paragraphs.

After the game verifier accepts both proofs and that $v$ is the successor of $u$ in the LOCALENDOFALINE graph, we are assured that $\mathbf{x}^{(\mathcal{B})}$ is indeed close to some Brouwer vertex $\mathbf{y}$. Furthermore, we know that there is an incoming Brouwer line segment from $\mathbf{s} = (\Pi(u), \Pi(v), 1, 0)$ and an outgoing Brouwer line segment to $\mathbf{t} = (\Pi(v), \Pi(v), 1, 0)$. We can locally compute $\mathbf{s}, \mathbf{y}, \mathbf{t}$ on $J^{(A,r)}$ with high probability; denote the resulting partial vectors $\mathbf{s}^{(\mathbf{a})}, \mathbf{y}^{(\mathbf{a})}, \mathbf{t}^{(\mathbf{a})}$.

Alternatively, consider the case where $\mathbf{y} = (\Pi(v), \Pi(v), 1, 0)$ (similarly for $(\Pi(u), \Pi(u), 0, 0)$), with an incoming Brouwer line segment from $\mathbf{s} = (\Pi(P^{\text{LOCAL}}(v)), \Pi(v), 1, 0)$ and outgoing Brouwer line segment to $\mathbf{t} = (\Pi(v), \Pi(v), 0, 0)$. By the Error-correction guarantee in Lemma 7.11, we can (with high probability over $r$) locally decode partial vectors $\mathbf{y}^{(\mathbf{a})}$ and $\mathbf{t}^{(\mathbf{a})}$ on $J^{(A,r)}$; furthermore, by the Extrapolation guarantee in the same lemma, we can locally compute $\mathbf{s}^{(\mathbf{a})}$.

It is left to show that we can also locally compute the more involved construction of displacement next to a Brouwer vertex.

We know $\|\mathbf{y} - \mathbf{t}\|_M = 1/2$ exactly, and we can estimate $\|\mathbf{s} - \mathbf{y}\|_M$ from $\|\mathbf{s}^{(\mathbf{a})} - \mathbf{y}^{(\mathbf{a})}\|_{J^{(A,r)}}$. Now, we can locally compute $\mathbf{z}_{(\mathbf{s} \to \mathbf{y})}$ and $\mathbf{z}_{(\mathbf{y} \to \mathbf{t})}$ on the coordinates $J^{(A,r)}$ for which we know the value of $\mathbf{s}, \mathbf{y}, \mathbf{t}$; denote those partial vectors $\mathbf{z}^{(\mathbf{a})}_{(\mathbf{s} \to \mathbf{y})}$ and $\mathbf{z}^{(\mathbf{a})}_{(\mathbf{y} \to \mathbf{t})}$, respectively.

Recall that

$$\Delta_{(\mathbf{s} \to \mathbf{y})}(\mathbf{x}) = \frac{(\mathbf{y} - \mathbf{s})}{\|\mathbf{s} - \mathbf{y}\|_M} \cdot (\mathbf{x} - \mathbf{s}) - \left(1 - \sqrt{h}\right)$$

$$\Delta_{(\mathbf{y} \to \mathbf{t})}(\mathbf{x}) = \sqrt{h} - \frac{(\mathbf{t} - \mathbf{y})}{\|\mathbf{y} - \mathbf{t}\|_M} \cdot (\mathbf{x} - \mathbf{y}).$$

We can use $\mathbf{x}^{(\mathbf{a})}, \mathbf{s}^{(\mathbf{a})}, \mathbf{y}^{(\mathbf{a})}, \mathbf{t}^{(\mathbf{a})}$ to locally compute estimates $\Delta^{(\mathbf{a})}_{(\mathbf{s} \to \mathbf{y})}(\mathbf{x})$ and $\Delta^{(\mathbf{a})}_{(\mathbf{y} \to \mathbf{t})}(\mathbf{x})$, and therefore also $\alpha^{(\mathbf{a})} \approx \alpha\left(\mathbf{x}^{(\mathcal{B})}\right)$ and $\mathbf{z}^{(\mathbf{a})} \triangleq \alpha^{(\mathbf{a})} \mathbf{z}^{(\mathbf{a})}_{(\mathbf{s} \to \mathbf{y})} + \left(1 - \alpha^{(\mathbf{a})}\right) \mathbf{z}^{(\mathbf{a})}_{(\mathbf{y} \to \mathbf{t})}$.

**Lemma 7.15.** *Let* $(\mathcal{A}, \mathcal{B})$ *be an* $(\epsilon_{\text{NASH}}, \epsilon_{\text{NASH}})$*-Well-Supported-WeakNash, and let* $\|\mathbf{x}^{(\mathcal{B})} - (\Pi(u), \Pi(v), 1, 0)\|_M \leq 2\sqrt{h}$ *for* $u, v \in V^{\text{LOCAL}}$ *such that* $v = S^{\text{LOCAL}}(u)$ *and* $u = P^{\text{LOCAL}}(v)$. *Then* $\|f^{(\mathbf{a})} - f(\mathbf{x}^{(\mathcal{B})})\|_M^2 = O(\epsilon_{\text{NASH}}/h)$.

*Proof.* By Lemmata 7.11 and 7.10, we have that for a $(1 - O(\epsilon_{\text{NASH}}))$-fraction of $(A, r)$'s and every strategy $\mathbf{a}$ in their supports, the game verifier recognizes that $\mathbf{x}^{(\mathbf{a})}$ is close to some Brouwer vertex $\mathbf{y} \triangleq (\Pi(u), \Pi(v), 1, 0)$. In particular, by the Error-correction desideratum in Lemma 7.11, the game verifier can construct $\mathbf{y}^{(\mathbf{a})}, \mathbf{s}^{(\mathbf{a})}, \mathbf{t}^{(\mathbf{a})} \in \{0, 1, \bot\}^M$ that, for a $(1 - O(\epsilon_{\text{NASH}}))$-fraction of $(A, r)$'s and every strategy $\mathbf{a}$ in their supports, are equal to $\mathbf{y}, \mathbf{s}, \mathbf{t}$ on $J^{(A,r)}$.

We also know $\|\mathbf{y} - \mathbf{t}\|_M = 1/2$ exactly, and by the Good Sample guarantee from Proposition 6.1,

$$\left|\|\mathbf{y}^{(\mathbf{a})} - \mathbf{t}^{(\mathbf{a})}\|_{J^{(A,r)}} - \|\mathbf{y} - \mathbf{t}\|_M\right|^2 = o(1).$$

Furthermore, as we argue in the proof of Lemma 7.14,

$$\left|\beta^{(\mathbf{a})}_{(\mathbf{s} \to \mathbf{y})} - \beta_{(\mathbf{s} \to \mathbf{y})}\left(\mathbf{x}^{(\mathcal{B})}\right)\right|^2 = O(\epsilon_{\text{NASH}})$$

$$\left|\beta^{(\mathbf{a})}_{(\mathbf{y} \to \mathbf{t})} - \beta_{(\mathbf{y} \to \mathbf{t})}\left(\mathbf{x}^{(\mathcal{B})}\right)\right|^2 = O(\epsilon_{\text{NASH}}). \tag{35}$$

Let $\Delta^{(\mathbf{a})}_{(\mathbf{s} \to \mathbf{y})} \triangleq \beta^{(\mathbf{a})}_{(\mathbf{s} \to \mathbf{y})} - \left(1 - \sqrt{h}\right)$, and $\Delta^{(\mathbf{a})}_{(\mathbf{y} \to \mathbf{t})} \triangleq \sqrt{h} - \beta^{(\mathbf{a})}_{(\mathbf{y} \to \mathbf{t})}$. If either quantity is negative, continue with the displacement close to a line as in Subsection 7.7. Recall that $f$ is $O(1)$-Lipschitz on $[-1, 2]^M$, and in particular near the interface between the line and vertex displacements (the hyperplanes defined by $\beta_{(\mathbf{s} \to \mathbf{y})}(\mathbf{x}) = 0$ and $\beta_{(\mathbf{y} \to \mathbf{t})}(\mathbf{x}) = 0$). Therefore, whenever all the parameters are computed approximately correctly, the displacement is also approximately correct - even if near the interface we use the vertex displacement instead of the line displacement or vice versa. We henceforth focus on $\Delta^{(\mathbf{a})}_{(\mathbf{s}\to\mathbf{y})}, \Delta^{(\mathbf{a})}_{(\mathbf{y}\to\mathbf{t})} \geq 0$.

Define
$$\alpha^{(\mathbf{a})} \triangleq \frac{\Delta^{(\mathbf{a})}_{(\mathbf{y}\to\mathbf{t})}}{\Delta^{(\mathbf{a})}_{(\mathbf{y}\to\mathbf{t})} + \Delta^{(\mathbf{a})}_{(\mathbf{s}\to\mathbf{y})}},$$

and finally also
$$\begin{aligned}
\mathbf{z}^{(\mathcal{B})} &\triangleq \alpha\left(\mathbf{x}^{(\mathcal{B})}\right) \mathbf{z}_{(\mathbf{s}\to\mathbf{y})} + \left(1 - \alpha\left(\mathbf{x}^{(\mathcal{B})}\right)\right) \mathbf{z}_{(\mathbf{y}\to\mathbf{t})} \\
\mathbf{z}^{(\mathbf{a})} &\triangleq \alpha^{(\mathbf{a})} \mathbf{z}^{(\mathbf{a})}_{(\mathbf{s}\to\mathbf{y})} + \left(1 - \alpha^{(\mathbf{a})}\right) \mathbf{z}^{(\mathbf{a})}_{(\mathbf{y}\to\mathbf{t})}.
\end{aligned} \tag{36}$$

We now consider two different cases depending on the value of $\left\|\mathbf{x}^{(\mathcal{B})} - \mathbf{z}^{(\mathcal{B})}\right\|_M$:

**Case $\left\|\mathbf{x}^{(\mathcal{B})} - \mathbf{z}^{(\mathcal{B})}\right\|_M > 10h$:**

The challenge is that when $\left\|\mathbf{x}^{(\mathcal{B})} - \mathbf{z}^{(\mathcal{B})}\right\|_M$ is huge, $\Delta_{(\mathbf{y}\to\mathbf{t})} + \Delta_{(\mathbf{s}\to\mathbf{y})}$ may be very small, which could lead to $\alpha^{(\mathbf{a})}$ being far from the true $\alpha\left(\mathbf{x}^{(\mathcal{B})}\right)$. Fortunately, in this case the true $\hat{g}\left(\mathbf{x}^{(\mathcal{B})}\right)$ is just the default displacement, $\delta\left(\mathbf{0}_{2m+1}, 1\right)$ - so we only have to argue that $f^{(\mathbf{a})}$ also applies the default displacement (for most players $(A, r)$).

Observe that if $\alpha\left(\mathbf{x}^{(\mathcal{B})}\right) = 1/2$, the point on $L_\mathbf{y}$ which is closest to $\mathbf{x}^{(\mathcal{B})}$ is indeed $\mathbf{z}^{(\mathcal{B})}$. In general, this is not true, but $\left\|\mathbf{x}^{(\mathcal{B})} - \mathbf{z}^{(\mathcal{B})}\right\|_M$ is at most $\sqrt{2}$-times larger than the distance from $\mathbf{x}^{(\mathcal{B})}$ to $L_\mathbf{y}$. In particular, for any $\mathbf{z}^{(\alpha^{(\mathbf{a})})} \triangleq \alpha^{(\mathbf{a})} \mathbf{z}_{(\mathbf{s}\to\mathbf{y})} + \left(1 - \alpha^{(\mathbf{a})}\right) \mathbf{z}_{(\mathbf{y}\to\mathbf{t})}$, we have that $\left\|\mathbf{x}^{(\mathcal{B})} - \mathbf{z}^{(\alpha^{(\mathbf{a})})}\right\|_M > 7h$.

Therefore by the Good Sample guarantee from Proposition 6.1 and Claim 7.5, for a $(1 - O\left(\epsilon_{\text{NASH}}\right))$-fraction of $(A, r)$'s and every strategy $\mathbf{a}$ in their supports, we have
$$\left\|\mathbf{x}^{(\mathbf{a})} - \mathbf{z}^{(\mathbf{a})}\right\|_{J^{(A,r)}} > 7h - O\left(\epsilon_{\text{NASH}}\right);$$

For all those $\mathbf{a}$'s, $f^{(\mathbf{a})}$ correctly implements the default displacement.

**Case $\left\|\mathbf{x}^{(\mathcal{B})} - \mathbf{z}^{(\mathcal{B})}\right\|_M \leq 10h$:**

From (35),
$$\begin{aligned}
\left|\Delta^{(\mathbf{a})}_{(\mathbf{s}\to\mathbf{y})} - \Delta_{(\mathbf{s}\to\mathbf{y})}\left(\mathbf{x}^{(\mathcal{B})}\right)\right|^2 &= O\left(\epsilon_{\text{NASH}}\right) \\
\left|\Delta^{(\mathbf{a})}_{(\mathbf{y}\to\mathbf{t})} - \Delta_{(\mathbf{y}\to\mathbf{t})}\left(\mathbf{x}^{(\mathcal{B})}\right)\right|^2 &= O\left(\epsilon_{\text{NASH}}\right).
\end{aligned}$$

Plugging into (4), we have that
$$\Delta^{(\mathbf{a})}_{(\mathbf{s}\to\mathbf{y})} + \Delta^{(\mathbf{a})}_{(\mathbf{y}\to\mathbf{t})} \geq \sqrt{h} - O(h),$$

and therefore also
$$\left|\alpha^{(\mathbf{a})} - \alpha\left(\mathbf{x}^{(\mathcal{B})}\right)\right|^2 = O\left(\epsilon_{\text{NASH}}/h\right).$$

Similarly, also[10]

$$\left\|\mathbf{z}^{(\mathbf{a})} - \mathbf{z}^{(\mathcal{B})}\right\|^2_{J^{(A,r)}} = O\left(\epsilon_{\text{NASH}}/h\right).$$

Finally, whenever $\mathbf{a}$ is $\epsilon_{\text{NASH}}$-optimal and satisfies all the above, we have

$$\left\|f^{(\mathbf{a})} - f\left(\mathbf{x}^{(\mathcal{B})}\right)\right\|^2_{J^{(A,r)}} = O\left(\epsilon_{\text{NASH}}/h\right).$$

□

## 8 From polymatrix to bimatrix

In this section we complete the proof of our main result, Theorem 1.2, by reducing from multiplayer polymatrix games (a-la Theorem 7.1) to two-player games.

### 8.1 From $(\sqrt{\epsilon} + \epsilon, \delta)$-Well-Supported-WeakNash to $(\epsilon, \delta)$-WeakNash

**Lemma 8.1.** *Consider a complete bipartite polymatrix game such that every two adjacent vertices play a bimatrix subgame with payoffs in $[0, 1/n_B]$, $[0, 1/n_A]$, where $n_A, n_B$ are respective the numbers of players on the two sides of the bipartite game. Given an $(\epsilon, \delta)$-WeakNash, we can construct in polynomial time a $(\sqrt{\epsilon} \cdot (\sqrt{\epsilon} + 5), \delta)$-Well-Supported-WeakNash.*

*Proof.* Let $(V_A; V_B)$ be the sets of players, where each $v \in V_A \cup V_B$ has utility $U^v$ and action set $\mathcal{S}^v$. Let $\mathbf{x} = (x_s^v) \in \Delta\left(\times_{v \in V_A \cup V_B}\mathcal{S}^v\right)$ be an $(\epsilon, \delta)$-WeakNash. Finally, let $U^v_{\max}(\mathbf{x}^{-v}) = \max_{s \in \mathcal{S}^v} U^v_s(\mathbf{x}^{-v})$. Since this is a polymatrix game, we can write

$$\forall v \in V_A \ U^v_s(\mathbf{x}) = \sum_{u \in V_B} U^{v,u}_s(\mathbf{x}^u),$$

$$\forall v \in V_B \ U^v_s(\mathbf{x}) = \sum_{u \in V_A} U^{v,u}_s(\mathbf{x}^u).$$

Let $k = k(\epsilon) > 0$ be some large number do be specified later. We construct our new approximate equilibrium as follows. For each of the $(1 - \delta)$ players who play $\epsilon$-optimally in $\mathbf{x}$, we take only the strategies that are within $\epsilon k$ of the optimum:

$$\hat{x}_s^v = \begin{cases} \frac{x_s^v}{1-z^v} & \text{if } U^v_s(\mathbf{x}^{-v}) \geq U^v_{\max}(\mathbf{x}^{-v}) - \epsilon k \\ 0 & \text{otherwise} \end{cases}$$

where $z^v$ is the total probability that player $v$ assigns to strategies that are more than $\epsilon k$ away from the optimum.

The above division is well-defined because for $k > 1$ and $v$ who plays $\epsilon$-optimally, $z^v$ is bounded away from 1. Moreover, the following claim from [32] formalizes the intuition that when $k$ is sufficiently large, the total weight on actions removed is small, so $\hat{\mathbf{x}}^v$ is close to $\mathbf{x}^v$:

*Claim* 8.2. ([32, Claim 6])

$$\forall v \in V_A \cup V_B \ \sum_{s \in \mathcal{S}^v} |\hat{x}_s^v - x_s^v| \leq \frac{2}{k-1}$$

---

[10]In fact, we actually have $\left\|\mathbf{z}^{(\mathbf{a})} - \mathbf{z}^{(\mathcal{B})}\right\|^2_{J^{(A,r)}} = O(\epsilon_{\text{NASH}})$, because $\alpha$ interpolates between $\mathbf{z}_{(\mathbf{s} \to \mathbf{y})}$ and $\mathbf{z}_{(\mathbf{y} \to \mathbf{t})}$, which are already at distance $O\left(\sqrt{h}\right)$ from each other.

Now, the total change to the expected payoff of player $v$ for each action $s$, is bounded by the total change in mixed strategies of its neighbors. In the following let $v \in V_A$; the analogous argument for the utility of players in $V_B$ follows with minor modifications.

$$\left|U_s^v\left(\mathbf{x}^{-v}\right) - U_s^v\left(\hat{\mathbf{x}}^{-v}\right)\right| \leq \sum_{u \in V_B} |U_s^{v,u}(\mathbf{x}^u) - U_s^{v,u}(\hat{\mathbf{x}}^u)|$$

$$\leq \frac{1}{n_B} \cdot \sum_{u \in V_B} \sum_{s \in \mathcal{S}^u} |\hat{x}_s^u - x_s^u| \leq \frac{2}{k-1}$$

It follows that $\hat{\mathbf{x}}$ is a $\left(k\epsilon + \frac{2}{k-1}, \delta\right)$-Well-Supported-WeakNash:

$$U_s^v(\hat{\mathbf{x}}^{-v}) \geq U_s^v(\mathbf{x}^{-v}) - \frac{2}{k-1} \geq U_{\max}^v(\mathbf{x}^{-v}) - \epsilon k - \frac{2}{k-1} \geq U_{\max}^v(\hat{\mathbf{x}}^{-v}) - \epsilon k - \frac{4}{k-1}$$

Finally, take $k = 1 + 1/\sqrt{\epsilon}$ to get that

$$k\epsilon + \frac{4}{k-1} \leq \sqrt{\epsilon} \cdot (\sqrt{\epsilon} + 5)$$

$\square$

## 8.2 From $(\epsilon, \delta)$-WeakNash to $\Theta(\epsilon \cdot \delta)$-ANE in a bimatrix game

**Lemma 8.3.** *Suppose we are given a complete bipartite polymatrix game between $n^{1/2+o(1)}$ vertices with $2^{n^{1/2+o(1)}}$ actions each; the payoffs in each bimatrix game are in $[0, 1/n_B], [0, 1/n_A]$, where $n_A, n_B$ denote the number of players on each side of the bipartite graph. Then we can construct a bimatrix game of size $2^{n^{1/2+o(1)}}$ such that every $\epsilon$-ANE of the new bimatrix game corresponds to a $(\delta, \epsilon_{\text{POLYMATRIX}})$-WeakNash of the polymatrix game, for sufficiently small $\epsilon = \Theta\left(\delta^2 \cdot \epsilon_{\text{POLYMATRIX}}^2\right)$*

*Proof.* The two players play three games simultaneously: the main game, which is the heart of the reduction; and two games based on a construction due to Althofer [4], which impose structural properties of any approximate Nash equilibrium.

**Main game** We let each of the two players "control" the vertices on one side of the bipartite graphical game.

Alice's actions correspond to a choice of vertex on her side of the polymatrix game, and an action for that player. Similarly, Bob's actions correspond to a a choice of vertex on his side, and a choice of action for that vertex. The utility for each player is the utility to the corresponding vertex from the induced subgame, scaled by a factor of $\lambda \cdot n_B$ or $\lambda \cdot n_A$ for Alice or Bob, respectively. Here $\lambda \gg \epsilon$ is a small constant do be defined later.

**Althofer games** In addition to the main game, we use a construction due to Althofer [4] to introduce two auxiliary games that force the player to spread their mixed strategies approximately evenly across all vertices. Althofer's original game is a one-sum game with payoffs in $\{0, 1\}$ and size $k \times \binom{k}{k/2}$. In each column, exactly half of the entries are 1's, and the rest are 0's. For example for $k = 4$, the payoff matrices are given by:

$$R = 1 - C = \begin{pmatrix} 1 & 1 & 1 & 0 & 0 & 0 \\ 1 & 0 & 0 & 0 & 1 & 1 \\ 0 & 1 & 0 & 1 & 0 & 1 \\ 0 & 0 & 1 & 1 & 1 & 0 \end{pmatrix}$$

The value of this game is $1/2$.

For our purposes, we consider two instantiations of Althofer's gadget: a "primal" Althofer game of size $n_A \times \binom{n_A}{n_A/2}$. and a "dual" Althofer game of size $\binom{n_B}{n_B/2} \times n_B$. In any (approximate) Nash, both Alice and Bob must mix (approximately) evenly among (almost) all of their actions.

**The final construction** Finally, we compose all three games together by identifying Alice's choice of vertex in the main game with her choice of row in the primal Althofer game, and Bob's choice of vertex in the main game with his choice of column in the dual Althofer game.

**Analysis** Alice's and Bob's respective mixed strategies induce a mixed strategy profile $(\mathcal{A}, \mathcal{B})$ on their vertices: Alice's vertex $(A, i)$'s action is drawn from Alice's action conditioned on picking $(A, i)$, and analogously for Bob's $(B, j)$. For any vertex that is never picked by its player, fix an arbitrary strategy. Our goal is to show that $(\mathcal{A}, \mathcal{B})$ is a $(\delta, \epsilon_{\text{POLYMATRIX}})$-WeakNash of the polymatrix game.

Let $U^{(A,i)}(\mathcal{A}[(A,i)], \mathcal{B})$ denote $(A, i)$'s expected utility when all players draw their strategies according to $(\mathcal{A}, \mathcal{B})$, and analogously for $U^{(B,j)}(\mathcal{B}[(B,j)], \mathcal{A})$.

By Lemma 2.1, in every $\lambda$-ANE (and in particular in any $\epsilon$-ANE), Alice's and Bob's respective marginal distributions over their vertices are $O(\lambda)$-close to uniform. Therefore, Alice's expected utilities from the main game satisfy

$$\left| U^A_{\text{MAIN}} - \lambda \mathsf{E}_{i \in [n_A]} U^{(A,i)}(\mathcal{A}[(A,i)], \mathcal{B}) \right| = O(\lambda^2).$$

Suppose by contradiction that Alice and Bob are in an $\epsilon$-ANE in the bimatrix game, but a $\delta$-fraction of Alice's vertices have an $\epsilon_{\text{POLYMATRIX}}$-deviating strategy in the polymatrix game. Let $(\widehat{\mathcal{A}}, \mathcal{B})$ denote the deviating mixed strategy profile (notice that the deviation of a player on Alice's side does not affect the utilities of other player on the same side). $(\widehat{\mathcal{A}}, \mathcal{B})$ induces, without changing the marginal distributions over vertices, a profile of mixed strategies for Alice and Bob; Alice's new expected utility from the main game, $\widehat{U^A_{\text{MAIN}}}$, also satisfies

$$\left| \widehat{U^A_{\text{MAIN}}} - \lambda \mathsf{E}_{i \in [n_A]} U^{(A,i)}(\widehat{\mathcal{A}}[(A,i)], \mathcal{B}) \right| = O(\lambda^2)$$

Therefore,

$$\begin{aligned}
\widehat{U^A_{\text{MAIN}}} &\geq \lambda \mathsf{E}_{i \in [n_A]} U^{(A,i)}(\widehat{\mathcal{A}}[(A,i)], \widehat{\mathcal{B}}) - O(\lambda^2) \\
&\geq \lambda \mathsf{E}_{i \in [n_A]} U^{(A,i)}(\mathcal{A}[(A,i)], \mathcal{B}) + \lambda \delta \cdot \epsilon_{\text{POLYMATRIX}} - O(\lambda^2) \\
&\geq U^A_{\text{MAIN}} + \lambda \delta \cdot \epsilon_{\text{POLYMATRIX}} - O(\lambda^2).
\end{aligned}$$

Set $\lambda \delta \cdot \epsilon_{\text{POLYMATRIX}} - O(\lambda^2) > \epsilon$. Since Alice and Bob have not changed their marginal vertex utilities, this is an $\epsilon$-deviating strategy. □